\documentclass[aip,jcp,amsmath,amssymb,twocolumn,superscriptaddress,floatfix,a4paper,reprint,longbibliography]{revtex4-2}
\usepackage[english]{babel}
\usepackage[utf8]{inputenc}
\usepackage{graphicx}
\usepackage{comment}
\usepackage{xcolor}
\usepackage[pdftex, pdftitle={Article}, pdfauthor={Author}]{hyperref} 

\begin{document}
\title{Fermionic-propagator and alternating-basis quantum Monte Carlo methods for correlated electrons on a lattice}

\author{Veljko Jankovi\'c}
    \email[]{veljko.jankovic@ipb.ac.rs}
    \affiliation{Institute of Physics Belgrade, University of Belgrade, Pregrevica 118, 11080 Belgrade, Serbia}
\author{Jak\v sa Vu\v ci\v cevi\'c}
    \email[]{jaksa.vucicevic@ipb.ac.rs}
    \affiliation{Institute of Physics Belgrade, University of Belgrade, Pregrevica 118, 11080 Belgrade, Serbia}

\begin{abstract}
Ultracold-atom simulations of the Hubbard model provide insights into the character of charge and spin correlations in and out of equilibrium.
The corresponding numerical simulations, on the other hand, remain a significant challenge.
We build on recent progress in the quantum Monte Carlo (QMC) simulation of electrons in continuous space, and apply similar ideas to the square-lattice Hubbard model.
We devise and benchmark two discrete-time QMC methods, namely the fermionic-propagator QMC (FPQMC) and the alternating-basis QMC (ABQMC).
In FPQMC, the time evolution is represented by snapshots in real space, whereas the snapshots in ABQMC alternate between real and reciprocal space.
The methods may be applied to study equilibrium properties within grand-canonical or canonical ensemble, external field quenches, and even the evolution of pure states.
Various real-space/reciprocal-space correlation functions are also within their reach.
Both methods deal with matrices of size equal to the number of particles (thus independent of the number of orbitals or time slices), which allows for cheap updates.
We benchmark the methods in relevant setups.
In equilibrium, the FPQMC method is found to have excellent average sign and, in some cases, yields correct results even with poor imaginary-time discretization.
ABQMC has significantly worse average sign, but also produces good results.
Out of equilibrium, FPQMC suffers from a strong dynamical sign problem.
On the contrary, in ABQMC, the sign problem is not time dependent.
Using ABQMC, we compute survival probabilities for several experimentally relevant pure states.
\end{abstract}

\maketitle

\section{Introduction}
Last two decades have witnessed remarkable developments in laser and ultracold-atom technologies that have enabled experimental studies of strongly correlated electrons in and out of equilibrium.~\cite{RevModPhys.86.779,CRPhysique.19.365}
Ultracold atoms in optical lattices~\cite{RevModPhys.80.885,CRPhysique.19.365} and optical-tweezers arrays~\cite{PhysRevLett.114.080402,PhysRevLett.128.223202,arxiv.2203.15023} have been used as quantum simulators of paradigmatic models of condensed-matter physics, such as the Hubbard model.~\cite{ProcRSocLondA.276.238,NatPhys.9.523}
Recent experiments with fermionic ultracold atoms have probed the equation of state,~\cite{PhysRevLett.116.175301} charge and spin correlation functions,~\cite{PhysRevX.7.031025,Science.353.1253,Science.353.1260,Science.365.251} as well as transport properties (by monitoring charge and spin diffusion~\cite{Science.363.379,NatPhys.8.213, NatCommun.10.1588,Science.363.383}).

These experimental achievements pose a significant challenge for the theory.
The level of difficulty, however, greatly depends on whether one computes instantaneous (equal-time) correlations, or the full time/frequency dependence of dynamical correlators.
The other factor is whether one considers thermal equilibrium or out-of-equilibrium setups.

Instantaneous correlators in equilibrium are the best-case scenario.
The average density, double occupancy, and correlations between particle or spin densities on adjacent sites, are still very important.
They serve as a thermometer: the temperature in cold-atom experiments cannot be measured directly, and is often gauged in comparison with numerical simulations.~\cite{Science.353.1253,Science.363.379}
For this kind of application, current state-of-the-art methods~\cite{PhysRevLett.47.1628,PhysRevB.26.5033,PhysRevD.24.2278,PhysRevB.24.4295,PhysRevLett.51.1900,PhysRevB.31.4403,PhysRevLett.56.2521,PhysRevB.40.506,PhysRevLett.74.3652,PhysRevLett.83.2777,PhysRevLett.90.136401,PhysRevB.80.075116,PhysRevX.5.041041,RevModPhys.83.349,JETPLett.80.61,PhysRevB.72.035122,EurophysLett.82.57003,PhysRevLett.81.2514,PhysRevLett.99.250201,EurophysLett.90.10004,PhysProcedia.6.95,PhysRevB.96.041105,PhysRevLett.97.187202,PhysRevA.84.053611,ComputPhysCommun.184.557,PhysRevResearch.4.043201} are often sufficient.
Equal-time multipoint density correlations are also of interest, as they hold information, e.g., about the emergence of string patterns~\cite{Science.365.251,PRXQuantum.2.020344,PhysRevB.104.075155} or the effect of holes on antiferromagnetic correlations.~\cite{Science.374.82,PhysRevResearch.3.033204,PhysRevA.106.013310}
However, measuring density at a larger number of points simultaneously is more difficult in many algorithms.
For example, in the Hirsch--Fye (HF) algorithm,~\cite{PhysRevLett.56.2521,PhD_Mikelsons} one cannot do this straight-forwardly, as the auxiliary Ising spins only distinguish between singly occupied and doubly occupied/empty sites.

Of even greater interest and much greater difficulty are the time-dependent correlations in equilibrium.
These pertain to studies of transport and hydrodynamics at the level of linear response theory.~\cite{PhysRevLett.123.036601,Science.366.6468,arxiv.2208.04047}
The limitations of current state-of-the-art methods here become starkly apparent.
If one is interested in long-wavelength behavior (as is precisely the case in the study of hydrodynamic properties~\cite{arxiv.2208.04047}), the lattices treated in the simulation need to be sufficiently large.
The finite-temperature Lanczos method~\cite{AdvPhys.49.1,Prelovsek2013} can only treat up to 20 lattice sites~\cite{PhysRevLett.123.036601,PhysRevB.102.115142} and is unsuitable for such applications.
QMC methods can treat up to 300 sites,~\cite{PhysRevB.80.075116,PhysRevX.5.041041} but only under certain conditions: doping away from half-filling leads to a significant sign problem which becomes more severe as the lattice size, inverse temperature, and coupling constant are increased.
Moreover, QMC methods are formulated in imaginary time, and require the ill-defined analytical continuation to reconstruct optical conductivity or any other real-frequency spectrum.~\cite{Science.366.6468}

Direct real-time calculations, regardless of proximity to the equilibrium, are the most difficult.~\cite{PhysRevB.79.035320,PhysRevB.79.153302,PhysRevB.81.085126,RevModPhys.83.349,PhysRevB.84.085134,PhysRevB.89.115139,PhysRevLett.115.266802,JChemPhys.154.184103,PhysRevB.90.155104,PhysRevB.95.165139,PhysRevB.85.235121,PhysRevB.90.125112,PhysRevLett.109.266403,PhysRevB.88.035129,PhysRevB.91.245154}
These present an alternative to analytical continuation in equilibrium calculations, but are necessary to describe non-equilibrium regimes, e.g., in external field quenches.~\cite{NatCommun.10.1588}
In the corresponding Kadanoff--Baym--Keldysh three- or two-piece contour formalism, QMC computations are plagued by the dynamical sign problem, that was so far overcome in only the smallest systems.~\cite{PhysRevLett.115.266802}
The time-dependent density matrix renormalization group~\cite{PhysRevLett.88.256403,PhysRevLett.93.076401,JPhysSocJpn.74.246} produces practically exact results, however, only in one-dimensional~\cite{PhysRevB.90.155104,PhysRevB.95.165139} or ladder systems.~\cite{PhysRevB.93.081107}
Simulations based on the nonequilibrium Green's functions formalism~\cite{Stefanucci-van-Leeuwen-book} are also possible, but are not numerically exact.
They can, however, treat much larger systems and much longer time scales than other real-time approaches.~\cite{PhysRevB.95.165139,JPhysCondensMatter.32.103001,PhysRevLett.124.076601,PhysRevB.105.165155}

The main goal of this work is to construct a numerically exact way to compute spatially resolved densities, in setups relevant for optical-lattice experiments.
This includes general multipoint correlators in real space and momentum space, and we focus on densities of charge and spin.
We are interested in both the equilibrium expectation values, and their time dependence in transient regimes.
(The latter can formally be used to access temporal correlations in equilibrium, as well.)

We take a largely unexplored QMC route~\cite{PhysRevLett.46.77,JStatPhys.27.731,PhysRep.127.233}
towards computation of correlation functions in the square-lattice Hubbard model.
Current state-of-the-art methods such as the continuous-time interaction-expansion (CT-INT),~\cite{RevModPhys.83.349,JETPLett.80.61,PhysRevB.72.035122} the continuous-time auxiliary-field (CT-AUX),~\cite{RevModPhys.83.349,EurophysLett.82.57003} and HF,~\cite{PhysRevLett.56.2521,PhD_Mikelsons} rely on computation of large matrix determinants, which in many cases presents the bottleneck of the algorithm.
In CT-INT and CT-AUX, the size of the matrices generally grows with coupling, inverse temperature, and lattice size.
In the HF, which is based on the Suzuki--Trotter decomposition (STD), the matrix size is fixed, yet presents a measure of the systematic error: the size of the matrix scales with both the number of time slices and the number of lattice sites.
A rather separate approach is possible where the size of the matrices scales only with the total number of electrons.
This approach builds on the path-integral MC (PIMC).~\cite{RevModPhys.67.279,ComputPhysCommun.63.415}
In PIMC, trajectories of individual electrons are tracked.
In continuous-space models, PIMC was used successfully even in the calculation of dynamical response functions.\cite{PhysRevLett.121.255001,PhysRevB.102.125150}
The downside is that the antisymmetry of electrons feeds into the overall sign of a given configuration, thus contributing to the sign problem.
A more sophisticated idea was put forward in Refs.~\onlinecite{JPhysSocJpn.53.963,JPhysAMathGen.38.6659,JPhysAMathTheor.40.7157}.
Namely, the propagation between two time-slices can be described by a single many-fermion propagator, which groups (blocks) all possible ways the electrons can go from one set of positions to another---including all possible exchanges.
The many-fermion propagator is evaluated as a determinant of a matrix of the size equal to the total number of electrons.
This scheme automatically eliminates one important source of the sign problem, and improves the average sign drastically.
Such \emph{permutation blocking} algorithms have been utilized with great success to compute thermodynamic quantities in continuous-space fermionic models.~\cite{PlasmaPhysControlFusion.43.743,ContribPlasmaPhys.51.687,NewJPhys.17.073017,JChemPhys.143.204101,PhysRevB.93.085102,PhysRep.744.1,PhysRevE.100.023307,JPhysAMathTheor.54.335001,ContribPlasmaPhys.61.e202100112}
Here, we investigate and test analogous formulations in the \emph{lattice models} of interest, and try generalizing the approach to real-time dynamics.

We develop and test two slightly different QMC methods.
The FPQMC is a real-space method similar to the permutation-blocking and fermionic-propagator PIMC respectively developed by Dornheim et al.~\cite{NewJPhys.17.073017} and Filinov et al.~\cite{ContribPlasmaPhys.61.e202100112} for continuous models.
On the other hand, the ABQMC method is formulated simultaneously in both real and reciprocal space, which makes measuring distance- and momentum-resolved quantities equally simple; The motion of electrons and their interactions are treated on equal footing.
Both methods are based on the STD and are straight-forwardly formulated along any part of, or the entire Kadanoff--Baym--Keldysh three-piece contour.
This allows access to both real- and imaginary-time correlation functions in and out of equilibrium.
Unlike CT-INT, CT-AUX and HF, our methods can be also used to treat canonical ensembles, as well as time evolution of pure states.
Our formulation ensures that the measurement of an arbitrary multipoint charge or spin correlation function is algorithmically trivial and cheap.

We perform benchmarks on several examples where numerically exact results are available.

In calculations of instantaneous correlators for the 2D Hubbard model in equilibrium, our main finding is that FPQMC method has a rather manageable sign problem.
The average sign is anti-correlated with coupling strength, which is in sharp contrast with some of the standard QMC methods.
More importantly, we find that the average sign drops off relatively slowly with the lattice size and the number of time slices---we have been able to converge results with as many as 8 time slices, or as many as 80 lattice sites.
At strong coupling and at half-filling, we find the average sign to be very close to 1.
In the temperature range relevant for optical-lattice experiments, we find that the average occupancy can be computed to a high accuracy with as few as 2 time slices; the double occupancy and the instantaneous spin-spin correlations require somewhat finer time discretization, but often not more than 6 time-slices in total.
We also document that FPQMC appears to be sign-problem-free for Hubbard chains.

However, in calculations of the time-evolving and spatially resolved density, we find that the FPQMC sign problem is mostly prohibitive of obtaining results.
Nevertheless, employing the ABQMC algorithm, we are able to compute survival probabilities of various pure states on $4\times 4$ clusters---in the ABQMC formulation, the sign-problem is manifestly independent of time and interaction strength, and one can scan the entire time evolution for multiple strengths of interaction in a single run.
The numerical results reveal several interesting trends.
Similarly to observations made in Ref.~\onlinecite{arxiv.2112.02903}, we find in general that the survival probability decays over longer times, when interactions are stronger.
At shorter times, we observe that the behavior depends strongly on the type of the initial state, likely related to the average density.

The paper is organized as follows.
The FPQMC and ABQMC methods are developed in Sec.~\ref{Sec:model_method} and applied to equilibrium and out-of-equilibrium setups in Sec.~\ref{Sec:numerics_overall}.
Section~\ref{Sec:relation_to_other_algorithms} discusses the FPQMC and ABQMC methods in light of other widely used QMC algorithms.
Section~\ref{Sec:Conclusion} summarizes our main findings and their implications, and discusses prospects for further work.

\section{Model and method}
\label{Sec:model_method}
\subsection{Hubbard Hamiltonian}
\label{Sec:Hubbard_model}
We study the Hubbard model on a square-lattice cluster containing $N_c=N_xN_y$ sites under periodic boundary conditions (PBC). The Hamiltonian reads as
\begin{equation}
\label{Eq:Hubbard-canonical}
    H=H_0+H_\mathrm{int}.
\end{equation}
The noninteracting (single-particle) part $H_0$ of the Hamiltonian describes a band of itinerant electrons
\begin{equation}
\label{Eq:H_0-canonical}
        H_0=-J\sum_{ \langle\mathbf{r},\mathbf{r}'\rangle\sigma}c_{\mathbf{r}\sigma}^\dagger c_{\mathbf{r}'\sigma}
        =\sum_{\mathbf{k}\sigma}\varepsilon_\mathbf{k}\:n_{\mathbf{k}\sigma},
\end{equation}
where $J$ is the hopping amplitude between the nearest-neighboring lattice sites $\mathbf{r}$ and $\mathbf{r}'$, while the operators $c_{\mathbf{r}\sigma}^\dagger$ $(c_{\mathbf{r}\sigma})$ create (destroy) an electron of spin $\sigma$ on lattice site $\mathbf{r}$. Under PBC, $H_0$ is diagonal in the momentum representation; the wave vector $\mathbf{k}$ may assume any of the $N_c$ allowed values in the first Brillouin zone of the lattice. The free-electron dispersion is given by $\varepsilon_\mathbf{k}=-2J(\cos k_x+\cos k_y)$. The density operator is $n_{\mathbf{k}\sigma}=c_{\mathbf{k}\sigma}^\dagger c_{\mathbf{k}\sigma}$ with $c_{\mathbf{k}\sigma}=\sum_\mathbf{r}\langle\mathbf{k}|\mathbf{r}\rangle c_{\mathbf{r}\sigma}$. The Hamiltonian of the on-site Hubbard interaction (two-particle part) reads as
\begin{equation}
\label{Eq:H_int}
    H_\mathrm{int}=U\sum_{\mathbf{r}}n_{\mathbf{r}\uparrow} n_{\mathbf{r}\downarrow}
\end{equation}
where $U$ is the interaction strength, while $n_{\mathbf{r}\sigma}=c_{\mathbf{r}\sigma}^\dagger c_{\mathbf{r}\sigma}$.

If the number of particles is not fixed, Eq.~\eqref{Eq:Hubbard-canonical} additionally features the chemical-potential term $-\mu\sum_{\mathbf{r}\sigma}n_{\mathbf{r}\sigma}=-\mu\sum_{\mathbf{k}\sigma}n_{\mathbf{k}\sigma}$ that can be added to either $H_0$ or $H_\mathrm{int}$. Here, since we develop a coordinate-space QMC method, we add it to $H_\mathrm{int}$.

\subsection{FPQMC method}
\label{Sec:QMC_all}
Finding viable approximations to the exponential of the form $e^{-\alpha H}$ is crucial to many QMC methods.
With $\alpha=1/T$ ($T$ denotes temperature), this is the Boltzmann operator, which will play a role whenever the system is in thermal equilibrium.
In the formulation of dynamical responses, the time-evolution operator will also have this form, with $\alpha=it$, where $t$ is the (real) time. 
One possible way to deal with these is the lowest-order STD~\cite{CommunMathPhys.51.183}
\begin{equation}
\label{Eq:Suzuki-Trotter-general}
 e^{-\alpha H}\approx\left(e^{-\Delta\alpha H_0}e^{-\Delta\alpha H_\mathrm{int}}\right)^{N_\alpha}
\end{equation}
where the interval of length $|\alpha|$ is divided into $N_\alpha$ subintervals of length $|\Delta\alpha|$ each, where $\Delta\alpha=\alpha/N_\alpha$.
The error of the approximation is of the order of $|\Delta\alpha|^2\|[H_0,H_\mathrm{int}]\|$, where the norm $\|[H_0,H_\mathrm{int}]\|$ may be defined as the largest (in modulus) eigenvalue of the commutator $[H_0,H_\mathrm{int}]$.~\cite{PhysRep.127.233}
The error can in principle be made arbitrarily small by choosing $N_\alpha$ large enough.
However, the situation is complicated by the fact that the RHS of Eq.~\eqref{Eq:Suzuki-Trotter-general} contains both single-particle and two-particle contributions.
The latter are diagonal in the coordinate representation, so that the spectral decomposition of $e^{-\Delta\alpha H_\mathrm{int}}$ is performed in terms of totally antisymmetric states in the coordinate representation, aka the Fock states,
\begin{equation}
\label{Eq:Psi-i}
    |\Psi_i\rangle=\prod_\sigma\prod_{j=1}^{N_\sigma}c_{\mathbf{r}^\sigma_{j}\sigma}^\dagger|\emptyset\rangle
\end{equation}
that contain $N_\sigma$ electrons of spin $\sigma$ whose positions $\mathbf{r}^\sigma_{1},\dots,\mathbf{r}^\sigma_{N_\sigma}$ are ordered according to a certain rule.
We define $\varepsilon_\mathrm{int}(\Psi_i)\equiv\langle\Psi_i|H_\mathrm{int}|\Psi_i\rangle$.
On the other hand, the matrix element of $e^{-\Delta\alpha H_0}$ between many-body states $|\Psi'_i\rangle$ and $|\Psi_i\rangle$ can be expressed in terms of determinants of single-electron propagators
\begin{eqnarray}
\langle\Psi'_i|e^{-\Delta\alpha H_0}|\Psi_i\rangle=\prod_\sigma\det S(\Psi'_i,\Psi_i,\Delta\alpha,\sigma)\label{Eq:free-prop-1}\\
\left[S(\Psi'_i,\Psi_i,\Delta\alpha,\sigma)\right]_{j_1j_2}=\langle\mathbf{r'}_{j_1}^{\sigma}|e^{-\Delta\alpha H_0}|\mathbf{r}_{j_2}^\sigma\rangle\label{Eq:free-prop-2}.
\end{eqnarray}
We provide a formal proof of Eqs.~\eqref{Eq:free-prop-1} and~\eqref{Eq:free-prop-2} in Appendix~\ref{App:single-particle-propagator}.
The same equations lie at the crux of conceptually similar permutation-blocking~\cite{NewJPhys.17.073017} and fermionic-propagator~\cite{ContribPlasmaPhys.61.e202100112} PIMC methods, which are formulated for continuous-space models.
When $\Delta\alpha$ is purely real (purely imaginary), the quantity on the right-hand side of Eq.~\eqref{Eq:free-prop-2} is the imaginary-time (real-time) lattice propagator of a free particle in the coordinate representation.~\cite{PhysRep.127.233}
Its explicit expressions are given in Appendix~\ref{App:lattice_propagator}.

\begin{figure}[ht!]
 \centering
 \includegraphics{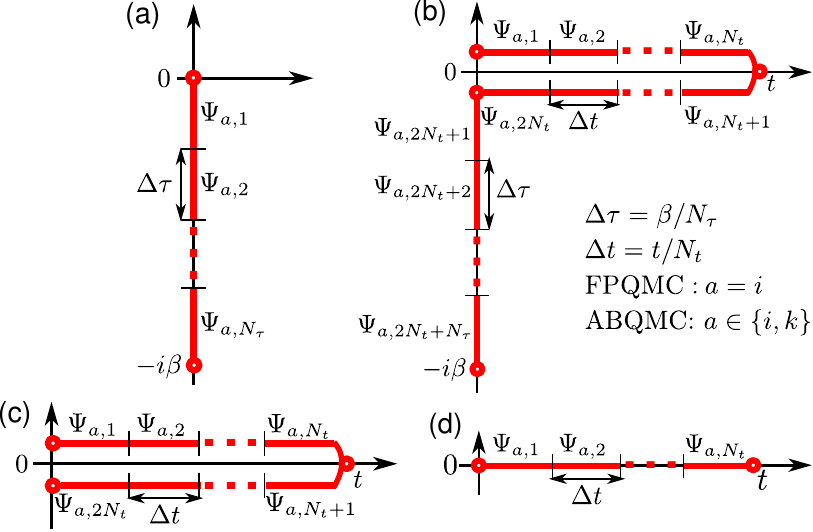}
 \caption{(Color online) Contours appropriate for computing (a) thermodynamic quantities at temperature $T=1/\beta$, (b) time-dependent quantities after quantum quench in which the Hamiltonian is suddenly changed from $H(0)$ at $t<0$ to $H$ at $t>0$, (c) time-dependent quantities during evolution from a pure state $|\psi(0)\rangle\equiv|\Psi_{i,1}\rangle$, (d) the survival probability of the initial pure state $|\psi(0)\rangle\equiv|\Psi_{i,1}\rangle$. In (a) and (b), the vertical part is divided into $N_\tau$ identical slices of length $\Delta\tau$. In (b)--(d), each horizontal line is divided into $N_t$ identical slices of length $\Delta t$.
 Within the FPQMC method, a many-body state in the coordinate representation $|\Psi_{i,l}\rangle$ [Eq.~\eqref{Eq:Psi-i}] is associated to each slice $l$. Within the ABQMC method, in addition to $|\Psi_{i,l}\rangle$, each slice $l$ features a many-body state in the momentum representation $|\Psi_{k,l}\rangle$ [Eq.~\eqref{Eq:Psi-k}].
 }
 \label{Fig:konture_040522}
\end{figure}

Further developments of the method somewhat depend on the physical situation of interest. We formulate the method first in equilibrium, and then in out-of-equilibrium situations. To facilitate discussion, in Figs.~\ref{Fig:konture_040522}(a)--\ref{Fig:konture_040522}(d) we summarize the contours appropriate for different situations we consider.

\subsubsection{FPQMC method for thermodynamic quantities}
\label{Sec:QMC_thermodynamics}
The equilibrium properties at temperature $T=\beta^{-1}$ follow from the partition function $Z\equiv\mathrm{Tr}\:e^{-\beta H}$, which may be computed by dividing the imaginary-time interval $[0,\beta]$ into $N_\tau$ slices of length $\Delta\tau\equiv\beta/N_\tau$, employing Eq.~\eqref{Eq:Suzuki-Trotter-general}, and inserting the spectral decompositions of $e^{-\Delta\tau H_\mathrm{int}}$.
The corresponding approximant for $Z$ reads as
\begin{equation}
    Z\approx\sum_{\mathcal{C}} \mathcal{D}_\beta(\mathcal{C},\Delta\tau)e^{-\Delta\tau\varepsilon_\mathrm{int}(\mathcal{C})}.
\end{equation}
The configuration
\begin{equation}
\label{Eq:def_C_eq}
\mathcal{C}=\left\{|\Psi_{i,1}\rangle,\dots,|\Psi_{i,N_\tau}\rangle\right\}    
\end{equation}
resides on the contour depicted in Fig.~\ref{Fig:konture_040522}(a) and consists of $N_\tau$ Fock states $|\Psi_{i,l}\rangle$ in the coordinate representation.
$\mathcal{D}_\beta(\mathcal{C},\Delta\tau)$ depends on the temperature and imaginary-time discretization through the imaginary-time step $\Delta\tau$
\begin{equation}
\label{Eq:D_C_eq}
\begin{split}
    \mathcal{D}_\beta(\mathcal{C},\Delta\tau)&\equiv\prod_{l=1}^{N_\tau}\langle\Psi_{i,l\oplus 1}|e^{-\Delta\tau H_0}|\Psi_{i,l}\rangle\\
    &=\prod_{l=1}^{N_\tau}\prod_\sigma\det S(\Psi_{i,l\oplus 1},\Psi_{i,l},\Delta\tau,\sigma)
\end{split}
\end{equation}
and is a product of $2N_\tau$ determinants of imaginary-time single-particle propagators on a lattice (this is emphasized by adding the subscript $\beta$).
The cyclic addition in Eq.~\eqref{Eq:D_C_eq} is the standard addition for $l=1,\dots,N_\tau-1$, while $N_\tau\oplus 1=1$.
The symbol $\varepsilon_\mathrm{int}(\mathcal{C})$ stands for
\begin{equation}
\label{Eq:total-energy-C}
    \varepsilon_\mathrm{int}(\mathcal{C})\equiv\sum_{l=1}^{N_\tau}\varepsilon_\mathrm{int}(\Psi_{i,l}).
\end{equation}

By virtue of the cyclic invariance under trace, the thermodynamic expectation value of an observable $A$ can be expressed as
\begin{equation}
\label{Eq:tmd_expt_value_general}
    \langle A\rangle=\frac{1}{N_\tau}\sum_{l=0}^{N_\tau-1}\frac{1}{Z}\mathrm{Tr}\left\{(e^{-\Delta\tau H})^l A(e^{-\Delta\tau H})^{N_\tau-l}\right\}.
\end{equation}
The FPQMC method is particularly suitable for observables diagonal in the coordinate representation (e.g., the interaction energy $H_\mathrm{int}$ or the real-space charge density $n_{\mathbf{r}\sigma}$). Such observables will be distinguished by adding the subscript $i$. Equation~\eqref{Eq:tmd_expt_value_general}, combined with the lowest-order STD [Eq.~\eqref{Eq:Suzuki-Trotter-general}], produces the following approximant for $\langle A_i\rangle$
\begin{equation}
\label{Eq:A_a_final}
    \langle A_i\rangle\approx\frac{\sum_\mathcal{C}\mathcal{D}_\beta(\mathcal{C},\Delta\tau)\:e^{-\Delta\tau\varepsilon_\mathrm{int}(\mathcal{C})}\frac{1}{N_\tau}\sum_{l=1}^{N_\tau}\mathcal{A}_i(\Psi_{i,l})}{\sum_\mathcal{C}\mathcal{D}_\beta(\mathcal{C},\Delta\tau)\:e^{-\Delta\tau\varepsilon_\mathrm{int}(\mathcal{C})}},
\end{equation}
where
\begin{equation}
\label{Eq:expect_value_A_slice}
    \mathcal{A}_i(\Psi_{i,l})\equiv\langle\Psi_{i,l}|A_i|\Psi_{i,l}\rangle.
\end{equation}
Defining the weight $w(\mathcal{C},\Delta\tau)$ of configuration $\mathcal{C}$ as
\begin{equation}
    w(\mathcal{C},\Delta\tau)\equiv|\mathcal{D}_\beta(\mathcal{C},\Delta\tau)|e^{-\Delta\tau\varepsilon_\mathrm{int}(\mathcal{C})},
\end{equation}
Eq.~\eqref{Eq:A_a_final} is rewritten as
\begin{equation}
\label{Eq:A-a-importance-sampling}
    \langle A_i\rangle\approx\frac{\left\langle\mathrm{sgn}(\mathcal{C})\frac{1}{N_\tau}\sum_{l=1}^{N_\tau}\mathcal{A}_i(\Psi_{i,l})\right\rangle_w}{\left\langle\mathrm{sgn}(\mathcal{C})\right\rangle_w}
\end{equation}
where $\langle\dots\rangle_w$ denotes the weighted average over all $\mathcal{C}$ with respect to the weight $w(\mathcal{C})$; $\mathrm{sgn}(\mathcal{C})\equiv\mathcal{D}_\beta(\mathcal{C},\Delta\tau)/|\mathcal{D}_\beta(\mathcal{C},\Delta\tau)|$ is the sign of configuration $\mathcal{C}$, while $|\langle\mathrm{sgn}\rangle|\equiv|\langle\mathrm{sgn}(\mathcal{C})\rangle_w|$ is the average sign of the QMC simulation.

By construction, our FPQMC approach yields exact results for the noninteracting electrons (ideal gas, $U=0$) and in the atomic limit ($J=0$). In both limits, due to $[H_0,H_\mathrm{int}]=0$, the FPQMC method with arbitrary $N_\tau$ should recover the exact results. However, the performance of the method, quantified through the average sign of the simulation, deteriorates with increasing $N_\tau$. For $N_\tau=1$, the FPQMC algorithm is sign-problem-free because it sums only diagonal elements $\langle\Psi_{i,1}|e^{-\beta H_0}|\Psi_{i,1}\rangle$ of the positive operator $e^{-\beta H_0}$. The sign problem is absent also for $N_\tau=2$ because $\mathcal{D}_\beta(\mathcal{C},\beta/2)$ is a square of a real number.

An important feature of the above-presented methodology is its direct applicability in both the grand-canonical and canonical ensemble.
The grand-canonical formulation is essential to current state-of-the-art approaches~\cite{RevModPhys.83.349} (e.g., CT-INT or CT-AUX) relying on the thermal Wick's theorem, which is not valid in the canonical ensemble.~\cite{Fetter-Walecka-book}
In the auxiliary-field QMC,~\cite{PhysRevLett.51.1900,PhysRevB.31.4403,PhysRevB.40.506,PhysRevLett.83.2777,PhysRevLett.90.136401} the Hubbard--Stratonovich transformation~\cite{PhysRevB.28.4059} decouples many-body propagators into sums (or integrals) over one-body operators whether the particle number is fixed or not.
Working in the grand-canonical ensemble is then analytically and computationally more convenient because the traces over \textit{all possible} fermion occupations result in determinants.~\cite{PhysRevB.31.4403,JChemPhys.153.204108}
In the canonical ensemble, the computation of traces over constrained fermion occupations is facilitated by observing that the Hubbard--Stratonovich decoupling produces an ensemble of noninteracting systems~\cite{JChemPhys.153.204108} to which theories developed for noninteracting systems in the canonical ensemble, such as particle projection~\cite{PhysRevB.68.045120,gilbreth2021reducing} or recursive methods,~\cite{JChemPhys.98.2484,PhysRevResearch.2.043206} can be applied. While such procedures may be numerically costly and/or unstable,~\cite{JChemPhys.153.204108} a very recent combination of the auxiliary-field QMC with the recursive auxiliary partition function formalism~\cite{PhysRevResearch.2.043206} is reported to be stable and scale favorably with the numbers of particles and available orbitals.~\cite{arxiv.2212.08654}
In contrast to all these approaches, the formulation of the FPQMC method does not depend on whether the electron number is fixed or not.
However, the selection of MC updates does depend on the ensemble we work in.
In the canonical ensemble, the updates should conserve the number of electrons;
In the grand-canonical ensemble, we need to include also the updates that insert/remove electrons.
Our MC updates, together with the procedure used to extract MC results and estimate their statistical error, are presented in great detail in Sec.~SI of the Supplementary Material.

\subsubsection{FPQMC method for time-dependent quantities}
\label{Sec:QMC_dynamics}
Ideally, numerical simulations of quench experiments such as those from Refs.~\onlinecite{Science.363.379,NatCommun.10.1588,Science.363.383} should provide the time-dependent expectation value {$\langle A(t)\rangle$} of an observable {$A$} at times $t>0$ after the Hamiltonian undergoes a sudden change from $H(0)$ for $t<0$ to $H$ for $t>0$.
Again, in many instances,~\cite{Science.363.379,Science.363.383} the experimentally measurable observable $A$ is diagonal in the coordinate representation, which will be emphasized by the subscript $i$.
The computation proceeds along the three-piece Kadanoff--Baym--Keldysh contour~\cite{RevModPhys.86.779}
\begin{equation}
\label{Eq:Kadanoff-Baym-general}
 \langle A_i(t)\rangle=\frac{\mathrm{Tr}\left(e^{-\beta H(0)}\:e^{iHt}\:A_i\:e^{-iHt}\right)}{\mathrm{Tr}\left(e^{-\beta H(0)}\:e^{iHt}\:e^{-iHt}\right)}
\end{equation}
where one may employ the above-outlined fermionic-propagator approach after dividing the whole contour into a number of slices, see Fig.~\ref{Fig:konture_040522}(b).
While $H$ is the Hubbard Hamiltonian given in Eqs.~\eqref{Eq:Hubbard-canonical}--\eqref{Eq:H_int}, $H(0)$ describes correlated electrons whose charge (or spin) density is spatially modulated by external fields.

The immediate complication (compared to the equilibrium case) is that there are now three operators (instead of one) that need to be decomposed via the STD.
A preset accuracy determined by the size of both $\Delta\tau$ and $\Delta t$ will therefore require a larger number of time-slices.
In turn, this will enlarge the configuration space to be sampled, and potentially worsen the sign problem in the MC summation.
Even worse, the individual terms in the denominator depend on time, so that the sign problem becomes time-dependent (dynamical).
The problem is expected to become worse at long times $t$, yet vanishes in the $t\to 0$ limit.

To simplify the task, and yet keep it relevant,
we consider the evolution from a pure state $|\psi(0)\rangle$ that is an eigenstate of real-space density operators $n_{\mathbf{r}\sigma}$, so that its most general form is given by Eq.~\eqref{Eq:Psi-i}.
Such a setup has been experimentally realized.~\cite{Science.363.379,Science.363.383,PhysRevLett.128.223202,arxiv.2203.15023}
Replacing $e^{-\beta H(0)}\to|\psi(0)\rangle\langle\psi(0)|$ in Eq.~\eqref{Eq:Kadanoff-Baym-general} leads to the expression for the time-dependent expectation value of observable $A_i$
\begin{equation}
\label{Eq:A_a_t_keldysh}
    \langle A_i(t)\rangle=\frac{\langle\psi(0)|e^{iHt}\:A_i\:e^{-iHt}|\psi(0)\rangle}{\langle\psi(0)|e^{iHt}\:e^{-iHt}|\psi(0)\rangle}.
\end{equation}
Here, the STD should be applied to both the forward and backward evolution operators, see Fig.~\ref{Fig:konture_040522}(c), which requires a larger number of time-slices to reach the desired accuracy (in terms of the systematic error).
Nevertheless, Eq.~\eqref{Eq:A_a_t_keldysh} is the simplest example on which the applicability of the real-time FPQMC method to follow the evolution of real-space observables may be examined.

Generally speaking, the symmetries of the model should be exploited to enable as efficient as possible MC evaluation of Eq.~\eqref{Eq:A_a_t_keldysh}.
Recent experimental~\cite{NatPhys.8.213} and theoretical~\cite{PhysRevLett.119.225302} studies have discussed the dynamical symmetry of the Hubbard model, according to which the temporal evolution of certain observables is identical for repulsive and attractive interactions of the same magnitude.
The symmetry relies on specific transformation laws of the Hamiltonian $H$, the initial state $|\psi(0)\rangle$, and the observable of interest $A_i$ under the combined action of two symmetry operations.
The first is the bipartite lattice symmetry or the $\pi$-boost~\cite{NatPhys.8.213} operation, which exploits the symmetry $\varepsilon_{\mathbf{k}}=-\varepsilon_{\mathbf{k}+(\pi,\pi)}$ of the free-electron dispersion and which is represented by the unitary operator $B$.
The second is the time reversal symmetry represented by the antiunitary operator $T$ ($TiT=-i$) that reverses electron spin and momentum according to $Tc_{\mathbf{r}\sigma}^{(\dagger)}T=(-1)^{\delta_{\sigma,\downarrow}}c_{\mathbf{r}\overline{\sigma}}^{(\dagger)}$ and $Tc_{\mathbf{k}\sigma}^{(\dagger)}T=(-1)^{\delta_{\sigma,\downarrow}}c_{-\mathbf{k},\overline{\sigma}}^{(\dagger)}$.
In Appendix~\ref{App:dynamical_symmetry}, we formulate our FPQMC method to evaluate Eq.~\eqref{Eq:A_a_t_keldysh} in a way that manifestly respects the dynamical symmetry of the model (each contribution to MC sums respects the symmetry).
Here, we only cite the final expression for the time-dependent expectation value of an observable $A_i$ that satisfies $TBA_iBT=A_i$ when the evolution starts from a state $|\psi(0)\rangle$ satisfying $TB|\psi(0)\rangle=e^{i\chi}|\psi(0)\rangle$
\begin{equation}
\begin{split}
\label{Eq:A_a_ABMC_symmetry}
    &\langle A_i(t)\rangle\approx\\&\frac{\sum_\mathcal{C}\mathcal{A}_i(\Psi_{i,N_t+1})\:\mathrm{Re}\{\mathcal{D}_{2t}(\mathcal{C},\Delta t)\}\cos[\Delta\varepsilon_\mathrm{int}(\mathcal{C})\Delta t]}{\sum_\mathcal{C}\mathrm{Re}\{\mathcal{D}_{2t}(\mathcal{C},\Delta t)\}\cos[\Delta\varepsilon_\mathrm{int}(\mathcal{C})\Delta t]}.
\end{split}
\end{equation}
Here, the configuration resides on the contour depicted in Fig.~\ref{Fig:konture_040522}(c), which is divided into $2N_t$ slices in total, and contains $2N_t-1$ independent states.
We assume that states $|\Psi_{i,l}\rangle$ for $l=1,\dots,N_t$ ($l=N_t+1,\dots,2N_t$) lie on the forward (backward) branch of the contour, while $|\Psi_{i,1}\rangle\equiv|\psi(0)\rangle$.
$\mathcal{A}_i(\Psi_{i,N_t+1})$ is defined as in Eq.~\eqref{Eq:expect_value_A_slice}, while
\begin{equation}
    \Delta\varepsilon_\mathrm{int}(\mathcal{C})\equiv\sum_{l=1}^{N_t}\left[\varepsilon_\mathrm{int}(\Psi_{i,l+N_t})-\varepsilon_\mathrm{int}(\Psi_{i,l})\right]\label{Eq:interaction_energy_difference_keldysh}.
\end{equation}
The symbol $\mathcal{D}_{2t}(\mathcal{C},\Delta t)$ stands for the following combination of forward and backward fermionic propagators (which is also emphasized by the subscript $2t$)
\begin{equation}
\label{Eq:D_2t_def}
\begin{split}
    \mathcal{D}_{2t}(\mathcal{C},\Delta t)=&\prod_{l=N_t+1}^{2N_t}\langle\Psi_{i,l\oplus 1}|e^{iH_0\Delta t}|\Psi_{i,l}\rangle\times\\&\prod_{l=1}^{N_t}\langle\Psi_{i,l\oplus 1}|e^{-iH_0\Delta t}|\Psi_{i,l}\rangle.
\end{split}
\end{equation}
The bipartite lattice symmetry guarantees that $\mathcal{D}_{2t}(\mathcal{C},\Delta t)=\mathcal{D}_{2t}(\mathcal{C},-\Delta t)$, see Eq.~\eqref{Eq:D_t_bipartite}.
The numerator and denominator of the RHS of Eq.~\eqref{Eq:A_a_ABMC_symmetry} are term-by-term invariant under the transformations $\Delta t\to-\Delta t$ and $\Delta\varepsilon_\mathrm{int}(\mathcal{C})\to-\Delta\varepsilon_\mathrm{int}(\mathcal{C})$ that respectively reflect the transformation properties under the time reversal symmetry and the fact that the dynamics of $\langle A_i(t)\rangle$ is identical in the repulsive and the attractive model.
Defining $w(\mathcal{C})\equiv|\mathrm{Re}\{\mathcal{D}_{2t}(\mathcal{C},\Delta t)\}|$ and $\mathrm{sgn}(\mathcal{C})\equiv\mathrm{Re}\{\mathcal{D}_{2t}(\mathcal{C},\Delta t)\}/|\mathrm{Re}\{\mathcal{D}_{2t}(\mathcal{C},\Delta t)\}|$, Eq.~\eqref{Eq:A_a_ABMC_symmetry} is recast as
\begin{equation}
\label{Eq:A_a_t_w_w}
 \langle A_i(t)\rangle\approx\frac{\langle\mathcal{A}_i(\Psi_{i,N_t+1})\:\mathrm{sgn}(\mathcal{C})\cos\left[\Delta\varepsilon_\mathrm{int}(\mathcal{C})\Delta t\right]\rangle_w}{\langle\mathrm{sgn}(\mathcal{C})\cos\left[\Delta\varepsilon_\mathrm{int}(\mathcal{C})\Delta t\right]\rangle_w}.
\end{equation}

The sign problem in the MC evaluation of Eq.~\eqref{Eq:A_a_t_w_w} is dynamical.
It generally becomes more serious with increasing time $t$ and interaction strength $|U|$.
Moreover, $w(\mathcal{C})$ also depends on both $t$ and $U$, meaning that MC evaluations for different $t$s and $U$s should be performed separately, using different Markov chains.
It is thus highly desirable to employ further symmetries in order to improve performance of the method.
Particularly relevant initial states $|\psi(0)\rangle$, from both experimental~\cite{Science.363.379,Science.363.383} and theoretical~\cite{NewJPhys.21.015003} viewpoint, are pure density-wave-like states.
Such states correspond to extreme spin-density wave (SDW) and charge-density wave (CDW) patterns, which one obtains by applying strong external density-modulating fields.
The SDW-like state can be written as
\begin{equation}
\label{Eq:sdw_state}
 |\psi_\mathrm{SDW}(\mathcal{G})\rangle=\prod_{\mathbf{r}_1\in\mathcal{G}}c_{\mathbf{r}_1\uparrow}^\dagger\prod_{\mathbf{r}_2\in\mathcal{U}\setminus\mathcal{G}}c_{\mathbf{r}_2\downarrow}^\dagger|\emptyset\rangle
\end{equation}
where $\mathcal{G}$ denotes the multitude of sites on which the electron spins are polarized up, while set $\mathcal{U}$ contains all sites of the cluster studied.
The electron spins on sites belonging to $\mathcal{U}\setminus\mathcal{G}$ are thus polarized down.
Such states have been experimentally realized in Ref.~\onlinecite{Science.363.383}.
The CDW-like states have also been realized in experiment,\cite{Science.363.379} and they read as
\begin{equation}
\label{Eq:cdw_state}
 |\psi_\mathrm{CDW}(\mathcal{G})\rangle=\prod_{\mathbf{r}\in\mathcal{G}}c_{\mathbf{r}\uparrow}^\dagger c_{\mathbf{r}\downarrow}^\dagger|\emptyset\rangle.
\end{equation}
The sites belonging to $\mathcal{G}$ are doubly occupied, while the remaining sites are empty.
The state $|\psi_\mathrm{CDW}(\mathcal{G})\rangle$ can be obtained from the corresponding $|\psi_\mathrm{SDW}(\mathcal{G})\rangle$ state by applying the partial particle--hole transformation that acts on spin-down electrons only
\begin{equation}
\label{Eq:relate_cdw_sdw}
 |\psi_\mathrm{SDW}(\mathcal{G})\rangle = \prod_{\mathbf{r}\in\mathcal{U}} \Big(c^\dagger_{\mathbf{r}\downarrow}(1-n_{\mathbf{r}\downarrow})+c_{\mathbf{r}\downarrow}n_{\mathbf{r}\downarrow}\Big)|\psi_\mathrm{CDW}(\mathcal{G})\rangle,
\end{equation}
see also Refs.~\onlinecite{ModPhysLett.4.759,PhysLettA.228.383}.
By combining the partial particle--hole, time-reversal, and bipartite lattice symmetries, the authors of Ref.~\onlinecite{NewJPhys.21.015003} have shown that the time evolution of spatially resolved charge and spin densities starting from states $|\psi_\mathrm{CDW}(\mathcal{G})\rangle$ and $|\psi_\mathrm{SDW}(\mathcal{G})\rangle$, respectively, obey
\begin{equation}
\begin{split}
\label{Eq:pops_cdw_sdw}
 &\langle\psi_\mathrm{CDW}(\mathcal{G})|e^{iHt}(n_{\mathbf{r}\uparrow}+n_{\mathbf{r}\downarrow}-1)e^{-iHt}|\psi_\mathrm{CDW}(\mathcal{G})\rangle=\\&\langle\psi_\mathrm{SDW}(\mathcal{G})|e^{iHt}(n_{\mathbf{r}\uparrow}-n_{\mathbf{r}\downarrow})e^{-iHt}|\psi_\mathrm{SDW}(\mathcal{G})\rangle.
\end{split}
\end{equation}
Equation~\eqref{Eq:pops_cdw_sdw} may be used to acquire additional statistics by combining the Markov chains for the two symmetry-related evolutions. The procedure is briefly described in Appendix~\ref{App:dynamical_symmetry} and applied to all corresponding computations presented in Sec.~\ref{Sec:numerics_apqmc_t_dep}. 

\subsubsection{ABQMC method for time-dependent quantities}
In this section, we develop the so-called alternating-basis QMC method, which is aimed at removing the dynamical character of the sign problem in real-time FPQMC simulations.
Moreover, using the ABQMC method, the results for different real times $t$ and different interactions $U$ may be obtained using just a single Markov chain, in contrast to the FPQMC method that employs separate chains for each $t$ and $U$.

Possible advantages of the ABQMC over the FPQMC method are most easily appreciated on the example of the survival probability of the initial state $|\psi(0)\rangle$
\begin{equation}
\label{Eq:P_t_def}
 P(t)=\left|\langle\psi(0)|e^{-iHt}|\psi(0)\rangle\right|^2,
\end{equation}
which is the probability of finding the system in its initial state after a time $t$ has passed.
Evaluating Eq.~\eqref{Eq:P_t_def} by any discrete-time QMC method necessitates only one STD, see Fig.~\ref{Fig:konture_040522}(d).
The survival probability is thus the simplest example on which the applicability of any QMC method to out-of-equilibrium setups can be systematically studied.

The FPQMC computation of $P(t)$ may proceed via the ratio
\begin{equation}
\label{Eq:P_t_ratio}
    \mathcal{R}(t)=\frac{\langle\psi(0)|e^{-iHt}|\psi(0)\rangle}{\langle\psi(0)|e^{-iH_0t}|\psi(0)\rangle}
\end{equation}
of the survival-probability amplitudes in the presence and absence of electron--electron interactions.
However, the average sign of the MC simulation of Eq.~\eqref{Eq:P_t_ratio} is proportional to the survival-probability amplitude of the noninteracting system, which generally decays very quickly to zero, especially for large clusters.~\cite{JChemPhys.154.184103}
This means that the dynamical sign problem in the FPQMC evaluation of Eq.~\eqref{Eq:P_t_ratio} may become very severe already at relatively short times $t$.

Instead of expressing the many-body free propagator $\langle\Psi'_i|e^{-iH_0\Delta t}|\Psi_i\rangle$ as a determinant of single-particle free propagators [Eqs.~\eqref{Eq:free-prop-1} and~\eqref{Eq:free-prop-2}], we could have introduced the spectral decomposition of $e^{-iH_0\Delta t}$ in terms of Fock states $|\Psi_k\rangle$ in the momentum representation. In analogy with Eq.~\eqref{Eq:Psi-i} such states are defined as
\begin{equation}
\label{Eq:Psi-k}
    |\Psi_k\rangle=\prod_\sigma\prod_{j=1}^{N_\sigma}c_{\mathbf{k}^\sigma_{j}\sigma}^\dagger|\emptyset\rangle.
\end{equation}
The state $|\Psi_k\rangle$ contains $N_\sigma$ electrons of spin $\sigma$ whose momenta $\mathbf{k}^\sigma_{1},\dots,\mathbf{k}^\sigma_{N_\sigma}$ are ordered according to a certain rule and we define $\varepsilon_0(\Psi_{k})\equiv\langle\Psi_k|H_0|\Psi_k\rangle$.
In this case, the final expression for the survival probability of state $|\psi(0)\rangle$ that satisfies $TB|\psi(0)\rangle=e^{i\chi}|\psi(0)\rangle$ reads as
\begin{equation}
\label{Eq:P-t-form-for-MC-final}
    P(t)\approx\left|\frac{\sum_\mathcal{C}\mathrm{Re}\{\mathcal{D}(\mathcal{C})\}\:\cos[\varepsilon_0(\mathcal{C})\Delta t]\:\cos[\varepsilon_\mathrm{int}(\mathcal{C})\Delta t]}{\sum_\mathcal{C}\mathrm{Re}\{\mathcal{D}(\mathcal{C})\}}\right|^2.
\end{equation}
A derivation of Eq.~\eqref{Eq:P-t-form-for-MC-final} is provided in Appendix~\ref{App:ABQMC_derive_survival}.
The MC evaluation of Eq.~\eqref{Eq:P-t-form-for-MC-final} should sample a much larger configuration space than the MC evaluation of Eq.~\eqref{Eq:P_t_ratio}.
The configuration $\mathcal{C}$ in Eq.~\eqref{Eq:P-t-form-for-MC-final} also resides on the contour depicted in Fig.~\ref{Fig:konture_040522}(d), but comprises $2N_t-1$ states in total: $N_t$ Fock states $|\Psi_{k,l}\rangle$ ($l=1,\dots,N_t$) and $N_t-1$ Fock states $|\Psi_{i,l}\rangle$ ($l=2,\dots,N_t$) (again, $|\Psi_{i,1}\rangle\equiv|\psi(0)\rangle$).
$\mathcal{D}(\mathcal{C})$ is the product of $2N_t$ Slater determinants
\begin{equation}
\label{Eq:D_C_abqmc}
    \mathcal{D}(\mathcal{C})=\prod_{l=1}^{N_t}\langle\Psi_{i,l\oplus 1}|\Psi_{k,l}\rangle\langle\Psi_{k,l}|\Psi_{i,l}\rangle
\end{equation}
that stem from the sequence of basis alternations between the momentum and coordinate eigenbasis.
Using the notation of Eqs.~\eqref{Eq:Psi-i} and~\eqref{Eq:Psi-k}, the most general Slater determinant $\langle\Psi_i|\Psi_k\rangle$ entering Eq.~\eqref{Eq:D_C_eq} is given as
\begin{eqnarray}
\langle\Psi_i|\Psi_k\rangle=\prod_\sigma\det \widetilde{S}(\Psi_i,\Psi_k,\sigma)\\
\left[\widetilde{S}(\Psi_i,\Psi_k,\sigma)\right]_{j_1j_2}=\langle\mathbf{r}_{j_1}^\sigma|\mathbf{k}_{j_2}^\sigma\rangle=\frac{\exp(i\mathbf{k}_{j_2}^\sigma\cdot\mathbf{r}_{j_1}^\sigma)}{\sqrt{N_c}},
\end{eqnarray} where $1\leq j_1,j_2\leq N_\sigma$. The symbol $\varepsilon_0(\mathcal{C})$ stands for [cf. Eq.~\eqref{Eq:total-energy-C}]
\begin{equation}
\label{Eq:varepsilon_0_C}
\varepsilon_0(\mathcal{C})\equiv\sum_{l=1}^{N_t}\varepsilon_0(\Psi_{k,l}).
\end{equation}
We note that each term in Eq.~\eqref{Eq:P-t-form-for-MC-final} is invariant under transformations $\varepsilon_0(\mathcal{C})\to-\varepsilon_0(\mathcal{C})$ and $\Delta t\to-\Delta t$, which reflect the action of the bipartite lattice symmetry and the time reversal symmetry, respectively.
Being term-by-term invariant under the transformation $\varepsilon_\mathrm{int}(\mathcal{C})\to-\varepsilon_\mathrm{int}(\mathcal{C})$, Eq.~\eqref{Eq:P-t-form-for-MC-final} explicitly satisfies the requirement that the dynamics of $P(t)$ for repulsive and attractive interactions of the same magnitude are identical.
Defining $w(\mathcal{C})\equiv|\mathrm{Re}\{\mathcal{D}(\mathcal{C})\}|$ and $\mathrm{sgn}(\mathcal{C})\equiv\mathrm{Re}\{\mathcal{D}(\mathcal{C})\}/|\mathrm{Re}\{\mathcal{D}(\mathcal{C})\}|$, Eq.~\eqref{Eq:P-t-form-for-MC-final} is recast as
\begin{equation}
\label{Eq:P_t_w_w}
 P(t)\approx\left|\frac{\left\langle\mathrm{sgn}(\mathcal{C})\:\cos[\varepsilon_0(\mathcal{C})\Delta t]\:\cos[\varepsilon_\mathrm{int}(\mathcal{C})\Delta t]\right\rangle_w}{\left\langle\mathrm{sgn(\mathcal{C})}\right\rangle_w}\right|^2.
\end{equation}
This choice for $w$ is optimal in the sense that it minimizes the variance of $\langle\mathrm{sgn}(\mathcal{C})\rangle_w$,~\cite{PhysRevE.49.3855} whose modulus is the average sign of the ABQMC simulation.
The sign problem encountered in the MC evaluation of Eq.~\eqref{Eq:P_t_w_w} does not depend on either time $t$ or interaction strength $U$, i.e., it is not dynamical.
The weight $w(\mathcal{C})$ in Eq.~\eqref{Eq:P_t_w_w} does not depend on either $\Delta t$ or any other property of configuration $\mathcal{C}$ ($\varepsilon_0,\varepsilon_\mathrm{int}$).
Therefore, the MC evaluation of Eq.~\eqref{Eq:P_t_w_w} may be performed simultaneously (using a single Markov chain) for any $U$ and any $t$.
This presents a technical advantage over the FPQMC method, which may be outweighed by the huge increase of the configuration space when going from FPQMC to ABQMC.
To somewhat reduce the dimension of the ABQMC configuration space and improve the sampling efficiency, we design the MC updates so as to respect the momentum conservation law throughout the real-time evolution.
The momentum conservation poses the restriction that all the momentum-space states $|\Psi_{k,l}\rangle$ have the same total electron momentum $\mathbf{K}\equiv\sum_{\mathbf{k}\sigma}\mathbf{k}\langle\Psi_{k,l}|n_{\mathbf{k}\sigma}|\Psi_{k,l}\rangle$ [modulo $(2\pi,2\pi)$].
The MC updates in the ABQMC method for the evaluation of the survival probability are presented in great detail in Sec.~SII of the Supplementary Material.

By relying on the partial particle--hole and bipartite lattice symmetries, in Appendix~\ref{App:ABQMC_derive_survival} we demonstrate that the dynamics of the survival probabilities of states $|\psi_\mathrm{SDW}(\mathcal{G})\rangle$ and $|\psi_\mathrm{CDW}(\mathcal{G})\rangle$ are identical, i.e.,
\begin{equation}
\begin{split}
 \label{Eq:equal_P_t}
 &\left|\langle\psi_\mathrm{SDW}(\mathcal{G})|e^{-iHt}|\psi_\mathrm{SDW}(\mathcal{G})\rangle\right|^2=\\&\left|\langle\psi_\mathrm{CDW}(\mathcal{G})|e^{-iHt}|\psi_\mathrm{CDW}(\mathcal{G})\rangle\right|^2.
\end{split}
\end{equation}
Evaluating Eq.~\eqref{Eq:P_t_w_w}, additional statistics can be acquired by combining the Markov chains for the $P(t)$ calculations starting from the two symmetry-related states $|\psi_\mathrm{CDW}(\mathcal{G})\rangle$ and $|\psi_\mathrm{SDW}(\mathcal{G})\rangle$. The procedure is similar to that described in Appendix~\ref{App:dynamical_symmetry} and we apply it in all corresponding computations presented in Sec.~\ref{Sec:numerics_t_dep}.

\section{Numerical results}
\label{Sec:numerics_overall}

We first apply the FPQMC method to equilibrium situations (the particle number is not fixed), see Sec.~\ref{Sec:numerics_eq}, and then to time-dependent local densities during evolution of pure states, see Sec.~\ref{Sec:numerics_apqmc_t_dep}. Section~\ref{Sec:numerics_t_dep} presents our ABQMC results for the survival probability of pure states. Our implementation of the ABQMC method on the full Kadanoff--Baym--Keldysh contour [Eq.~\eqref{Eq:Kadanoff-Baym-general}] is benchmarked in Sec.~SVII of the Supplementary Material.

\subsection{Equilibrium results: Equation of state}
\label{Sec:numerics_eq}
\begin{figure}[ht!]
 \centering
 \includegraphics[width=0.9\columnwidth]{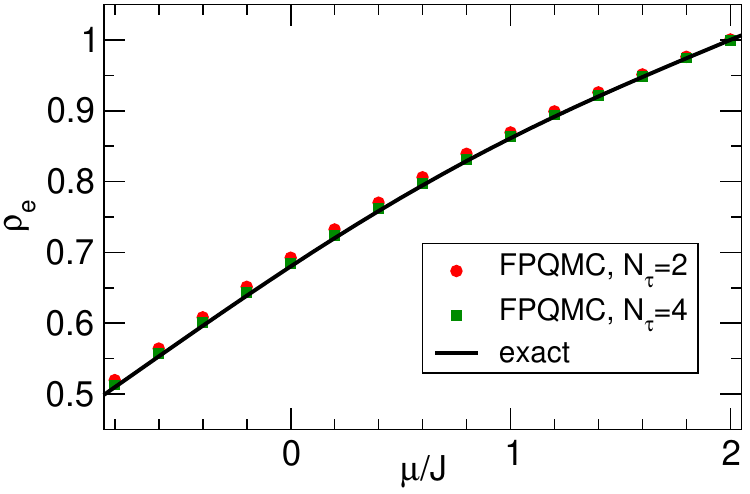}
 \caption{(Color online) Equation of state $\rho_e(\mu)$ for the Hubbard dimer with $T/J=1$, $U/J=4$. Full red circles (green squares) are the results of FPQMC simulations employing $N_\tau=2$ ($N_\tau=4$) imaginary-time slices, while the solid black line is computed using the exact diagonalization. The estimated statistical error of the FPQMC data is in all cases smaller than the symbol size.}
 \label{Fig:state_equation_dimer_091221}
\end{figure}

\begin{figure}[ht!]
    \centering
    \includegraphics[width=0.9\columnwidth]{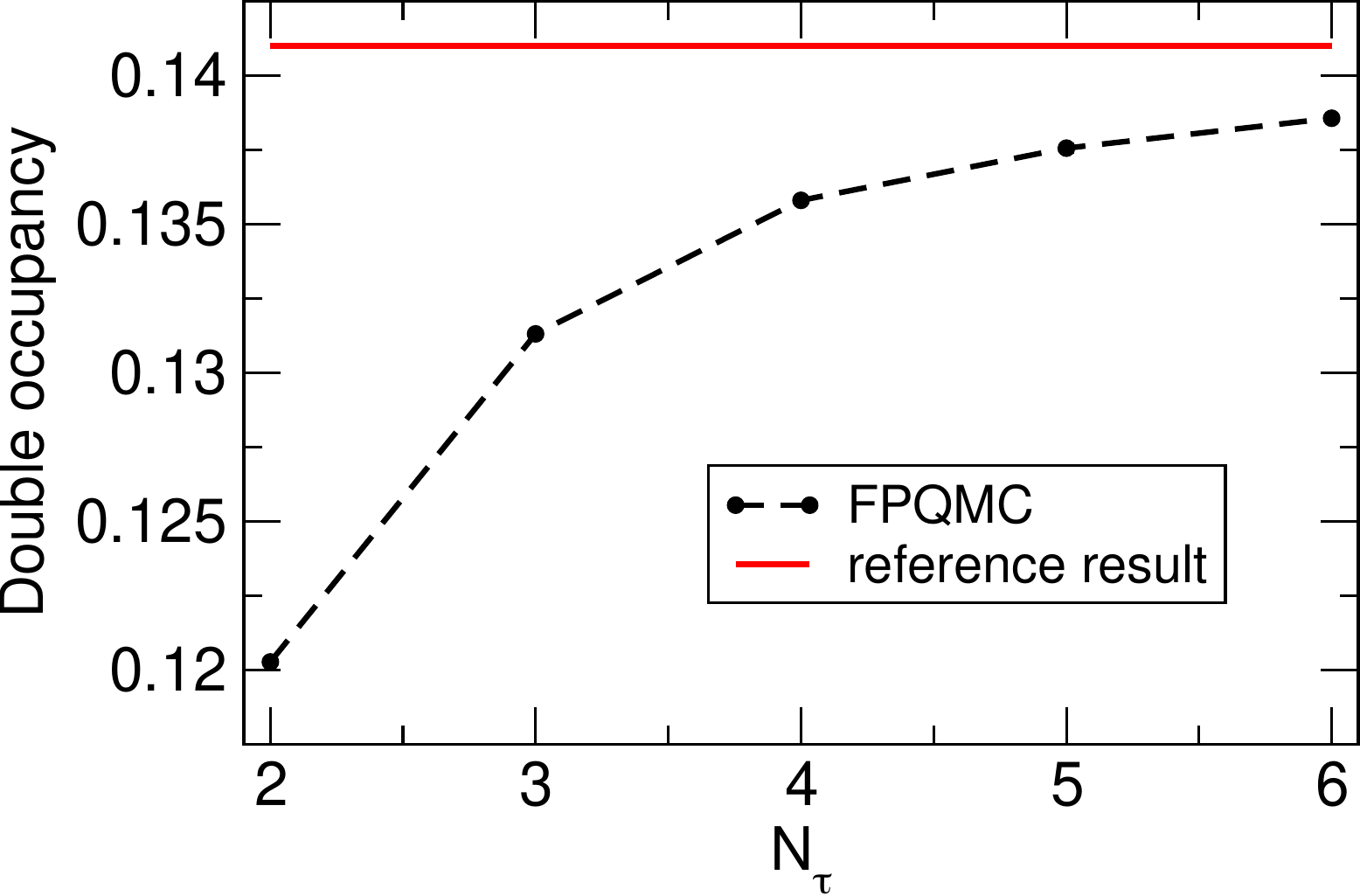}
    \caption{(Color online) Double occupancy of the $N_c=20$-site Hubbard chain at half-filling ($\mu=U/2$, $\rho_e=1$) as a function of the number $N_\tau$ of imaginary-time slices. The remaining parameters are $U/J=3$ and $T/J=1$. Dotted line connecting full symbols (FPQMC results) serves as a guide to the eye. Reference result is taken from Ref.~\onlinecite{PhysRevB.101.035149}. Relative deviation of the FPQMC result with $N_\tau=6$ from the reference result is around 2\%. Statistical error bars of the FPQMC results are smaller than the symbol size.}
    \label{Fig:docc_1d_hubbard}
\end{figure}

We start by considering the Hubbard dimer, the minimal model capturing the subtle interplay between electron delocalization and electron--electron interaction.~\cite{EurPhysJB.36.445}
We opt for moderate temperature $T/J=1$ and interaction $U/J=4$, so that the expected number of imaginary-time slices needed to obtain convergent FPQMC results is not very large. 
Figure~\ref{Fig:state_equation_dimer_091221} presents the equation of state (i.e., the dependence of the electron density $\rho_e=\left\langle\widehat{N}_\uparrow+\widehat{N}_\downarrow\right\rangle/N_c$ on the chemical potential $\mu$) for a range of $\mu$ below the half-filling.
Here, $\widehat{N}_\uparrow$ and $\widehat{N}_\downarrow$ are the operators of the total number of spin-up and spin-down electrons, respectively.
Figure~\ref{Fig:state_equation_dimer_091221} suggests that already $N_\tau=2$ imaginary-time slices suffice to obtain very good results in the considered range of $\mu$, while increasing $N_\tau$ from 2 to 4 somewhat improves the accuracy of the FPQMC results.
It is interesting that, irrespective of the value of $N_\tau$, FPQMC simulations on the dimer are manifestly sign-problem-free.
First, the one-dimensional imaginary-time propagator defined in Eq.~\eqref{Eq:def_I_z_l} is positive, $\mathcal{I}(J\Delta\tau,l)=\left[e^{J\Delta\tau}+(-1)^l e^{-J\Delta\tau}\right]/2$ for both $l=0$ and 1.
Second, the configuration containing two electrons of the same spin is of weight $\cosh^2(J\Delta\tau)-\sinh^2(J\Delta\tau)\equiv 1$, implying that weights of all configurations are positive. 
Furthermore, our results on longer chains suggest that FPQMC simulations of one-dimensional lattice fermions do not display sign problem.
While similar statements have been repeated for continuum one-dimensional models of both noninteracting~\cite{JPhysSocJpn.53.963,JPhysAMathGen.38.6659} and interacting fermions,~\cite{JPhysAMathTheor.40.7157} there is, to the best of our knowledge, no rigorous proof that the sign problem is absent from coordinate-space QMC simulations of one-dimensional fermionic systems.
While we do not provide such a proof either, Fig.~\ref{Fig:docc_1d_hubbard} is an illustrative example showing how the FPQMC results for the double occupancy $\sum_\mathbf{r}\langle n_{\mathbf{r}\uparrow}n_{\mathbf{r}\downarrow}\rangle/N_c$ of the Hubbard chain at half-filling approach the reference result (taken from Ref.~\onlinecite{PhysRevB.101.035149}) as the imaginary-time discretization becomes finer.
For all $N_\tau$s considered, the average sign of FPQMC simulations is $|\langle\mathrm{sgn}\rangle|=1$.

\begin{figure}[ht!]
 \centering
 \includegraphics[width=0.9\columnwidth]{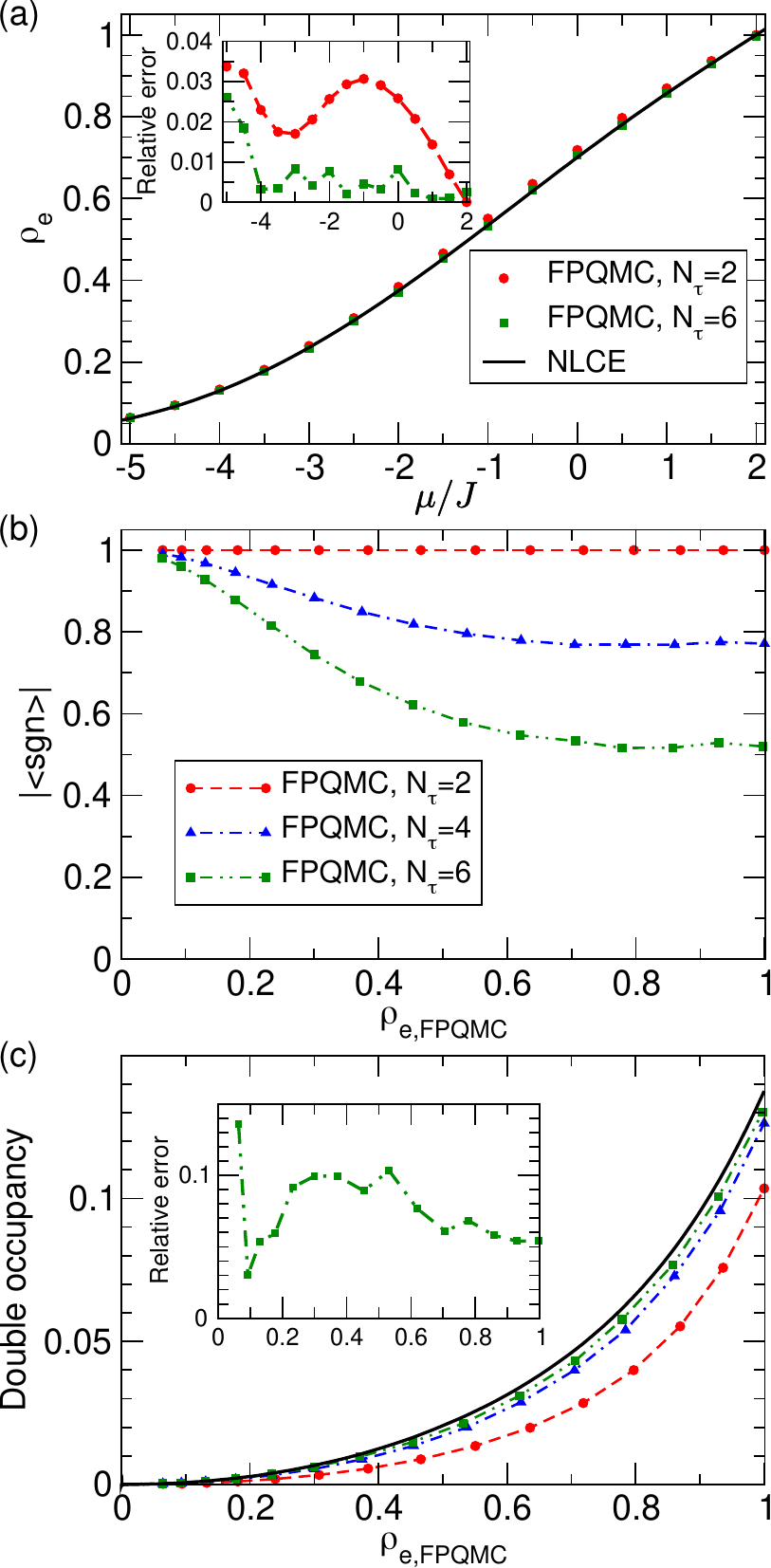}
 \caption{(Color online) (a) Equation of state $\rho_e(\mu)$ for the Hubbard model on a $4\times 4$ cluster with the following values of model parameters: $U/J=4$, $T/J=1.0408$.
 (b) The average sign as a function of the FPQMC estimate $\rho_{e,\mathrm{FPQMC}}$ of the electron density for different values of $N_\tau$. The dashed lines are guides to the eye.
 (c) The double occupancy $\sum_\mathbf{r}\langle n_{\mathbf{r}\uparrow}n_{\mathbf{r}\downarrow}\rangle/N_c$ as a function of the FPQMC estimate $\rho_{e,\mathrm{FPQMC}}$ of the electron density for different values of $N_\tau$.
 In (a) and (c), full symbols represent FPQMC results, the solid line shows the NLCE data taken from Ref.~\onlinecite{PhysRevA.84.053611}, while the insets show the relative deviation of FPQMC results from the reference NLCE results.
 The estimated statistical error of the FPQMC data is in all cases smaller than the symbol size.}
 \label{Fig:state_equation_4x4_weak_U_091221}
\end{figure}

We now apply the FPQMC method to evaluate the equation of state on larger clusters. 
We focus on a $4\times 4$ cluster, which may already be representative of the thermodynamic limit at $T/J\gtrsim 1$.~\cite{PhysRevLett.123.036601} 
We compare our $\rho_e(\mu)$ results with the results of the numerical linked-cluster expansion (NLCE) method.~\cite{PhysRevLett.97.187202,PhysRevA.84.053611,ComputPhysCommun.184.557}
The NLCE results are numerically exact and converged with respect to the control parameter, i.e., the maximal cluster-size used. NLCE is commonly used to benchmark methods and understand experimental data.~\cite{PhysRevLett.116.175301,PhysRevX.7.031025}
Again, we keep $U/J=4$, but we take $T/J=1.0408$ to be able to compare results to the data of Ref.~\onlinecite{PhysRevA.84.053611}.
Figure~\ref{Fig:state_equation_4x4_weak_U_091221}(a) reveals that the FPQMC results with only $N_\tau=2$ imaginary-time slices agree very well (within a couple of percent) with the NLCE results over a wide range of chemical potentials. 
This is a highly striking observation, especially keeping in mind that the FPQMC method with $N_\tau=2$ is sign-problem-free, see Fig.~\ref{Fig:state_equation_4x4_weak_U_091221}(b). 
It is unclear whether other STD-based methods would reach here the same level of accuracy with only two imaginary-time slices (and without sign problem).
This may be a specific property of the FPQMC method.
Finer imaginary-time discretization introduces the sign problem, see Fig.~\ref{Fig:state_equation_4x4_weak_U_091221}(b), which becomes more pronounced as the density is increased and reaches a plateau for $\rho_e\gtrsim 0.8$.
Still, the sign problem remains manageable.
Increasing $N_\tau$ for 2 to 6 somewhat improves the agreement of the density $\rho_e$ [the inset of Fig.~\ref{Fig:state_equation_4x4_weak_U_091221}(a)] and considerably improves the agreement of the double occupancy [Fig.~\ref{Fig:state_equation_4x4_weak_U_091221}(c)] with the referent NLCE results.
Still, comparing the insets of Figs.~\ref{Fig:state_equation_4x4_weak_U_091221}(a) and~\ref{Fig:state_equation_4x4_weak_U_091221}(c), we observe that the agreement between FPQMC ($N_\tau=6$) and NLCE results for $\rho_e$ is significantly better than for the double occupancy.
The systematic error in FPQMC comes from the time-discretization and the finite size of the system.
At $N_\tau=6$, it is not a priori clear which error contributes more, but it appears most likely that the time-discretization error is dominant.
In any case, the reason why systematic error is greater for the double occupancy than for the average density could be that the double occupancy contains more detailed information about the correlations in the system.
This might be an indication that measurement of multipoint density correlations will generally be more difficult---it may require a finer time resolution and/or greater lattice size.

The average sign above the half-filling $\rho_e = 1$ mirrors that below the half-filling.
The particle--hole symmetry ensures that $\rho_e(\mu)=2-\rho_e(U-\mu)$, but that it also governs the average sign is not immediately obvious from the construction of the method.
A formal demonstration of the electron-doping--hole-doping symmetry of the average sign is, however, possible (see Appendix~\ref{App:full_ph_symmetric_sign}).
Note that we restrict our density calculations to $\rho_e<1$ because in this case the numerical effort to manipulate the determinants [Eqs.~\eqref{Eq:free-prop-2} and~\eqref{Eq:D_C_eq}] is lower (size of the corresponding matrices is given by the number of particles of a given spin).
The performance of the FPQMC algorithm to compute $\rho_e(\mu)$ (average time needed to propose/accept an MC update and acceptance rates of individual MC updates) is discussed in Sec.~SIII of the Supplementary Material.

\begin{figure}[ht!]
 \centering
 \includegraphics[width=0.9\columnwidth]{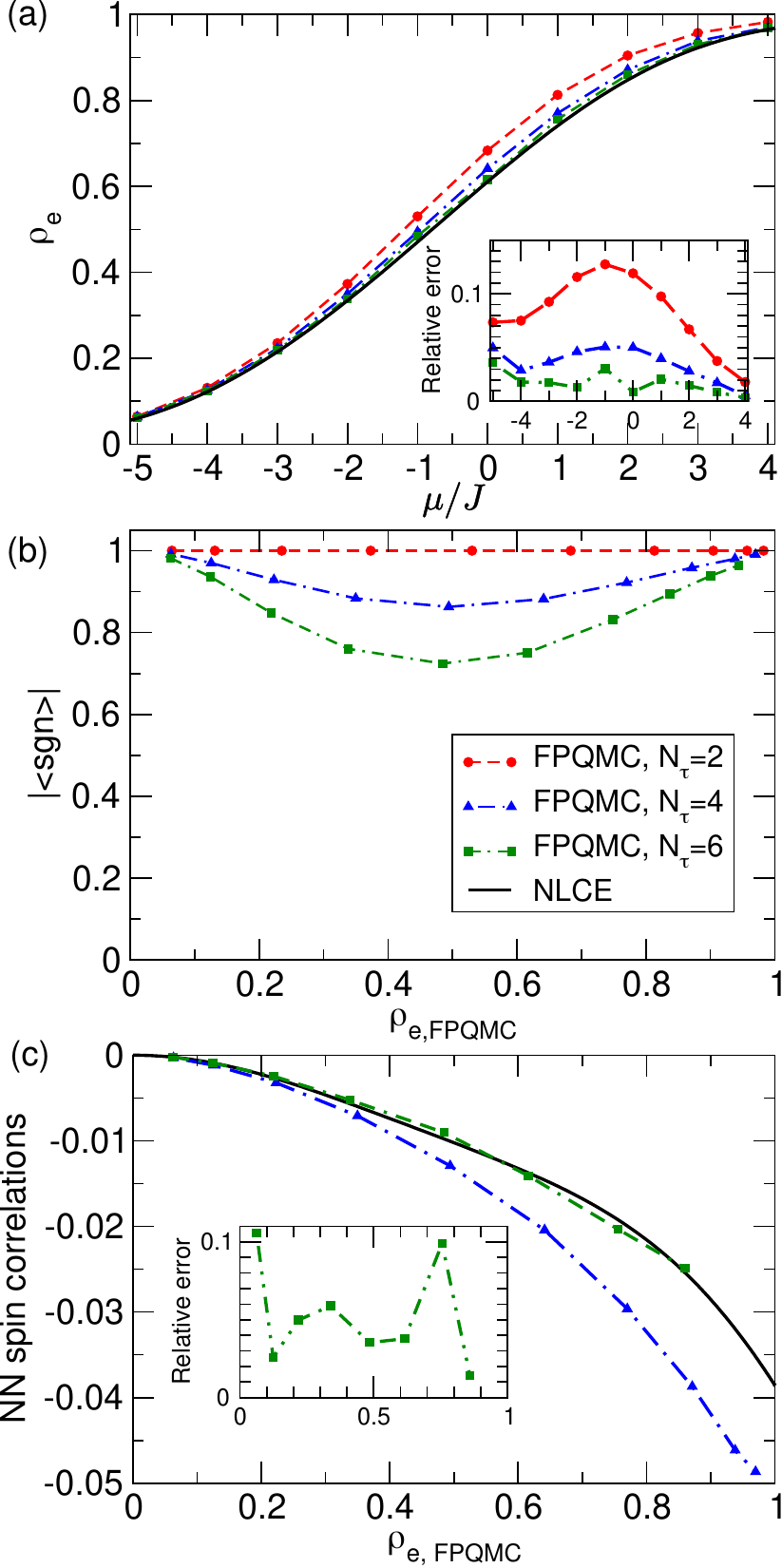}
 \caption{(Color online) (a) Equation of state $\rho_e(\mu)$ for the Hubbard model on a $4\times 4$ cluster with the following values of model parameters: $U/J=24$, $T/J=1.0408$.
 (b) The average sign as a function of the FPQMC estimate $\rho_{e,\mathrm{FPQMC}}$ of the electron density for different values of $N_\tau$.
 (c) Nearest-neighbor spin correlations $\sum_{\mathbf{r}\boldsymbol{\delta}}\langle S_\mathbf{r}^zS_{\mathbf{r}+\boldsymbol{\delta}}^z\rangle/N_c$ as a function of the FPQMC estimate $\rho_{e,\mathrm{FPQMC}}$ of the electron density for $N_\tau=4$ and 6.
 In (a) and (c), full symbols represent FPQMC results, the solid line shows the NLCE data taken from Ref.~\onlinecite{PhysRevA.84.053611}, while the insets show the relative deviation of FPQMC results from the reference NLCE results.
 The dashed or dash-dotted lines connecting the symbols serve as guides to the eye.
 The estimated statistical error of the FPQMC data is in all cases smaller than the symbol size.}
 \label{Fig:state_equation_4x4_strong_U_101221}
\end{figure}

We further benchmark our method in the case of very strong coupling, $U/J=24$ and, again, $T/J=1.0408$.
Figure~\ref{Fig:state_equation_4x4_strong_U_101221}(a) compares the FPQMC results on a $4\times 4$ cluster using $N_\tau=2,4$, and 6 imaginary-time slices with the NLCE results.
At extremely low fillings $\rho_e\lesssim 0.1$, the relative importance of the interaction term with respect to the kinetic term is quite small, and taking only $N_\tau=2$ suffices to reach a very good agreement between the FPQMC and NLCE results, see the inset of Fig.~\ref{Fig:state_equation_4x4_strong_U_101221}(a).
As the filling is increased, the interaction effects become increasingly important, and it is necessary to increase $N_\tau$ in order to accurately describe the competition between the kinetic and interaction terms.
In the inset of Fig.~\ref{Fig:state_equation_4x4_strong_U_101221}(a), we see that $N_\tau=6$ is sufficient to reach an excellent (within a couple of percent) agreement between FPQMC and NLCE results over a broad range of fillings.
At very high fillings $\rho_e\gtrsim 0.9$ and for $N_\tau=6$, our MC updates that insert/remove particles have very low acceptance rates, which may lead to a slow sampling of the configuration space.
It is for this reason that FPQMC results with $N_\tau=6$ do not significantly improve over $N_\tau=4$ results in this parameter regime.
For $N_\tau=6$, an inefficient sampling near the half-filling also renders the corresponding results for the nearest-neighbor spin correlations $\sum_{\mathbf{r}\boldsymbol{\delta}}\langle S_\mathbf{r}^zS_{\mathbf{r}+\boldsymbol{\delta}}^z\rangle/N_c$ inaccurate, so that they are not displayed in Fig.~\ref{Fig:state_equation_4x4_strong_U_101221}(c).
Here, vector $\boldsymbol{\delta}$ connects nearest-neighboring sites, while $S_\mathbf{r}^z=(n_{\mathbf{r}\uparrow}-n_{\mathbf{r}\downarrow})/2$ is the operator of $z$ projection of the local spin.
At lower fillings $\rho_e\lesssim 0.8$, the agreement between our FPQMC results with $N_\tau=6$ and the NLCE results is good, while decreasing $N_\tau$ from 6 to 4 severely deteriorates the quality of the FPQMC results.      

At this strong coupling, the dependence of the average sign on the density is somewhat modified, see Fig.~\ref{Fig:state_equation_4x4_strong_U_101221}(b).
The minimal sign is no longer reached around half-filling but at quarter filling $\rho_e\sim 0.5$, around which $|\langle\mathrm{sgn}\rangle|$ appears to be symmetric. Comparing Fig.~\ref{Fig:state_equation_4x4_strong_U_101221}(b) to Fig.~\ref{Fig:state_equation_4x4_weak_U_091221}(b), we see that the average sign does not become smaller with increasing interaction, in sharp contrast with interaction-expansion-based methods such as CT-INT~\cite{JETPLett.80.61,PhysRevB.72.035122} or configuration PIMC.~\cite{ContribPlasmaPhys.51.687}

\begin{figure}[ht!]
 \centering
 \includegraphics[width=0.9\columnwidth]{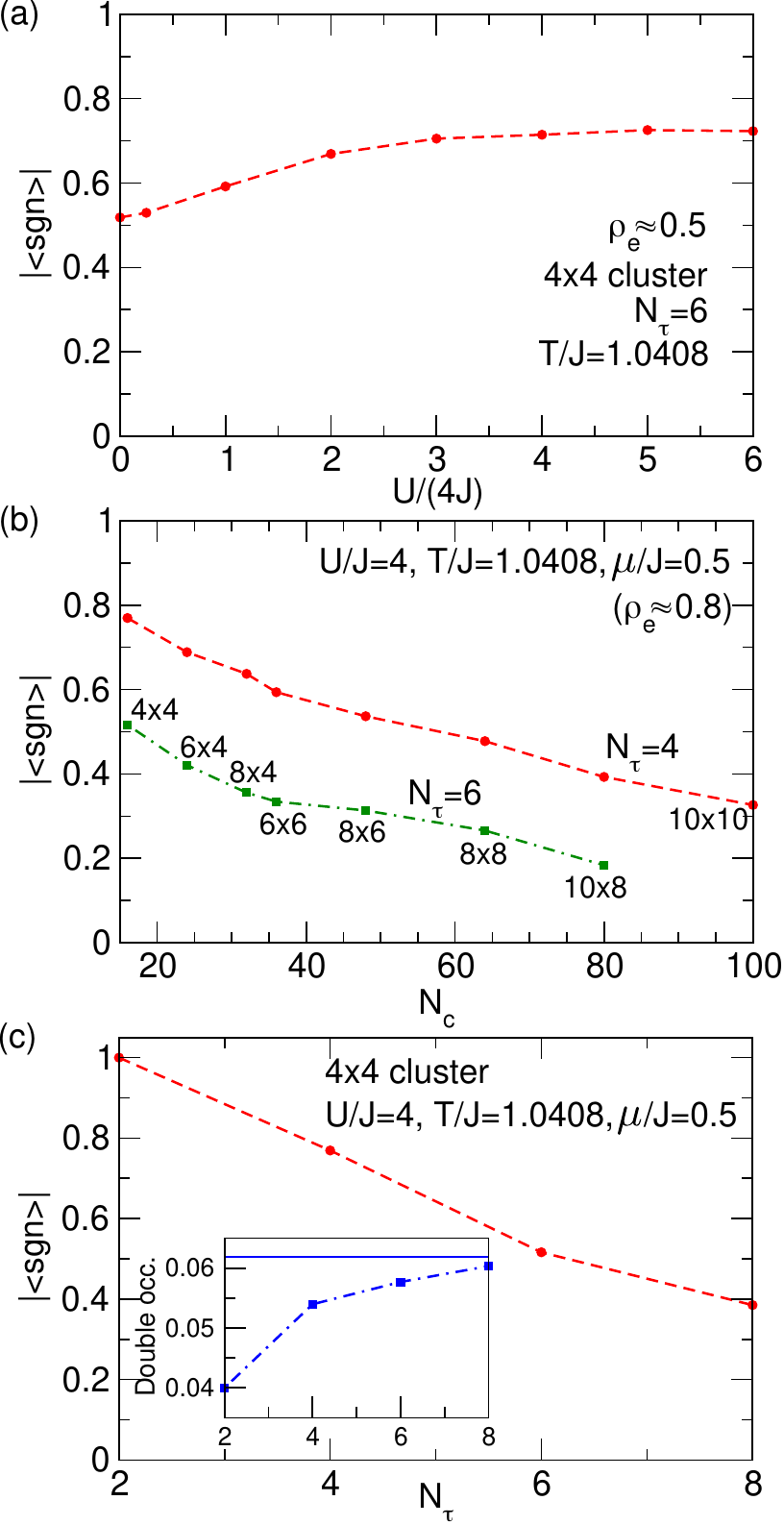}
 \caption{(Color online) (a) Average sign as a function of the ratio between the typical interaction and kinetic energies. Full symbols are results of FPQMC computations on a $4\times 4$ cluster with $N_\tau=6$, the temperature is fixed to $T/J=1.0408$, and the chemical potential at each $U$ is chosen such that $\rho_e\approx 0.5$. (b) Average sign as a function of the cluster size $N_c$ for the values of model parameters summarized in the figure. The FPQMC results (full symbols) are obtained using $N_\tau=4$ and 6. (c) Average sign as a function of $N_\tau$. The FPQMC results (full symbols) are obtained on a $4\times 4$ cluster for the values of model parameters summarized in the main part of the figure. The inset shows how the FPQMC result for double occupancy approaches the referent NLCE result as the imaginary-time discretization becomes finer.}
 \label{Fig:Fig_misc}
\end{figure}

To better understand the relation between the average sign and the interaction, in Fig.~\ref{Fig:Fig_misc}(a) we plot $|\langle\mathrm{sgn}\rangle|$ as a function of the ratio $U/(4J)$ of the typical interaction and kinetic energy.
We take $N_\tau=6$ and adjust the chemical potential using the data from Ref.~\onlinecite{PhysRevA.84.053611} so that $\rho_e\approx 0.5$.
We see that $|\langle\mathrm{sgn}\rangle|$ monotonically increases with the interaction and reaches a plateau at very strong interactions.
This is different from interaction-expansion-based QMC methods, whose sign problem becomes more pronounced as the interaction is increased.
Also, for weak interactions, the performance of the FPQMC method deteriorates at high density, see Fig.~\ref{Fig:state_equation_4x4_weak_U_091221}(b), while methods such as CT-INT become problematic at low densities.
The FPQMC method could thus become a method of choice to study the regimes of moderate coupling and temperature, which is highly relevant for optical-lattice experiments.
Figure~\ref{Fig:Fig_misc}(b) shows the decrease of the average sign with the cluster size $N_c$ in the weak-coupling and moderate-temperature regime at filling $\rho_e\approx 0.8$.
We observe that for both $N_\tau=4$ and $N_\tau=6$ the average sign decreases linearly with $N_c$.
For $N_\tau=6$, we observe that the decrease for $N_c\lesssim 40$ is somewhat faster than the decrease for $N_c\gtrsim 40$.
We, however, note that the acceptance rates of our MC updates strongly decrease with $N_c$ and that this decrease is more pronounced for finer imaginary-time discretizations.
That is why we were not able to obtain any meaningful result for the $10\times 10$ cluster with $N_\tau=6$.
At fixed cluster size and filling, the average sign decreases linearly with $N_\tau$, see the main part of Fig.~\ref{Fig:Fig_misc}(c), while the double occupancy tends to the referent NLCE value, see the inset of Fig.~\ref{Fig:Fig_misc}(c).

\begin{figure}
    \centering
    \includegraphics[width=0.9\columnwidth]{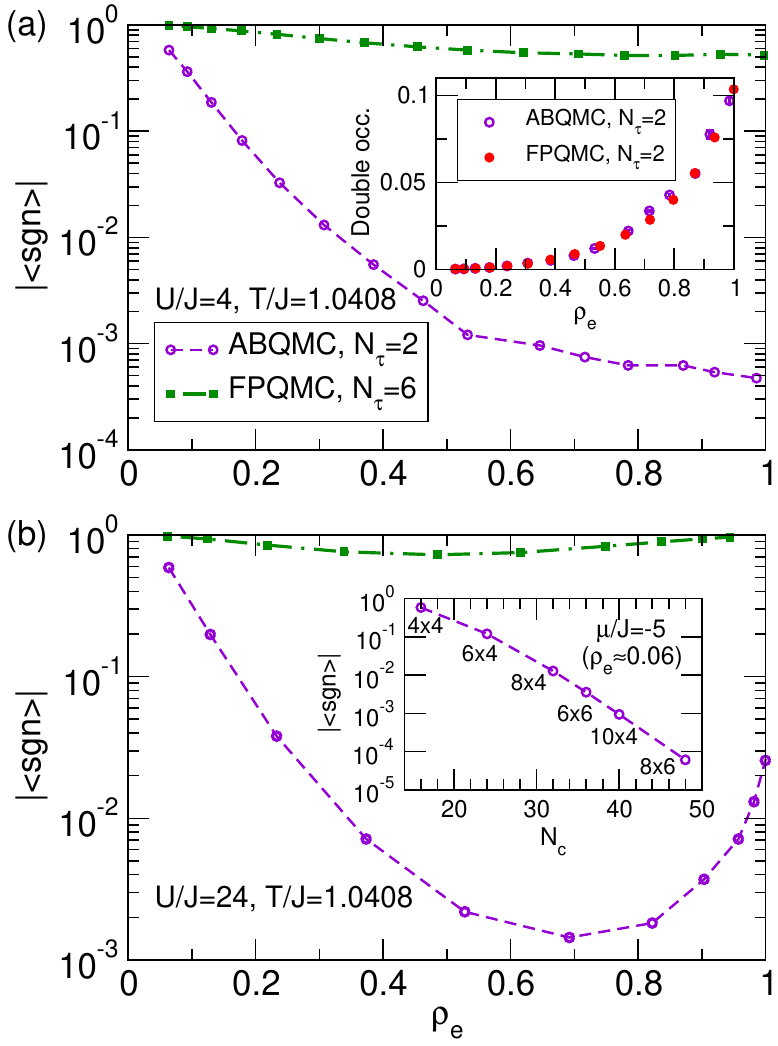}
    \caption{(Color online) Average sign as a function of the electron density in ABQMC simulations with $N_\tau=2$ (open circles) and FPQMC simulations with $N_\tau=6$ (full squares) for $T/J=1.0408$ and (a) $U/J=4$ and (b) $U/J=24$.
    The inset in panel (a) compares ABQMC (open circles) and FPQMC (full circles) results for the double occupancy as a function of $\rho_e$ (both methods employ $N_\tau=2$).
    The inset in panel (b) shows the average sign of ABQMC simulations with $N_\tau=2$ as a function of cluster size $N_c$ at low density ($\rho_e\approx 0.06,\mu/J=-5$).}
    \label{Fig:abqmc_vs_fpqmc_161222}
\end{figure}

In Sec.~SIV of the Supplementary Material we provide an implementation
of the ABQMC method in the equilibrium setup.
Figures~\ref{Fig:abqmc_vs_fpqmc_161222}(a) and~\ref{Fig:abqmc_vs_fpqmc_161222}(b), which deal with the same parameter regimes as Figs.~\ref{Fig:state_equation_4x4_weak_U_091221} and~\ref{Fig:state_equation_4x4_strong_U_101221}, respectively, clearly illustrate the advantages of the fermionic-propagator approach with respect to the alternating-basis approach in equilibrium.
The average sign of ABQMC simulations with only 2 imaginary-time slices is orders of magnitude smaller than the sign of FPQMC simulations with three times finer imaginary-time discretization. 
Since FPQMC and ABQMC methods are related by an exact transformation, they should produce the same results for thermodynamic quantities (assuming that $N_\tau$ is the same in both methods).
This is shown in the inset of Fig.~\ref{Fig:abqmc_vs_fpqmc_161222}(a) on the example of the double occupancy.
The inset of Fig.~\ref{Fig:abqmc_vs_fpqmc_161222}(b) suggests that the average sign decreases exponentially with the cluster size $N_c$.
Overall, our current implementation of the ABQMC method in equilibrium cannot be used to simulate larger clusters with a finer imaginary-time discretization.

\subsection{Time-dependent results using FPQMC method: Local charge and spin densities}
\label{Sec:numerics_apqmc_t_dep}
\subsubsection{Benchmarks on small clusters}
\begin{figure}
    \centering
    \includegraphics{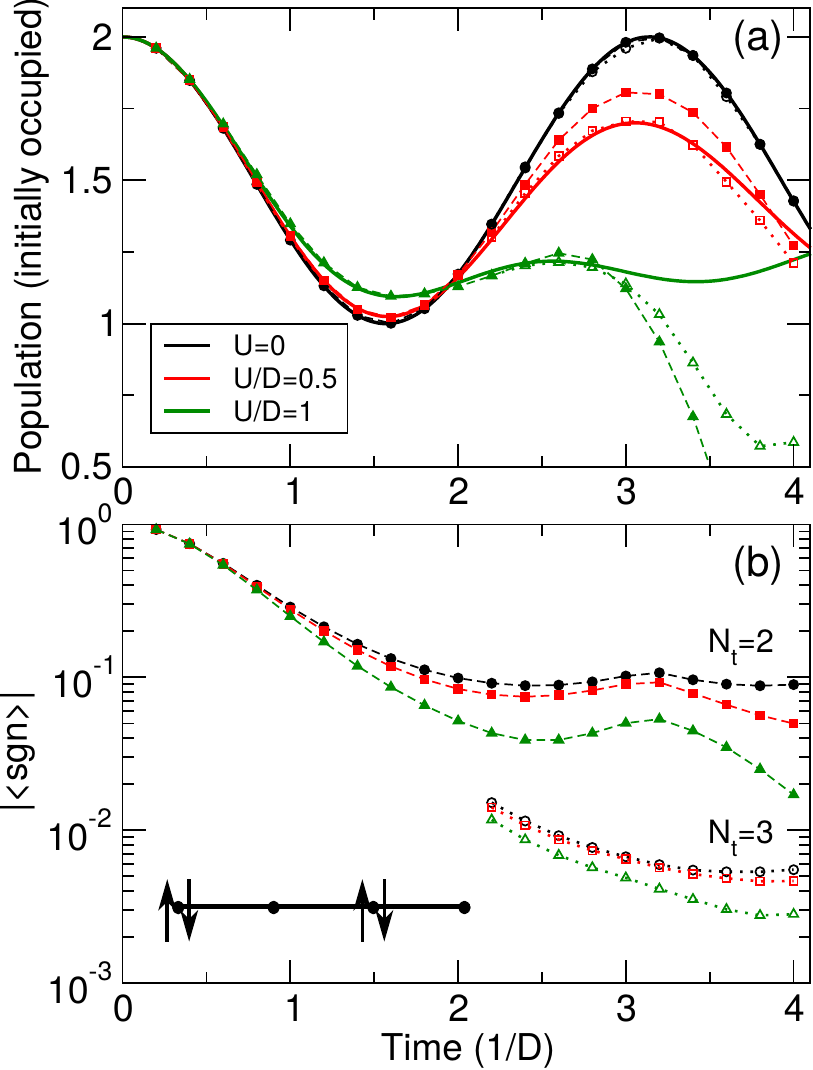}
    \caption{(Color online) (a) Time-dependent population of sites occupied in the initial CDW state of a tetramer for different interaction strengths. Solid lines represent exact results, full symbols connected by dashed lines are FPQMC results using $N_t=2$ real-time slices, while empty symbols connected by dotted lines are FPQMC results using $N_t=3$ real-time slices. The initial CDW state is schematically depicted in panel (b). (b) Time-dependent average sign of the FPQMC simulation using $N_t=2$ (full symbols connected by dashed lines) and $N_t=3$ (empty symbols connected by dotted lines) for different interaction strengths. In (a) and (b), FPQMC simulations using $N_t=3$ real-time slices are carried out only for $2<Dt\leq 4$.}
    \label{Fig:Fig_tetramer_090922}
\end{figure}

In Figs.~\ref{Fig:Fig_tetramer_090922}(a) and~\ref{Fig:Fig_tetramer_090922}(b) we benchmark our FPQMC method for time-dependent local densities on the example of the CDW state of the Hubbard tetramer, see the inset of Fig.~\ref{Fig:Fig_tetramer_090922}(b).
We follow the evolution of local charge densities on initially occupied sites for different ratios $U/D$, where $D$ is the half-bandwidth of the free-electron band ($D=2J$ for the tetramer).
For all the interaction strengths considered, taking $N_t=2$ real-time slices on each branch (4 slices in total) is sufficient to accurately describe evolution of local densities up to times $Dt\sim 2$, see full symbols in Fig.~\ref{Fig:Fig_tetramer_090922}(a).
At longer times, $2<Dt\leq 4$, taking $N_t=3$ improves results obtained using $N_t=2$, compare empty to full symbols in Fig.~\ref{Fig:Fig_tetramer_090922}(a).
Nevertheless, for the strongest interaction considered ($U/D=1$), 6 real-time slices are not sufficient to bring the FPQMC result closer to the exact result at times $3\leq Dt\leq 4$.
The average sign strongly depends on time and it drops by an order of magnitude upon increasing $N_t$ from 2 to 3, see Fig.~\ref{Fig:Fig_tetramer_090922}(b).
In spite of this, the discrepancy between the $N_t=3$ result and the exact result for $U/D=1$ cannot be ascribed to statistical errors, but to the systematic error of the FPQMC method (the minimum $N_t$ needed to obtain results with certain systematic error increases with both time and interaction strength).   

\subsubsection{Results on larger clusters}
\begin{figure*}[t]
 \centering
 \includegraphics{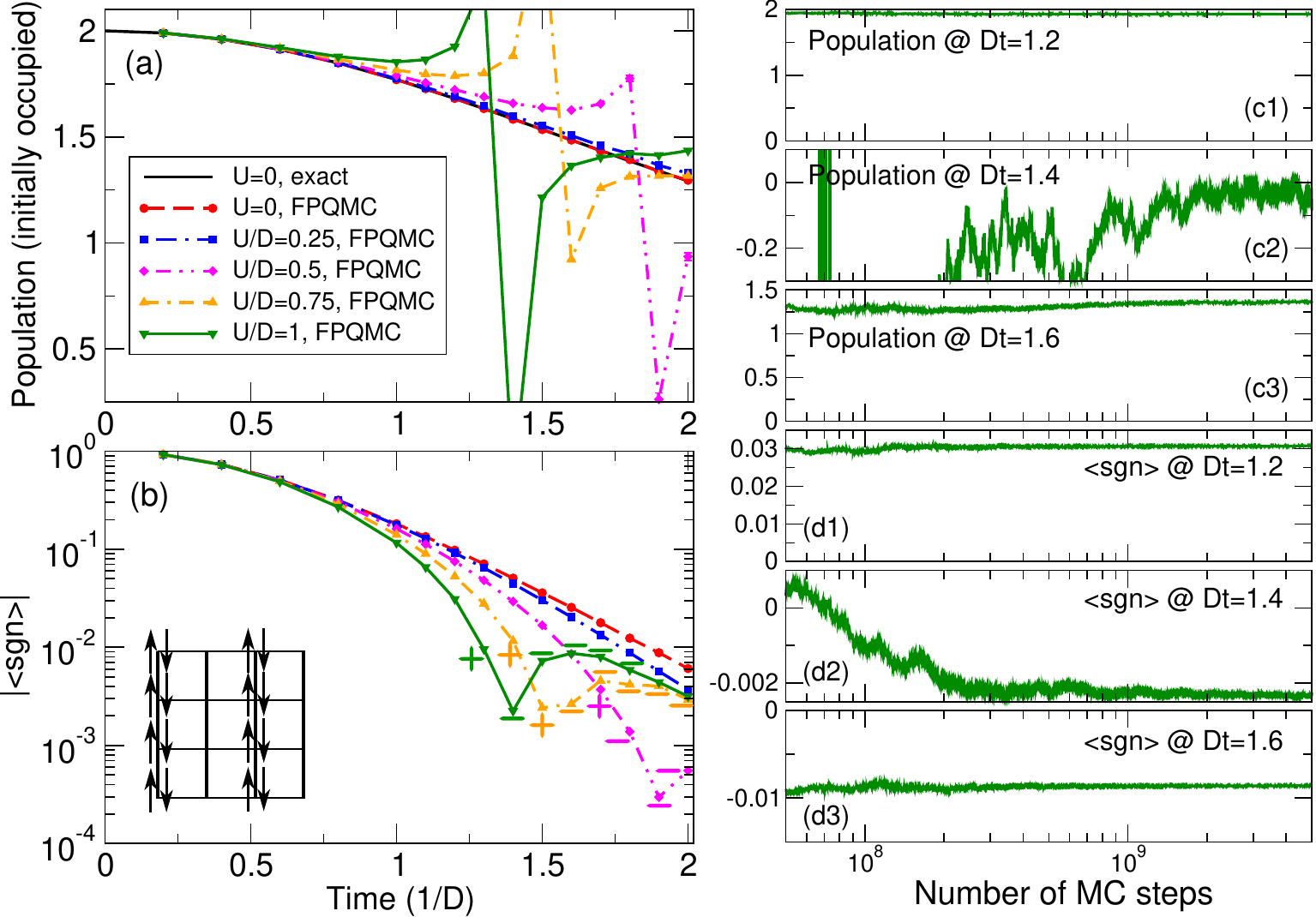}
 \caption{(Color online) (a) Time-dependent population of sites occupied in the initial CDW state of a $4\times 4$ cluster, which is schematically depicted in the inset of panel (b). FPQMC results using $N_t=2$ real-time slices (4 slices in total) are shown for five different interaction strengths (symbols) and compared with the noninteracting result (solid line). (b) Magnitude of the average sign as a function of time for different interaction strengths. Color code is the same as in panel (a). For $U/D=0.5,0.75,$ and 1 and $Dt\geq 1.2$, symbols "$+$" and "$-$" next to each point specify whether $\langle\mathrm{sgn}\rangle$ is positive or negative. (c) MC series for the population of initially occupied sites for $U/D=1$ and (c1) $Dt=1.2$, (c2) $Dt=1.4$, and (c3) $Dt=1.6$. (d) MC series for $\langle\mathrm{sgn}\rangle$ for $U/D=1$ and (d1) $Dt=1.2$, (d2) $Dt=1.4$, and (d3) $Dt=1.6$. Note the logarithmic scale on the abscissa in (c) and (d).}
 \label{Fig:Fig_ab}
\end{figure*}

Figure~\ref{Fig:Fig_ab}(a) summarizes the evolution of local charge densities on initially occupied sites of a half-filled $4\times 4$ cluster, on which the electrons are initially arranged as depicted in the inset of Fig.~\ref{Fig:Fig_ab}(b).
This state is representative of a CDW pattern formed by applying strong external density-modulating fields with wave vector $\mathbf{q}=(\pi,0)$.
The FPQMC method employs 4 real-time slices in total, i.e., the forward and backward branch are divided into $N_t=2$ identical slices each.
On the basis of the $N_t=2$ results in Fig.~\ref{Fig:Fig_tetramer_090922}(a), we present the FPQMC dynamics up to maximum time $Dt_\mathrm{max}=2$.
The extent of the dynamical sign problem is shown in Fig.~\ref{Fig:Fig_ab}(b).

At shortest times, $Dt\lesssim 1$, the results for all the interactions considered do not significantly differ from the noninteracting result.
The same also holds for the average sign.
As expected, the decrease of $|\langle\mathrm{sgn}\rangle|$ with time becomes more rapid as the interaction $U$ and time discretization $\Delta t=t/N_t$ are increased.
The oscillatory nature of $\langle\mathrm{sgn}\rangle$ as a function of time [see Eq.~\ref{Eq:A_a_t_w_w}] is correlated with the discontinuities in time-dependent populations observed in Fig.~\ref{Fig:Fig_ab}(a) for $U/D\geq 0.5$.
Namely, at shortest times and for all the interactions considered, $\langle\mathrm{sgn}\rangle$ is positive, while for sufficiently strong interactions it becomes negative at longer times.
This change is indicated in Fig.~\ref{Fig:Fig_ab}(b) by placing symbols "$+$" and "$-$" next to each relevant point.
We now see that the discontinuities in populations occur precisely around instants at which $\langle\mathrm{sgn}\rangle$ turns from positive to negative values.
Focusing on $U/D=1$, in Figs.~\ref{Fig:Fig_ab}(c1)--\ref{Fig:Fig_ab}(c3) we show MC series for the population of initially occupied sites at instants before [(c1)] and after [(c2),~(c3)] $\langle\mathrm{sgn}\rangle$ passes through zero.
The corresponding series for $\langle\mathrm{sgn}\rangle$ are presented in Figs.~\ref{Fig:Fig_ab}(d1)--\ref{Fig:Fig_ab}(d3).
Well before [Figs.~\ref{Fig:Fig_ab}(c1) and~\ref{Fig:Fig_ab}(d1)] and after [Figs.~\ref{Fig:Fig_ab}(c3) and~\ref{Fig:Fig_ab}(d3)] $\langle\mathrm{sgn}\rangle$ changes sign, the convergence with the number of MC steps is excellent, while it is somewhat slower close to the positive-to-negative transition point, see Figs.~\ref{Fig:Fig_ab}(c2) and~\ref{Fig:Fig_ab}(d2).
Still, the convergence at $Dt=1.4$ cannot be denied, albeit the statistical error of the population is larger than at $Dt=1.2$ and 1.6.
At longer times $Dt\geq 1.5$, when $\langle\mathrm{sgn}\rangle$ is negative and of appreciable magnitude, the population again falls in the physical range $[0,2]$. Nevertheless, at such long times, the systematic error may be large due to the coarse real-time discretization.

In Sec.~SV of the Supplementary Material we discuss FPQMC results for the dynamics of local charge densities starting from some other initial states.

\begin{figure*}[t]
 \centering
 \includegraphics[width=1.55\columnwidth]{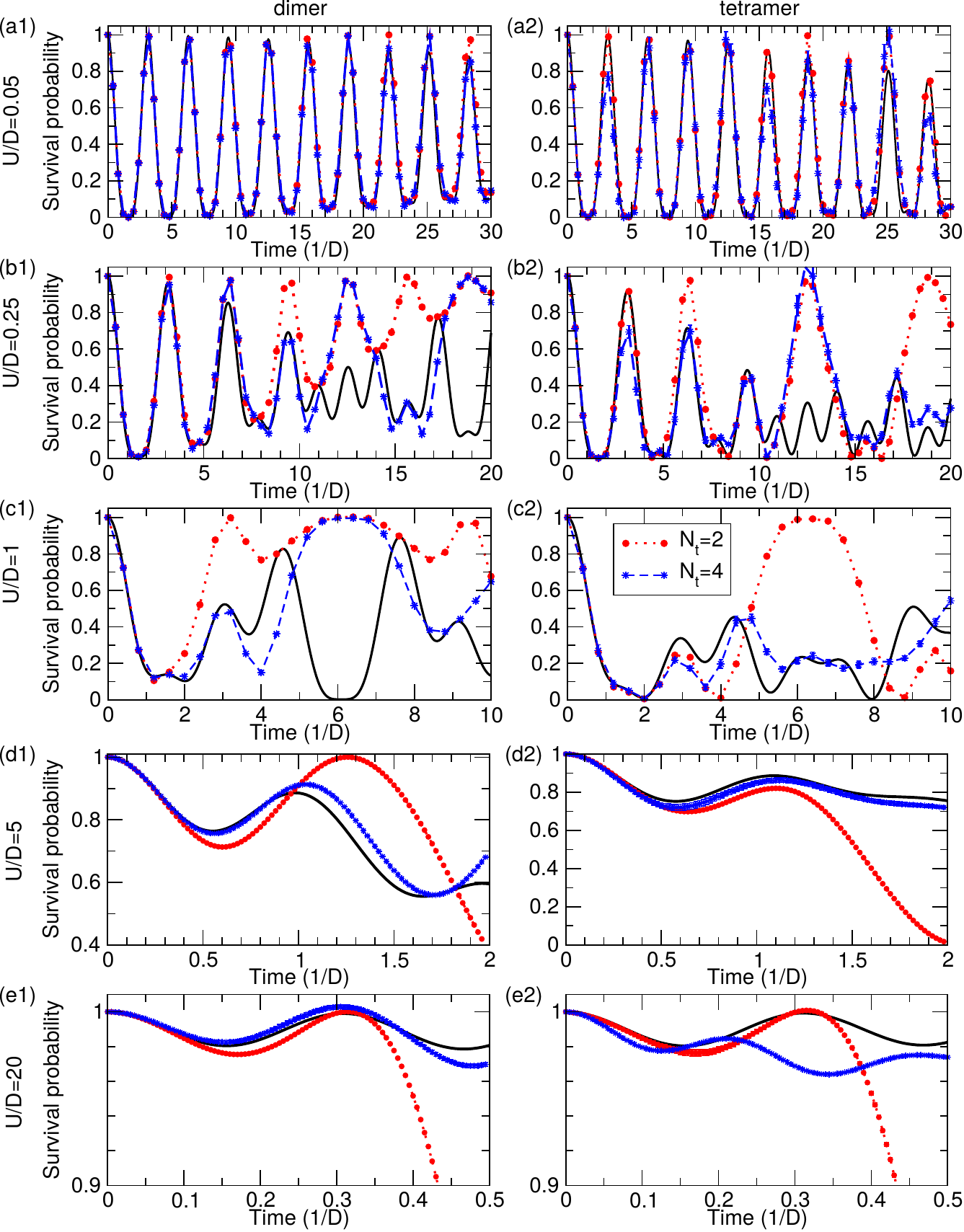}
 \caption{(Color online) Time dependence of the survival probability of the initial state $|\psi_\mathrm{CDW}\rangle$ or $|\psi_\mathrm{SDW}\rangle$ (see Table~\ref{Table:small_systems_scheme}) for the dimer [(a1)--(e1)] and tetramer [(a2)--(e2)] for five different interaction strengths starting from the noninteracting limit and approaching the atomic limit: $U/D=0.05$ [(a1) and (a2)], $U/D=0.25$ [(b1) and (b2)], $U/D=1$ [(c1) and (c2)], $U/D=5$ [(d1) and (d2)], and $U/D=20$ [(e1) and (e2)]. The ABQMC results with $N_t=2$ (red full circles) and $N_t=4$ (blue stars) are compared with the exact result (black solid lines). The dotted/dashed lines connecting subsequent circles/stars are guides to the eye. In most cases, the MC error bars are smaller than the linear size of the symbols.
 }
 \label{Fig:figure_dimer_tetramer}
\end{figure*}

\subsection{Time-dependent results using ABQMC method: Survival probability}

\label{Sec:numerics_t_dep}
\subsubsection{Benchmarks on small clusters}

We first benchmark our ABQMC method for the survival probability on Hubbard dimers and tetramers.
The initial states are schematically summarized in Table~\ref{Table:small_systems_scheme}. In both cases, we are at half-filling.

\begin{table}[htbp!]
 \begin{tabular}{c|c|c}
  system & $|\psi_\mathrm{CDW}\rangle$ & $|\psi_\mathrm{SDW}\rangle$\\
  \hline\hline
  dimer & \parbox[c]{8em}{
      \includegraphics[width=0.4in]{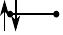}} & \parbox[c]{8em}{
      \includegraphics[width=0.4in]{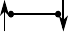}}\\
  \hline
  tetramer & \parbox[c]{8em}{
      \includegraphics[width=1.0in]{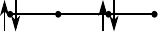}} & \parbox[c]{8em}{
      \includegraphics[width=1.0in]{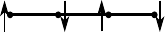}}
 \end{tabular}
 \caption{Schematic representations of the initial states of small systems on which the ABQMC method for $P(t)$ is benchmarked.}
 \label{Table:small_systems_scheme}
\end{table}

Figures~\ref{Fig:figure_dimer_tetramer}(a1)--\ref{Fig:figure_dimer_tetramer}(e2) present time evolution of the survival probability of the initial CDW-like and SDW-like states depicted in Table~\ref{Table:small_systems_scheme} for the dimer (left panels, $D=J$) and tetramer (right panels, $D=2J$) for different values of $U/D$ starting from the limit of weakly nonideal gas ($U/D=0.05$) and approaching the atomic limit ($U/D=20$). The results are obtained using $N_t=2$ (full red circles) and $N_t=4$ (blue stars) real-time slices and contrasted with the exact result (solid black lines). The ABQMC results with $N_t=2$ agree both qualitatively (oscillatory behavior) and quantitatively with the exact result up to $t_\mathrm{max}\sim 1/U$. Increasing $N_t$ from 2 to 4 may help decrease the deviation of the ABQMC data from the exact result at later times. Even when finer real-time discretization does not lead to better quantitative agreement, it may still help the ABQMC method qualitatively reproduce gross features of the exact result. The converged values of $|\langle\mathrm{sgn}\rangle|$ for the dimer and tetramer for $N_t=2,3,4$ are summarized in Table~\ref{Table:small_systems_av_sgn}. For the dimer, increasing $N_t$ by one reduces $|\langle\mathrm{sgn}\rangle|$ by a factor of 2. In contrast, in the case of the tetramer, increasing $N_t$ by one reduces $|\langle\mathrm{sgn}\rangle|$ by almost an order of magnitude.

\begin{table}
 \begin{tabular}{c|c|c|c}
  system & $N_t=2$ & $N_t=3$ & $N_t=4$\\
  \hline\hline
  dimer & 1/2 & 1/4 & 1/8\\
  \hline
  tetramer & 1/8 & $2.4\times 10^{-2}$ & $3\times 10^{-3}$
 \end{tabular}
 \caption{Modulus of the average sign for ABQMC simulations of $P(t)$ on dimer and tetramer with $N_t=2,3,$ and 4.}
 \label{Table:small_systems_av_sgn}
\end{table}

\subsubsection{Results on larger clusters}
\begin{figure*}[t]
 \centering
 \includegraphics{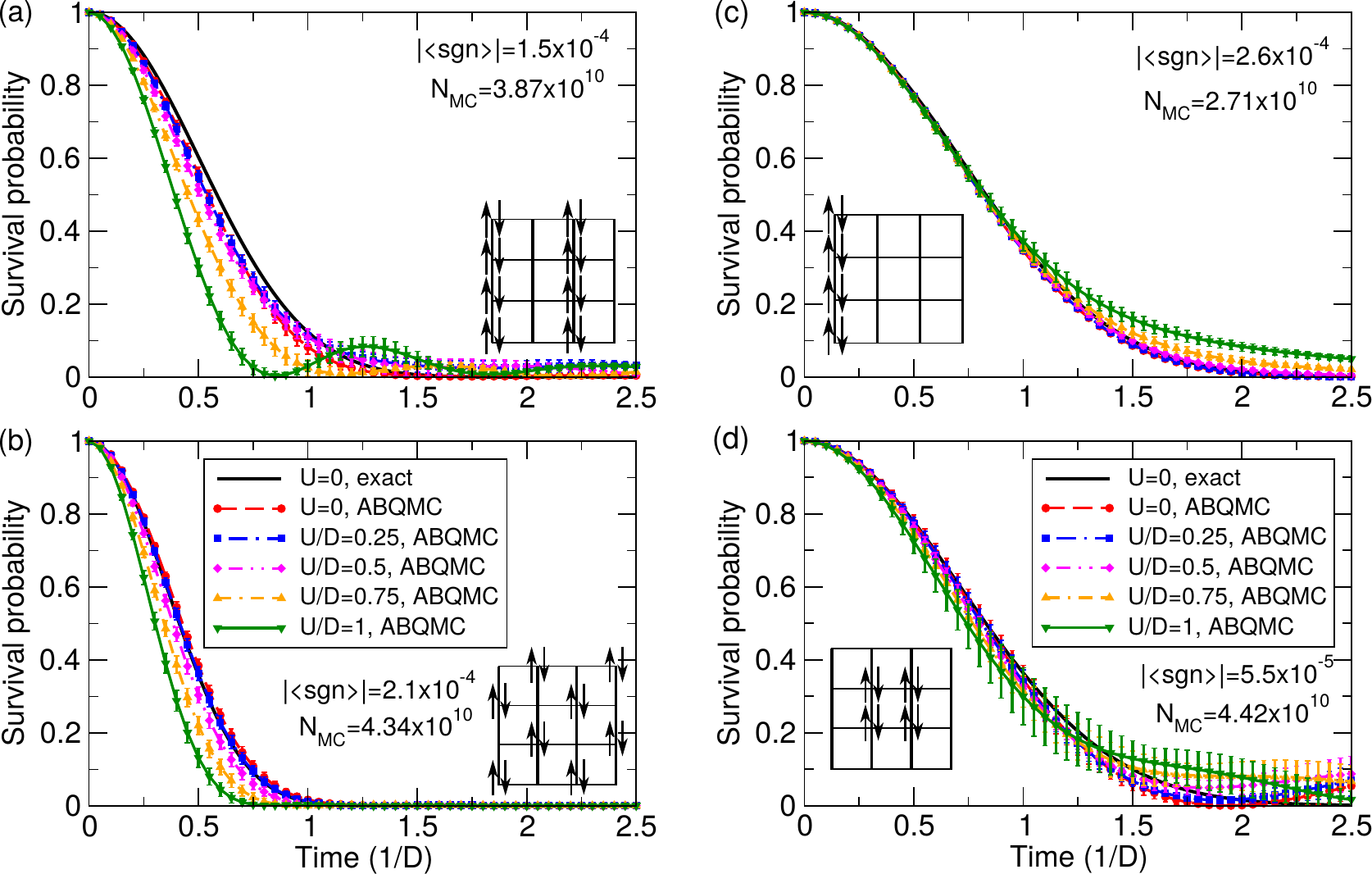}
 \caption{(Color online) Survival-probability dynamics of
the 16-electron states [in (a) and (b)] and 8-electron states [in (c) and (d)] that are schematically depicted in the respective insets. ABQMC results are shown for five different interaction strengths (symbols) and compared with the noninteracting result (solid line). We cite the converged value of the average sign $|\langle\mathrm{sgn}\rangle|$, as well as the total number $N_\mathrm{MC}$ of MC steps completed.}
 \label{Fig:Figs6-7}
\end{figure*}

We move on to discuss the survival-probability dynamics of different 16-electron and 8-electron states on a $4\times 4$ cluster. Figures~\ref{Fig:Figs6-7}(a) and~\ref{Fig:Figs6-7}(b) present $P(t)$ for 16-electron states schematically depicted in their respective insets.
These states are representative of CDW patterns formed by applying strong external density-modulating fields with wave vectors $\mathbf{q}=(\pi,0)$ in Fig.~\ref{Fig:Figs6-7}(a) and $\mathbf{q}=(\pi,\pi)$ in Fig.~\ref{Fig:Figs6-7}(b).
Figures~\ref{Fig:Figs6-7}(c) and~\ref{Fig:Figs6-7}(d) present $P(t)$ for 8-electron states schematically depicted in their respective insets. The ABQMC method employs $N_t=2$ real-time slices. The results are shown up to the maximum time $Dt_\mathrm{max}=2.5$, which we chose on the basis of the results presented in Fig.~\ref{Fig:figure_dimer_tetramer}(c2).

\begin{figure}
 \includegraphics[width=0.9\columnwidth]{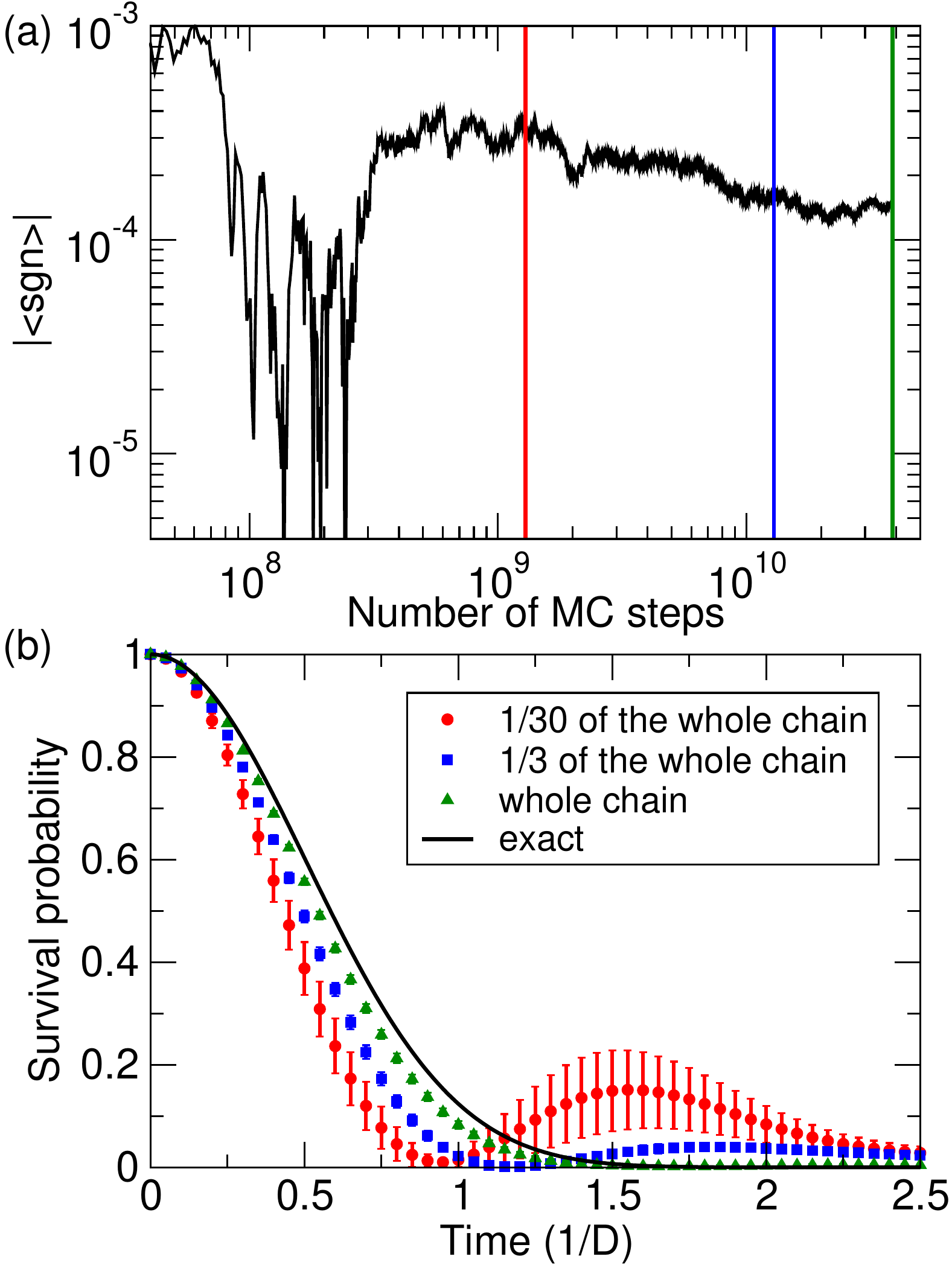}
 \caption{(Color online) (a) Average sign as a function of the number of MC steps in the ABQMC simulation of $P(t)$ for the 16-electron initial state schematically depicted in Fig.~\ref{Fig:Figs6-7}(a). (b) Time dependence of the survival probability for $U=0$ extracted using the first $1/30$ of the total number of MC steps completed ($1.29\times 10^9$ steps, full red circles), the first $1/3$ of the total number of MC steps completed ($1.29\times 10^{10}$ steps, full blue squares), and all the MC steps completed ($3.87\times 10^{10}$ steps, full green up-triangles). These results are compared to the exact result in the noninteracting limit, which are represented by the solid line. The vertical lines in (a), whose colors match the colors of the symbols in (b), denote the ending points of the simulations.}
 \label{Fig:U_0_interference_161221}
\end{figure}

As a sensibility check of our ABQMC results, we first compare the exact result in the noninteracting limit, see solid lines in Figs.~\ref{Fig:Figs6-7}(a)--\ref{Fig:Figs6-7}(d), with the corresponding ABQMC prediction, see full circles in Figs.~\ref{Fig:Figs6-7}(a)--\ref{Fig:Figs6-7}(d). While the exact and ABQMC result agree quite well in Figs.~\ref{Fig:Figs6-7}(b) and~\ref{Fig:Figs6-7}(c), the agreement in Figs.~\ref{Fig:Figs6-7}(a) and~\ref{Fig:Figs6-7}(d) is not perfect. Since no systematic errors are expected in ABQMC at $U=0$, the discrepancy must be due to statistical error. We confirm this expectation in Fig.~\ref{Fig:U_0_interference_161221} where we see that the obtained curve tends to the exact one with the increasing number of MC steps. The average sign cited in Fig.~\ref{Fig:Figs6-7}(d) suggests that more MC steps are needed to obtain fully converged results. Even though the converged average sign in Figs.~\ref{Fig:Figs6-7}(a)--\ref{Fig:Figs6-7}(c) is of the same order of magnitude, we find that the rate of convergence depends on both the number and the initial configuration of electrons.
 
In Figs.~\ref{Fig:Figs6-7}(a)--\ref{Fig:Figs6-7}(d), we observe that weak interactions ($U/D\lesssim 0.5$) do not cause any significant departure of $P(t)$ from the corresponding noninteracting result. On the other hand, the effect of somewhat stronger interactions on $P(t)$ depends crucially on the filling.
In the 16-electron case, the increasing interactions speed up the initial decay of $P$, see Figs.~\ref{Fig:Figs6-7}(a) and~\ref{Fig:Figs6-7}(b), while in the 8-electron case interactions have little effect at $Dt<1$, see Figs.~\ref{Fig:Figs6-7}(c) and~\ref{Fig:Figs6-7}(d).
This we attribute to the essential difference in the overall electron density and the relative role of the interaction term in the Hamiltonian.
In the 16-electron case, starting from the moderate coupling $U/D\sim 1$, there is a clear revival of the initial state in Fig.~\ref{Fig:Figs6-7}(a), while no such a revival is observed in Fig.~\ref{Fig:Figs6-7}(b). Furthermore, the memory loss of the initial density-wave pattern is more rapid in Fig.~\ref{Fig:Figs6-7}(a) than in Fig.~\ref{Fig:Figs6-7}(b), even at $U=0$. The revival of the inital state is observed in the 8-electron case as well: at $t<1/D$ there is barely any effect of the interaction, yet at longer times it boosts $P$. However, in contrast to the 16-electron case, the results in Figs.~\ref{Fig:Figs6-7}(c) and~\ref{Fig:Figs6-7}(d) exhibit a weaker dependence of the survival-probability dynamics on the initial density-wave pattern. Indeed, the exact results in the noninteracting case are identical for both patterns in Figs.~\ref{Fig:Figs6-7}(c) and~\ref{Fig:Figs6-7}(d). Except in the case of the $(\pi,\pi)$ wave, the interactions lead to a persistence of the initial pattern at longer times, $t>1/D$.
The precise form of temporal correlations that develop due to interactions apparently depends on the initial spatial arrangement of the electrons.

Section~SVI of the Supplementary Material presents additional ABQMC results for the time-dependent survival probability.

\section{Relation to other algorithms}
\label{Sec:relation_to_other_algorithms}
As mentioned in the Introduction, a variant of the FPQMC method was first proposed by De Raedt and Lagendijk in 1980s.~\cite{PhysRevLett.46.77,JStatPhys.27.731,PhysRep.127.233}
They, however, explicitly retain permutation operators appearing in Eq.~\eqref{Eq:permutation-ops-appendix} in their final expression for $Z$, see, e.g., Eq.~(3) in Ref.~\onlinecite{PhysRevLett.46.77} or Eqs.~(4.13)~and~(4.14) in Ref.~\onlinecite{PhysRep.127.233}.
On the other hand, we analytically perform summation over permutation operators, thus grouping individual contributions into determinants.
This is much more efficient [as factorial number of terms is captured in only $O(N^3)$ steps, or even faster] and greatly improves the average sign (cancellations between different permutations are already contained in the determinant, see Fig.~\ref{Fig:abqmc_vs_fpqmc_161222}).
The approach followed by De Raedt and Lagendijk later became known as permutation-sampling QMC, and the route followed by us is known as antisymmetric-propagator QMC,~\cite{JPhysSocJpn.53.963,JPhysAMathGen.38.6659} permutation-blocking QMC,~\cite{NewJPhys.17.073017} or fermionic-propagator QMC.~\cite{ContribPlasmaPhys.61.e202100112} The analytical summation over permutation operators entering Eq.~\eqref{Eq:permutation-ops-appendix} was first performed by Takahashi and Imada.~\cite{JPhysSocJpn.53.963}

Our FPQMC method employs the lowest-order STD [Eq.~\eqref{Eq:Suzuki-Trotter-general}], which was also used in the permutation-sampling QMC method of De Raedt and Lagendijk.~\cite{PhysRevLett.46.77,JStatPhys.27.731,PhysRep.127.233}
The maximum number of imaginary-time slices $N_\tau$ they could use was limited by the acceptance rates of MC updates, which decrease quickly with increasing $N_\tau$ and the cluster size $N_c$.
In our present implementation of FPQMC, we encounter the same issue, and our sampling becomes prohibitively inefficient when the total number of time slices is greater than 6--8, depending on the cluster size.
To circumvent this issue, the fermionic-propagator idea was combined with higher-order STDs~\cite{CommunMathPhys.51.183,JPhysSocJpn.53.3765,PhysLettA.146.319,JChemPhys.130.204109} and more advanced sampling techniques~\cite{PhysLettA.238.253,PhysRevLett.96.070601} to simulate equilibrium properties of continuum models of interacting fermions in the canonical~\cite{NewJPhys.17.073017,PhysRevB.93.085102} and grand-canonical~\cite{ContribPlasmaPhys.61.e202100112} ensemble.
More recent algorithmic developments enabled simulations with as much as 2000 imaginary-time slices,~\cite{ContribPlasmaPhys.59.e201800157} which is a great improvement.
Whether similar ideas can be applied to lattice systems to improve the efficiency of sampling is currently unclear.
Generally, more sophisticated STD schemes have been regarded as not useful in lattice-model applications.~\cite{PhysRep.220.53}
It is important to note that the success of the antisymmetric-propagator algorithms in continuous systems relies on weak degeneracy.
This corresponds to extremely low occupancy regime in lattice models, and it is precisely in this regime that our FPQMC method has the average sign close to 1 [see Figs.~\ref{Fig:state_equation_4x4_weak_U_091221}(b) and~\ref{Fig:state_equation_4x4_strong_U_101221}(b)], and the sampling is most efficient [see Sec.~SIII of the Supplementary Material]. Near half-filling, lattice models present a fundamentally different physics, which may ultimately require a substantially different algorithmic approach.

We further emphasize that the low acceptance rates and the resulting inefficiency of sampling that we encounter is directly related to the discrete nature of space in our model.
Some strategies for treating the analogous problem in continuous-space models may not be applicable here.
For example, in continuous-space models, acceptance rates of individual updates can be adjusted by moving electrons over shorter distances, so that the new configuration weight is less likely to be substantially different from the old one.
In contrast, in lattice models, electronic coordinates are discrete, and the minimum distance the electrons may cover is set by the lattice constant; In most cases moving a single electron by a single lattice spacing in a single time slice is sufficient to drastically reduce the configuration weight.
There is no general rule on how electrons should be moved to ensure that the new configuration weight is close to the original one.
This is particularly true for the updates that insert/remove a particle, and the problem becomes more pronounced with increasing $N_\tau$.
When each of the $N_\tau$ states $|\Psi_{i,l}\rangle$ [see Eq.~\eqref{Eq:def_C_eq}] is changed to $|\Psi'_{i,l}\rangle$, the chances that at least one of $\langle\Psi'_{i,l\oplus 1}|e^{-\Delta\tau H_0}|\Psi'_{i,l}\rangle$ is much smaller than $\langle\Psi_{i,l\oplus 1}|e^{-\Delta\tau H_0}|\Psi_{i,l}\rangle$ [see Eq.~\eqref{Eq:D_C_eq}] increase with $N_\tau$.
Our configuration weight is appreciable only in small mutually disconnected regions of the configuration space, the movement between which is difficult.
In Sec.~\ref{Sec:Conclusion}, we touch upon possible strategies to improve sampling of such a structured configuration space.

It is also important to compare our methods to the HF QMC method,~\cite{PhysRevLett.56.2521,PhD_Mikelsons} which is a well-established STD-based method for the treatment of the Hubbard model. The HF method is manifestly sign-problem-free, but only at particle--hole symmetry. The sign problem can become severe away from half-filling, or on lattices other than the simple square lattice with no longer-range hoppings. On the other hand, our FPQMC method is nearly sign-problem-free at low occupancy, but also near the half-filling, albeit only at strong coupling [see Figs.~\ref{Fig:state_equation_4x4_weak_U_091221}(b) and~\ref{Fig:state_equation_4x4_strong_U_101221}(b)]. The other important difference is that matrices manipulated in HF are of the size $N_c N_\tau$, while in FPQMC, the matrices are of the size $<2N_c$, i.e., given by the number of particles. Algorithmic complexity of the individual MC step in FPQMC scales only linearly with $N_\tau$, while in HF, the MC step may go as $O(N_\tau^2)$ [determinant is $O(N^3)$, but fast updates $O(N^2)$ are possible when the determinant is not calculated from scratch~\cite{PhD_Mikelsons}]. Low cost of individual steps in FPQMC has allowed us to perform as many as $\sim 10^{10}$ MC steps in some calculations. This advantage, however, weighs against an increased configuration space to be sampled. In HF the number of possible configurations is $2^{N_c N_\tau}$ (space is spanned by $N_c N_\tau$ auxiliary Ising spins), while in FPQMC it is $4^{N_c N_\tau}$ (although, symmetries can be used to significantly reduce the number of possible configurations). The ABQMC method manipulates matrices of the same size as does FPQMC, but twice the number, and the configuration space is a priori even bigger ($16^{N_c N_\tau}$). Our methods also have a technical advantage that the measurements of multipoint charge and spin correlation functions are algorithmically trivial and cheap. Especially in ABQMC, the densities in both coordinate and momentum space can be simply read off the configuration. This is not possible in HF where the auxiliary Ising spin only distinguishes between singly-occupied and doubly-occupied/empty site.
Most importantly, the ABQMC/FPQMC methods can be readily applied to canonical ensembles and pure states, which may not be possible with the HF method.
However, the HF is commonly used with tens of time slices, for lattice sizes of order $N_c=100-200$; in FPQMC, algorithmic developments related to configuration updates are necessary before it can become a viable alternative to HF in any wide range of applications.

Finally, we are unaware of any numerically exact method for large lattice systems which can treat the full Kadanoff--Baym--Keldysh contour, and yield real-time correlation functions. Our ABQMC method represents an interesting example of a real-time QMC method with manifestly no dynamical sign problem. However, the average sign is generally poor. To push ABQMC to larger number of time slices (as needed for calculation of the time-dependence of observables) and lattices larger than $4\times 4$ will require further work, and most likely, conceptually new ideas.

\section{Summary and outlook}
\label{Sec:Conclusion}

We revisit one of the earliest proposals for a QMC treatment of the Hubbard model, namely the permutation-sampling QMC method developed in Refs.~\onlinecite{PhysRevLett.46.77,JStatPhys.27.731,PhysRep.127.233}.
Motivated by recent progress in the analogous approach to continuous space models, we group all permutations into a determinant, which is known as the antisymmetric-propagator,~\cite{JPhysSocJpn.53.963} permutation-blocking,~\cite{NewJPhys.17.073017} fermionic-propagator~\cite{ContribPlasmaPhys.61.e202100112} idea.
We devise and implement two slightly different QMC methods.
Depending on the details of the STD scheme, we distinguish between (1) the FPQMC method, where snapshots are given by real-space Fock states, and determinants represent antisymmetric propagators between those states, and (2) the ABQMC method, where slices alternate between real and reciprocal space representation, and determinants are simple Slater determinants.
We thoroughly benchmark both methods against the available numerically exact data, and then use ABQMC to obtain some new results in the real-time domain.

The FPQMC method exhibits several promising properties.
The average sign can be close to 1, and does not drop off rapidly with either the size of the system, or the number of time slices.
In 1D, the method appears to be sign-problem-free.
At present, the limiting factor is not the average sign, but the ability to sample the large configuration space.
At discretizations finer than $N_\tau=6-8$, further algorithmic developments are necessary.
Nevertheless, our calculations show that excellent results for instantaneous correlators can be obtained with very few time slices, and efficiently.
Average density, double occupancy, and antiferromagnetic correlations can already be computed with high accuracy, at temperature and coupling strengths relevant for optical-lattice experiments.
The FPQMC method is promising for further applications in equilibrium setups.
In real-time applications, however, the sign problem in FPQMC is severe.

On the other hand, the ABQMC method has a significant sign problem in equilibrium applications, yet has some advantages in real-time applications.
In ABQMC, the sign problem is manifestly time-independent, and calculations can be performed for multiple times and coupling strengths with a single Markov chain.
We use this method to compute time-dependent survival probabilities of different density modulated states and identify several trends.
The relevant transient regime is short, and, based on benchmarks, we estimate the systematic error due to the time discretization here to be small.
Our results reveal that interactions speed up the initial decay of the survival probability, but facilitate a persistence of the initial charge-pattern at longer times.
Additionally, we observe a characteristic value of the coupling constant, $U\sim 0.5 D$, below which the interaction makes no visible effect on time-evolution.
These findings bare qualitative predictions for future ultracold-atom experiments, but are limited to dynamics at the shortest wave-lengths, as dictated by the maximal size of the lattice that we can treat.
We finally note that, within the ABQMC method, uniform currents, which are diagonal in the momentum representation, may be straightforwardly treated. 

There is room for improvements in both the ABQMC and FPQMC methods.
We already utilize several symmetries of the Hubbard model to improve efficiency and enforce some physical properties of solutions, but more symmetries can certainly be uncovered in the configuration spaces.
Further grouping of configurations connected by symmetries can be used to alleviate some of the sign problem or improve efficiency.
Also, sampling schemes may be improved along the lines of the recently proposed many-configuration Markov-chain MC, which visits an arbitrary number of configurations at every MC step.~\cite{simkovic2021manyconfiguration}
Moreover, a better insight in the symmetries of the configuration space may make a deterministic, structured sampling (along the lines of quasi-MC methods~\cite{BullAmerMathSoc.84.957, PhysRevLett.125.047702,PhysRevB.103.155104}) superior to the standard pseudo-random sampling.

\section*{Supplementary Material}
See the supplementary material for (i) a detailed description of MC updates within the FPQMC method, (ii) a detailed description of MC updates within the ABQMC method for time-dependent survival probability, (iii) details on the performance of the FPQMC method in equilibrium calculations, (iv) discussion on the applicability of the ABQMC method in equilibrium calculations, (v) additional FPQMC calculations of time-dependent local densities, (vi) additional ABQMC calculations of time-dependent survival probability, and (vii) formulation and benchmarks of ABQMC method in quench setups (on the full three-piece Kadanoff--Baym--Keldysh contour). 

\section*{Authors' Contributions}
J.V. conceived the research. V.J. developed the formalism and computational codes under the guidance of J.V., conducted all numerical simulations, analyzed their results, and prepared the initial version of the manuscript. Both authors contributed to the submitted version of the manuscript.

\acknowledgments
We acknowledge funding provided by the Institute of Physics Belgrade, through the grant by the Ministry of Education, Science, and Technological Development of the Republic of Serbia, as well as by the Science Fund of the Republic of Serbia, under the Key2SM project (PROMIS program, Grant No. 6066160). Numerical simulations were performed on the PARADOX-IV supercomputing facility at the Scientific
Computing Laboratory, National Center of Excellence for the Study of Complex Systems, Institute of Physics Belgrade.

\section*{Data availability}
The data that support the findings of this study are available from the corresponding author upon reasonable request.

\widetext
\appendix
\section{Many-body propagator as a determinant of single-particle propagators}
\label{App:single-particle-propagator}
The demonstration of Eqs.~\eqref{Eq:free-prop-1} and~\eqref{Eq:free-prop-2} can be conducted for each spin component separately.
We thus fix the spin index $\sigma$ and further omit it from the definition of the many-fermion state $|\Psi_i\rangle$ [Eq.~\eqref{Eq:Psi-i}].
Since $H_0$ is diagonal in the momentum representation, we express the state $|\Psi_i\rangle$ in the momentum representation
\begin{equation}
\label{Eq:i-to-k}
    |\Psi_i\rangle=\sum_{\{\mathbf{k}_j\}}\left(\prod_{l=1}^N\langle\mathbf{k}_l|\mathbf{r}_l\rangle c_{\mathbf{k}_l}^\dagger\right)|\emptyset\rangle
\end{equation}
and similarly for $|\Psi'_i\rangle$.
While the positions $\mathbf{r}_1,\dots,\mathbf{r}_N$ are ordered according to a certain rule, the wave vectors $\mathbf{k}_1,\dots,\mathbf{k}_N$ entering Eq.~\eqref{Eq:i-to-k} are not ordered, and there is no restriction on the sum over them.
We have
\begin{equation}
\label{Eq:sp-propagator-appendix}
    \begin{split}
        \langle\Psi'_i|e^{-\Delta\alpha H_0}|\Psi_i\rangle=&\sum_{\{\mathbf{k}'_l\}}\sum_{\{\mathbf{k}_l\}}e^{-\Delta\alpha\varepsilon_{\mathbf{k}_1}}\dots e^{-\Delta\alpha\varepsilon_{\mathbf{k}_N}}\times\\
        &\langle\mathbf{r}'_N|\mathbf{k}'_N\rangle\dots\langle \mathbf{r}'_1|\mathbf{k}'_1\rangle\langle\mathbf{k}_1|\mathbf{r}_1\rangle\langle\mathbf{k}_N|\mathbf{r}_N\rangle\times\\
        &\langle\emptyset|c_{\mathbf{k}'_N}\dots c_{\mathbf{k}'_1}c_{\mathbf{k}_1}^\dagger\dots c_{\mathbf{k}_N}^\dagger|\emptyset\rangle.
    \end{split}
\end{equation}
The sums over $\{\mathbf{k}'_l\}$ are eliminated by employing the identity~\cite{PhysRep.127.233}
\begin{equation}
\label{Eq:identity-appendix}
\begin{split}
    &\langle\emptyset|c_{\mathbf{k}'_N}\dots c_{\mathbf{k}'_1}c_{\mathbf{k}_1}^\dagger\dots c_{\mathbf{k}_N}^\dagger|\emptyset\rangle=\\
    &\sum_\mathcal{P}\mathrm{sgn}(\mathcal{P})\:\delta(\mathbf{k}'_1,\mathbf{k}_{\mathcal{P}(1)})\dots\delta(\mathbf{k}'_N,\mathbf{k}_{\mathcal{P}(N)})
\end{split}
\end{equation}
where the permutation operator $\mathcal{P}$ acts on the set of indices $\{1,\dots,N\}$, while $\mathrm{sgn}(\mathcal{P})=\pm 1$ is the permutation parity.
We then observe that
\begin{equation}
\label{Eq:inverse-permutation-appendix}
    \prod_{l=1}^N \langle\mathbf{r}'_l|\mathbf{k}_{\mathcal{P}(l)}\rangle=\prod_{l=1}^N \langle\mathbf{r}'_{\mathcal{P}^{-1}(l)}|\mathbf{k}_{l}\rangle,
\end{equation}
which permits us to perform the sums over individual $\mathbf{k}_l$s independently.
Combining Eqs.~\eqref{Eq:sp-propagator-appendix}--\eqref{Eq:inverse-permutation-appendix} and changing the permutation variable $\mathcal{P}'=\mathcal{P}^{-1}$ we eventually obtain
\begin{equation}
\label{Eq:permutation-ops-appendix}
 \begin{split}
    \langle\Psi'_i|e^{-\Delta\alpha H_0}|\Psi_i\rangle&=\sum_{\mathcal{P}'}\mathrm{sgn}(\mathcal{P}')\prod_{l=1}^N\langle\mathbf{r}'_{\mathcal{P}'(l)}|e^{-\Delta\alpha H_0}|\mathbf{r}_l\rangle\\
    &=\det S(\Psi'_i,\Psi_i,\Delta\alpha)
 \end{split}
\end{equation}
where matrix $S(\Psi'_i,\Psi_i,\Delta\alpha)$ (here without the spin index) is defined in Eq.~\eqref{Eq:free-prop-2}.
\section{Propagator of a free particle on the square lattice}
\label{App:lattice_propagator}
Here, we provide the expressions for the propagator of a free particle on the square lattice in imaginary [$\Delta\alpha=\Delta\tau$ in Eq.~\eqref{Eq:free-prop-2}] and real [$\Delta\alpha=i\Delta t$ in Eq.~\eqref{Eq:free-prop-2}] time. In imaginary time,
\begin{equation}
    \langle\mathbf{r}'|e^{-\Delta\tau H_0}|\mathbf{r}\rangle=\mathcal{I}(2J\Delta\tau,r'_x-r_x)\mathcal{I}(2J\Delta\tau,r'_y-r_y)
\end{equation}
where the one-dimensional imaginary-time propagator ($l$ is an integer)
\begin{equation}
\label{Eq:def_I_z_l}
    \mathcal{I}(z,l)=\frac{1}{N}\sum_{j=0}^{N-1}\cos\left(\frac{2\pi jl}{N}\right)\exp\left(z\cos\left(\frac{2\pi j}{N}\right)\right)
\end{equation}
is related to the modified Bessel function of the first kind $I_l(z)$ via
\begin{equation}
    \lim_{N\to\infty}\mathcal{I}(z,l)=\frac{1}{\pi}\int_0^\pi d\theta\:\cos(l\theta)\:e^{z\cos\theta}=I_l(z).
\end{equation}
In real time,
\begin{equation}
    \langle\mathbf{r}'|e^{-i\Delta t H_0}|\mathbf{r}\rangle=\mathcal{J}(2J\Delta t,r'_x-r_x)\mathcal{J}(2J\Delta t,r'_y-r_y)
\end{equation}
where the one-dimensional real-time propagator ($l$ is an integer)
\begin{equation}
    \mathcal{J}(z,l)=\frac{1}{N}\sum_{j=0}^{N-1}\cos\left(\frac{2\pi jl}{N}\right)\exp\left(iz\cos\left(\frac{2\pi j}{N}\right)\right)
\end{equation}
is related to the Bessel function of the first kind $J_l(z)$ via
\begin{equation}
    \lim_{N\to\infty}\mathcal{J}(z,l)=\frac{1}{\pi}\int_0^\pi d\theta\:\cos(l\theta)\:e^{iz\cos\theta}=i^l J_l(z).
\end{equation}
For finite $N$, $\mathcal{J}(z,2l)$ is purely real, while $\mathcal{J}(z,2l+1)$ is purely imaginary.

\section{Derivation of the FPQMC formulae that manifestly respect the dynamical symmetry of the Hubbard model}
\label{App:dynamical_symmetry}
Here, we derive the FPQMC expression for the time-dependent expectation value of a local observable [Eq.~\eqref{Eq:A_a_t_keldysh}] that manifestly respects the dynamical symmetry of the Hubbard model.

We start by defining the operation of the bipartite lattice symmetry, which is represented by a unitary, hermitean, and involutive operator $B$ ($B^\dagger=B=B^{-1}$) whose action on electron creation and annihilation operators in the real space is given as
\begin{equation}
\label{Eq:bipartite_real}
    B c_{\mathbf{r}\sigma}^{\left(\dagger\right)} B = (-1)^{r_x+r_y} c_{\mathbf{r}\sigma}^{\left(\dagger\right)}.
\end{equation}
In the momentum space, $B$ is actually the so-called $\pi$-boost~\cite{NatPhys.8.213}
\begin{equation}
\label{Eq:bipartite_momentum}
    B c_{\mathbf{k}\sigma}^{\left(\dagger\right)} B = c_{\mathbf{k}+\mathbf{Q},\sigma}^{\left(\dagger\right)}
\end{equation}
that increases the electronic momentum by $\mathbf{Q}=(\pi,\pi)$. The time reversal operator $T$ is an antiunitary (unitary and antilinear), involutive, and hermitean operator whose action on electron creation and annihilation operators in the real space is given as
\begin{equation}
\label{Eq:time_reversal_real}
 Tc_{\mathbf{r}\uparrow}^{\left(\dagger\right)}T=c_{\mathbf{r}\downarrow}^{\left(\dagger\right)},\quad
 Tc_{\mathbf{r}\downarrow}^{\left(\dagger\right)}T=-c_{\mathbf{r}\uparrow}^{\left(\dagger\right)},
\end{equation}
while the corresponding relations in the momentum space read as
\begin{equation}
\label{Eq:time_reversal_momentum}
 Tc_{\mathbf{k}\uparrow}^{\left(\dagger\right)}T=c_{-\mathbf{k}\downarrow}^{\left(\dagger\right)},\quad
 Tc_{\mathbf{k}\downarrow}^{\left(\dagger\right)}T=-c_{-\mathbf{k}\uparrow}^{\left(\dagger\right)}.
\end{equation}
Using Eqs.~\eqref{Eq:bipartite_real}--\eqref{Eq:time_reversal_momentum}, it follows that
\begin{equation}
\label{Eq:combined_symmetry_H_0_H_int}
 BH_0B=-H_0,\quad BH_\mathrm{int}B=H_\mathrm{int},\quad TH_0T=H_0,\quad TH_\mathrm{int}T=H_\mathrm{int}.
\end{equation}
In Sec.~\ref{Sec:QMC_dynamics}, we assumed that the initial state $|\psi(0)\rangle$ is an eigenstate of local density operators $n_{\mathbf{r}\sigma}$, which means that $B|\psi(0)\rangle=e^{i\chi_B}|\psi(0)\rangle$, see Eq.~\eqref{Eq:bipartite_real}.

The denominator of Eq.~\eqref{Eq:A_a_t_keldysh}
\begin{equation}
    A_\mathrm{den}(t)=\langle\psi(0)|e^{iHt}\:e^{-iHt}|\psi(0)\rangle
\end{equation}
is purely real,
$A_\mathrm{den}(t)=A_\mathrm{den}(t)^*$,
so that
\begin{equation}
\begin{split}
    A_\mathrm{den}(t)\approx&\frac{1}{2}\left\langle\psi(0)\left|\left(e^{iH_0\Delta t}e^{iH_\mathrm{int}\Delta t}\right)^{N_t}\left(e^{-iH_0\Delta t}e^{-iH_\mathrm{int}\Delta t}\right)^{N_t}\right|\psi(0)\right\rangle+\\
    &\frac{1}{2}\left\langle\psi(0)\left|\left(e^{iH_\mathrm{int}\Delta t}e^{iH_0\Delta t}\right)^{N_t}\left(e^{-iH_\mathrm{int}\Delta t}e^{-iH_0\Delta t}\right)^{N_t}\right|\psi(0)\right\rangle.
\end{split}
\end{equation}
We thus obtain
\begin{equation}
\label{Eq:A_den_reality}
    A_\mathrm{den}(t)\approx\sum_\mathcal{C}\left\{\mathrm{Re}\{\mathcal{D}_{2t}(\mathcal{C},\Delta t)\}\cos[\Delta\varepsilon_\mathrm{int}(\mathcal{C})\Delta t]-\mathrm{Im}\{\mathcal{D}_{2t}(\mathcal{C},\Delta t)\}\sin[\Delta\varepsilon_\mathrm{int}(\mathcal{C})\Delta t]\right\},
\end{equation}
where configuration $\mathcal{C}$ consists of $2N_t-1$ independent states $|\Psi_{i,2}\rangle,\dots,|\Psi_{i,2N_t}\rangle$, $|\Psi_{i,1}\rangle\equiv|\psi(0)\rangle$, while $\mathcal{D}_{2t}(\mathcal{C},\Delta t)$ and $\Delta\varepsilon_\mathrm{int}(\mathcal{C})$ are defined in Eqs.~\eqref{Eq:D_2t_def} and~\eqref{Eq:interaction_energy_difference_keldysh}, respectively.
The denominator is also invariant under time reversal, $A_\mathrm{den}(t)=A_\mathrm{den}(-t)$, which is not a consequence of a specific behavior of the initial state under time reversal, but rather follows from $A_\mathrm{den}(t)\equiv\langle\psi(0)|\psi(0)\rangle$. In other words, Eq.~\eqref{Eq:A_den_reality} should contain only contributions invariant under the transformation $\Delta t\to-\Delta t$. Using the bipartite lattice symmetry, under which $B|\Psi_{i,l}\rangle=e^{i\chi_l}|\Psi_{i,l}\rangle$, we obtain
\begin{equation}
\begin{split}
\label{Eq:D_t_bipartite}
    \mathcal{D}_{2t}(\mathcal{C},-\Delta t)&=\prod_{l=N_t+1}^{2N_t}\langle\Psi_{i,l\oplus 1}|BBe^{-iH_0\Delta t}BB|\Psi_{i,l}\rangle\prod_{l=1}^{N_t}\langle\Psi_{i,l\oplus 1}|BBe^{iH_0\Delta t}BB|\Psi_{i,l}\rangle\\
    &=\prod_{l=N_t+1}^{2N_t}\langle\Psi_{i,l\oplus 1}|e^{iH_0\Delta t}|\Psi_{i,l}\rangle\prod_{l=1}^{N_t}\langle\Psi_{i,l\oplus 1}|e^{-iH_0\Delta t}|\Psi_{i,l}\rangle=\mathcal{D}_{2t}(\mathcal{C},\Delta t).
\end{split}
\end{equation}
Equation~\eqref{Eq:A_den_reality} then reduces to
\begin{equation}
   A_\mathrm{den}(t)=\sum_\mathcal{C}\mathrm{Re}\{\mathcal{D}_{2t}(\mathcal{C},\Delta t)\}\cos[\Delta\varepsilon_\mathrm{int}(\mathcal{C})\Delta t].
\end{equation}

We now turn to the numerator of Eq.~\eqref{Eq:A_a_t_keldysh}
\begin{equation}
    A_\mathrm{num}(t)=\langle\psi(0)|e^{iHt}\:A_i\:e^{-iHt}|\psi(0)\rangle,
\end{equation}
which is also purely real, $A_\mathrm{num}(t)=A_\mathrm{num}(t)^*$, so that
\begin{equation}
\label{Eq:A_num_reality}
    A_\mathrm{num}(t)\approx\sum_\mathcal{C}\mathcal{A}_i(\Psi_{i,N_t+1})\left\{\mathrm{Re}\{\mathcal{D}_{2t}(\mathcal{C},\Delta t)\}\cos[\Delta\varepsilon_\mathrm{int}(\mathcal{C})\Delta t]-\mathrm{Im}\{\mathcal{D}_{2t}(\mathcal{C},\Delta t)\}\sin[\Delta\varepsilon_\mathrm{int}(\mathcal{C})\Delta t]\right\}.
\end{equation}
In the following discussion, we assume that the time reversal operation changes $|\psi(0)\rangle$ by a phase factor, $T|\psi(0)\rangle=e^{i\chi_T}|\psi(0)\rangle$. This, combined with $B|\psi(0)\rangle=e^{i\chi_B}|\psi(0)\rangle$, gives the assumption on $|\psi(0)\rangle$ that is mentioned before Eq.~\eqref{Eq:A_a_ABMC_symmetry}. We furthermore assume that $TBA_iBT=A_i$. Under these assumptions, the numerator is invariant under time reversal, $A_\mathrm{num}(-t)=A_\mathrm{num}(t)$, meaning that Eq.~\eqref{Eq:A_num_reality} should contain only contributions invariant under the transformation $\Delta t\to -\Delta t$. Using Eq.~\eqref{Eq:D_t_bipartite}, Eq.~\eqref{Eq:A_num_reality} reduces to
\begin{equation}
\label{Eq:A_num_final}
  A_\mathrm{num}(t)\approx\sum_\mathcal{C}\mathcal{A}_i(\Psi_{i,N_t+1})\:\mathrm{Re}\{\mathcal{D}_{2t}(\mathcal{C},\Delta t)\}\cos[\Delta\varepsilon_\mathrm{int}(\mathcal{C})\Delta t],
\end{equation}
and Eq.~\eqref{Eq:A_a_t_w_w} follows immediately.

An example of the initial state $|\psi(0)\rangle$ and the observable $A_i$ that satisfy $TB|\psi(0)\rangle=e^{i\chi}|\psi(0)\rangle$ and $TBA_iBT=A_i$ are the CDW state $|\psi_\mathrm{CDW}\rangle$ [Eq.~\eqref{Eq:cdw_state}] and the local charge density $A_i=\sum_\sigma n_{\mathbf{r}\sigma}$. While the time-reversal operation may change a general SDW state [Eq.~\eqref{Eq:sdw_state}] by more than a phase factor, Eq.~\eqref{Eq:A_num_final} is still applicable when the observable of interest is the local spin density $A_i=n_{\mathbf{r}\uparrow}-n_{\mathbf{r}\downarrow}$. This follows from the transformation law $T(n_{\mathbf{r}\uparrow}-n_{\mathbf{r}\downarrow})T=n_{\mathbf{r}\downarrow}-n_{\mathbf{r}\uparrow}$ and the fact that the roles of spin-up and spin-down electrons in the state $T|\psi_\mathrm{SDW}\rangle$ are exchanged with respect to the state $|\psi_\mathrm{SDW}\rangle$.

We now explain how we use Eq.~\eqref{Eq:pops_cdw_sdw} to enlarge statistics in computations of time-dependent local spin (charge) densities when the evolution starts from state $|\psi_\mathrm{SDW}\rangle$ in Eq.~\eqref{Eq:sdw_state} [$|\psi_\mathrm{CDW}\rangle$ in Eq.~\eqref{Eq:cdw_state}]. Let us limit the discussion to the spin (charge) density at fixed position $\mathbf{r}$. Suppose that we obtained Markov chains (of length $N_\mathrm{CDW}$) $\{\mathcal{N}_1^\mathrm{CDW}(t),\dots,\mathcal{N}_{N_\mathrm{CDW}}^\mathrm{CDW}(t)\}$ and $\{\mathcal{D}_1^\mathrm{CDW},\dots,\mathcal{D}_{N_\mathrm{CDW}}^\mathrm{CDW}\}$ for the numerator and denominator. Suppose also that we obtained Markov chains (of length $N_\mathrm{SDW}$) $\{\mathcal{N}_1^\mathrm{SDW}(t),\dots,\mathcal{N}_{N_\mathrm{SDW}}^\mathrm{SDW}(t)\}$ and $\{\mathcal{D}_1^\mathrm{SDW},\dots,\mathcal{D}_{N_\mathrm{SDW}}^\mathrm{SDW}\}$ for the numerator and denominator. Using these Markov chains, we found that the best result for the time-dependent local spin (charge) density is obtained by joining them into one Markov chain $\{\mathcal{N}_1^\mathrm{SDW}(t),\dots,\mathcal{N}_{N_\mathrm{SDW}}^\mathrm{SDW}(t),\mathcal{N}_1^\mathrm{CDW}(t),\dots,\mathcal{N}_{N_\mathrm{CDW}}^\mathrm{CDW}(t)\}$ of length $N_\mathrm{SDW}+N_\mathrm{CDW}$ for the numerator, and another Markov chain $\{\mathcal{D}_1^\mathrm{SDW},\dots,\mathcal{D}_{N_\mathrm{SDW}}^\mathrm{SDW},\mathcal{D}_1^\mathrm{CDW},\dots,\mathcal{D}_{N_\mathrm{CDW}}^\mathrm{CDW}\}$ of length $N_\mathrm{SDW}+N_\mathrm{CDW}$ for the denominator. If individual chain lengths $N_\mathrm{CDW}$ and $N_\mathrm{SDW}$ are sufficiently large, the manner in which the chains are joined is immaterial; here, we append the CDW chain to the SDW chain, and we note that other joining possibilities lead to the same final result (within the statistical error bars). To further reduce statistical error bars, we also combine SDW+CDW chains at all positions $\mathbf{r}$ that have the same spin (charge) density by the symmetry of the initial state.

\section{Derivation of the ABQMC formula for the survival probability}
\label{App:ABQMC_derive_survival}
We start from the survival-probability amplitude
\begin{equation}
    A_P(t)=\frac{\langle\psi(0)|e^{-iHt}|\psi(0)\rangle}{\langle\psi(0)|\psi(0)\rangle}
\end{equation}
whose numerator can be expressed as
\begin{equation}
\label{Eq:survival_prob_ampl_abqmc_1}
\begin{split}
    \langle\psi(0)|e^{-iHt}|\psi(0)\rangle&\approx\frac{1}{2}\left\langle\psi(0)\left|\left(e^{-iH_0\Delta t}e^{-iH_\mathrm{int}\Delta t}\right)^{N_t}\right|\psi(0)\right\rangle+
        \frac{1}{2}\left\langle\psi(0)\left|\left(e^{-iH_\mathrm{int}\Delta t}e^{-iH_0\Delta t}\right)^{N_t}\right|\psi(0)\right\rangle\\&=\sum_{\Psi_{i,2}\dots\Psi_{i,N_t}}\mathrm{Re}\left\{\prod_{l=1}^{N_t}\langle\Psi_{i,l\oplus 1}|e^{-iH_0\Delta t}|\Psi_{i,l}\rangle\right\} e^{-i\varepsilon_\mathrm{int}(\mathcal{C})\Delta t}\\
    &=\sum_{\Psi_{i,2}\dots\Psi_{i,N_t}}\sum_{\Psi_{k,1}\dots\Psi_{k,N_t}}\mathrm{Re}\left\{\prod_{l=1}^{N_t}\langle\Psi_{i,l\oplus 1}|\Psi_{k,l}\rangle\langle\Psi_{k,l}|\Psi_{i,l}\rangle e^{-i\varepsilon_0(\mathcal{C})\Delta t}\right\}e^{-i\varepsilon_\mathrm{int}(\mathcal{C})\Delta t}\\
    &=\sum_\mathcal{C}\left\{\mathrm{Re}\{\mathcal{D}(\mathcal{C})\}\cos[\varepsilon_0(\mathcal{C})\Delta t]+\mathrm{Im}\{\mathcal{D}(\mathcal{C})\}\sin[\varepsilon_0(\mathcal{C})\Delta t]\right\}e^{-i\varepsilon_\mathrm{int}(\mathcal{C})\Delta t}.
\end{split}
\end{equation}
In going from the second to the third line of Eq.~\eqref{Eq:survival_prob_ampl_abqmc_1}, we introduced spectral decompositions of $N_t$ factors $e^{-iH_0\Delta t}$. The configuration $\mathcal{C}$ entering the last line of Eq.~\eqref{Eq:survival_prob_ampl_abqmc_1} consists of $N_t-1$ independent states $|\Psi_{i,2}\rangle,\dots,|\Psi_{i,N_t}\rangle$ in the coordinate representation and $N_t$ independent states $|\Psi_{k,1}\rangle,\dots,|\Psi_{k,N_t}\rangle$ in the momentum representation, while $|\Psi_{i,1}\rangle\equiv|\psi(0)\rangle$. $\mathcal{D}(\mathcal{C})$ and $\varepsilon_0(\mathcal{C})$ are defined in Eqs.~\eqref{Eq:D_C_abqmc} and~\eqref{Eq:varepsilon_0_C}, respectively.
By virtue of the bipartite lattice symmetry, under which $\mathcal{D(\mathcal{C})}$ remains invariant, while $\varepsilon_0(\mathcal{C})$ changes sign, the summand containing $\sin[\varepsilon_0(\mathcal{C})\Delta t]$ in Eq.~\eqref{Eq:survival_prob_ampl_abqmc_1} vanishes, so that
\begin{equation}
\begin{split}
\label{Eq:survival_two_halves}
 \langle\psi(0)|e^{-iHt}|\psi(0)\rangle\approx\sum_\mathcal{C}\mathrm{Re}\{\mathcal{D}(\mathcal{C})\}\:\cos[\varepsilon_0(\mathcal{C})\Delta t]\:e^{-i\varepsilon_\mathrm{int}(\mathcal{C})\Delta t}.
 \end{split}
\end{equation}
This form should be used, e.g., when $|\psi(0)\rangle$ is the SDW state defined in Eq.~\eqref{Eq:sdw_state}. When the initial state is the CDW state defined in Eq.~\eqref{Eq:cdw_state}, $T|\psi(0)\rangle=e^{i\chi_T}|\psi(0)\rangle$, $\langle\psi(0)|e^{-iHt}|\psi(0)\rangle$ is purely real, so that
\begin{equation}
\label{Eq:survival_prob_ampl_final}
   \langle\psi(0)|e^{-iHt}|\psi(0)\rangle\approx\sum_\mathcal{C}\mathrm{Re}\{\mathcal{D}(C)\}\cos[\varepsilon_0(\mathcal{C})\Delta t]\cos[\varepsilon_\mathrm{int}(\mathcal{C})\Delta t].
\end{equation}
Equation~\eqref{Eq:P-t-form-for-MC-final} then follows by combining Eq.~\eqref{Eq:survival_prob_ampl_final} with $\langle\psi(0)|\psi(0)\rangle=\sum_\mathcal{C}\mathrm{Re}\{\mathcal{D}(\mathcal{C})\}$.

We now provide a formal demonstration of Eq.~\eqref{Eq:equal_P_t}. The partial particle--hole transformation is represented by a unitary, hermitean, and involutive operator $P$ ($P^\dagger=P=P^{-1}$) whose action on electron creation and annihilation operators in the real space is given as~\cite{ModPhysLett.4.759,PhysLettA.228.383}
\begin{eqnarray}
 Pc_{\mathbf{r}\uparrow}P=c_{\mathbf{r}\uparrow},\quad Pc_{\mathbf{r}\uparrow}^\dagger P=c_{\mathbf{r}\uparrow}^\dagger\label{Eq:P_real_1},\\
 Pc_{\mathbf{r}\downarrow}P=(-1)^{r_x+r_y}c_{\mathbf{r}\downarrow}^\dagger,\quad Pc_{\mathbf{r}\downarrow}^\dagger P=(-1)^{r_x+r_y}c_{\mathbf{r}\downarrow}\label{Eq:P_real_2}.
\end{eqnarray}
The interaction Hamiltonian $H_\mathrm{int}$ thus transforms under the partial particle--hole transformation as $PH_\mathrm{int}P=U\widehat{N}_\uparrow-H_\mathrm{int}$. 
The action of the partial particle--hole transformation in the momentum space reads as [$\mathbf{Q}=(\pi,\pi)$]
\begin{eqnarray}
 Pc_{\mathbf{k}\uparrow}P=c_{\mathbf{k}\uparrow},\quad Pc_{\mathbf{k}\uparrow}^\dagger P=c_{\mathbf{k}\uparrow}^\dagger,\\
 Pc_{\mathbf{k}\downarrow}P=c_{\mathbf{Q}-\mathbf{k},\downarrow}^\dagger,\quad Pc_{\mathbf{k}\downarrow}^\dagger P=c_{\mathbf{Q}-\mathbf{k},\downarrow}.
\end{eqnarray}
The kinetic energy, therefore, remains invariant under the partial particle--hole transformation, i.e., $PH_0P=H_0$. Equations~\eqref{Eq:P_real_1} and~\eqref{Eq:P_real_2} imply that $P|\emptyset\rangle=\prod_{\mathbf{r}\in\mathcal{U}}c_{\mathbf{r}\downarrow}^\dagger|\emptyset\rangle$. We then find that $P|\psi_\mathrm{CDW}\rangle=|\psi_{\mathrm{SDW}}\rangle$, i.e., the partial particle--hole transformation transforms the CDW state defined in Eq.~\eqref{Eq:cdw_state} into the SDW state defined in Eq.~\eqref{Eq:sdw_state} and vice versa.~\cite{NewJPhys.21.015003} The states $|\psi_\mathrm{CDW}\rangle$ and $|\psi_\mathrm{SDW}\rangle$ have the same number of spin-up electrons, while their numbers of spin-down electrons add to $N_c$. Using the combination of the partial particle--hole transformation $P$ and bipartite lattice transformation $B$ defined in Appendix~\ref{App:dynamical_symmetry}, one obtains
\begin{equation}
\label{Eq:prob_amps_cdw_sdw}
 \langle\psi_\mathrm{CDW}|e^{-iHt}|\psi_\mathrm{CDW}\rangle=e^{-iN_\uparrow(\psi)Ut}\langle\psi_\mathrm{SDW}|e^{-iHt}|\psi_\mathrm{SDW}\rangle^*,
\end{equation}
where $N_\uparrow(\psi)=\langle\psi_\mathrm{CDW}|\widehat{N}_\uparrow|\psi_\mathrm{CDW}\rangle=\langle\psi_\mathrm{SDW}|\widehat{N}_\uparrow|\psi_\mathrm{SDW}\rangle$ is the total number of spin-up electrons in CDW and SDW states. Equation~\eqref{Eq:equal_P_t} then follows immediately from Eq.~\eqref{Eq:prob_amps_cdw_sdw}.

A similar procedure to that described in Appendix~\ref{App:dynamical_symmetry} is used to combine Markov chains for the survival probabilities of the CDW and SDW states related by the dynamical symmetry in Eq.~\eqref{Eq:prob_amps_cdw_sdw}.

\section{Using the particle--hole symmetry to discuss the average sign of the FPQMC method for chemical potentials $\mu$ and $U-\mu$}
\label{App:full_ph_symmetric_sign}
The (full) particle--hole transformation is represented by a unitary, hermitean, and involutive operator $P_f$ ($P_f^\dagger=P_f=P_f^{-1}$) whose action on electron creation and annihilation operators in the real space is defined as~\cite{ModPhysLett.4.759,PhysLettA.228.383}
\begin{equation}
 P_fc_{\mathbf{r}\sigma}P_f=(-1)^{r_x+r_y}c_{\mathrm{r}\sigma}^\dagger.
\end{equation}
The corresponding formula in the momentum space reads as
\begin{equation}
 P_fc_{\mathbf{k}\sigma}P_f=c_{\mathbf{Q}-\mathbf{k},\sigma}^\dagger.
\end{equation}
Let us fix $J,U,T$, and $N_\tau$ and compute the equation of state $\rho_e(\mu)$ using Eq.~\eqref{Eq:A_a_final} in which $\mathcal{A}_i(\Psi_{i,l})=[N_\uparrow(\mathcal{C})+N_\downarrow(\mathcal{C})]/N_c$.
It is convenient to make the $\mu$-dependence in $\varepsilon_\mathrm{int}(\mathcal{C},\mu)$ explicit. In the sums entering Eq.~\eqref{Eq:A_a_final} we make the substitution
\begin{equation}
\mathcal{C}\to\mathcal{C}'=\{|\Phi_{i,l}\rangle=P_f|\Psi_{i,l}\rangle|l=1,\dots,N_\tau\}
\end{equation}
under which
\begin{eqnarray}
 \mathcal{D}_\beta(\mathcal{C},\Delta\tau)=\mathcal{D}_\beta(\mathcal{C}',\Delta\tau),\\
 \varepsilon_\mathrm{int}(\mathcal{C},\mu)=\varepsilon_\mathrm{int}(\mathcal{C}',U-\mu)+(U-2\mu)N_\tau N_c,\\
 N_\sigma(\mathcal{C}')=N_c-N_\sigma(\mathcal{C}).
\end{eqnarray}
It then follows that
\begin{equation}
\begin{split}
 &\frac{\sum_\mathcal{C}\mathcal{D}_\beta(\mathcal{C},\Delta\tau)\:e^{-\Delta\tau\varepsilon_\mathrm{int}(\mathcal{C},\mu)}[N_\uparrow(\mathcal{C})+N_\downarrow(\mathcal{C})]/N_c}{\sum_\mathcal{C}\mathcal{D}_\beta(\mathcal{C},\Delta\tau)\:e^{-\Delta\tau\varepsilon_\mathrm{int}(\mathcal{C},\mu)}}=\\&2-\frac{\sum_{\mathcal{C}'}\mathcal{D}_\beta(\mathcal{C}',\Delta\tau)\:e^{-\Delta\tau\varepsilon_\mathrm{int}(\mathcal{C}',U-\mu)}[N_\uparrow(\mathcal{C}')+N_\downarrow(\mathcal{C}')]/N_c}{\sum_{\mathcal{C}'}\mathcal{D}_\beta(\mathcal{C}',\Delta\tau)\:e^{-\Delta\tau\varepsilon_\mathrm{int}(\mathcal{C}',U-\mu)}}.
\end{split}
\end{equation}
The FPQMC simulations of the ratios in the last equation are performed for chemical potentials $\mu$ and $U-\mu$, which are symmetric with respect to the chemical potential $U/2$ at the half-filling. Since ${\varepsilon}_\mathrm{int}(\mathcal{C},\mu)$ and ${\varepsilon}_\mathrm{int}(\mathcal{C}',U-\mu)$ differ by a constant additive factor, the corresponding configuration weights differ by a constant multiplicative factor, and the average signs of the two FPQMC simulations are thus mutually equal.

\bibliography{apssamp}
\end{document}


\title{Supplementary Material for:
Fermionic-propagator and alternating-basis quantum Monte Carlo methods for correlated electrons on a lattice}

\author{Veljko Jankovi\'c}
    \email[]{veljko.jankovic@ipb.ac.rs}
    \affiliation{Institute of Physics Belgrade, University of Belgrade, Pregrevica 118, 11080 Belgrade, Serbia}
\author{Jak\v sa Vu\v ci\v cevi\'c}
    \email[]{jaksa.vucicevic@ipb.ac.rs}
    \affiliation{Institute of Physics Belgrade, University of Belgrade, Pregrevica 118, 11080 Belgrade, Serbia}


\maketitle

\section{FPQMC method: Monte Carlo updates}
\label{Sec:MC_updates}
Here, we present the Monte Carlo updates we use to move through the configuration space of our FPQMC method.

The configuration space is sampled through Markov chains starting from an, in principle arbitrary, configuration $\mathcal{C}_0$. The Metropolis--Hastings algorithm is used to determine the probability of transferring from configuration $\mathcal{C}_n$ at Monte Carlo step $n$ to configuration $\mathcal{C}_{n+1}$ at the subsequent Monte Carlo step $n+1$. The transition probability $W_{n\to n+1}$ is
\begin{equation}
 W_{n\to n+1}=W_{n\to n+1}^\mathrm{prop}W_{n\to n+1}^\mathrm{acc}
\end{equation}
where $W_{n\to n+1}^\mathrm{prop}$ is the probability of proposing the update from configuration $\mathcal{C}_n$ to configuration $\mathcal{C}_{n+1}$, while $W_{n\to n+1}^\mathrm{acc}$ determines the probability with which such a proposal is accepted. The Metropolis--Hastings acceptance rate reads as
\begin{equation}
 W_{n\to n+1}^\mathrm{acc}=\min\{1,R_{n\to n+1}\}
\end{equation}
where the acceptance ratio $R_{n\to n+1}=1/R_{n+1\to n}$ depends on the weights of the configurations involved, as well as on the proposal probabilities in both directions $\mathcal{C}_n\leftrightarrow\mathcal{C}_{n+1}$ in the following manner
\begin{equation}
 R_{n\to n+1}=\frac{w(\mathcal{C}_{n+1})W_{n+1\to n}^\mathrm{prop}}{w(\mathcal{C}_n)W_{n\to n+1}^\mathrm{prop}}.
\end{equation}
\subsection{Updates that conserve the number of particles}
\begin{enumerate}
    \item \texttt{change\_r\_local}\\
    We randomly choose one real-space state $|\Psi_{i,l_0}^n\rangle$ [in all the time-dependent computations we perform, $l_0\neq 1$ due to $|\Psi_{i,1}^n\rangle\equiv|\psi(0)\rangle$ at each Monte Carlo step $n$] and move an arbitrarily chosen electron (spin $\sigma$, position $\mathbf{r}_j^\sigma$) to a new position $\mathbf{s}_j^\sigma$ under the condition that the state $(\sigma,\mathbf{s}_j^\sigma)$ is unoccupied in $|\Psi_{i,l_0}^n\rangle$.
 
 The inverse move proceeds in the same manner as described above. The ratio of proposal weights is $\displaystyle{\frac{W^\mathrm{prop}_{n+1\to n}}{W^\mathrm{prop}_{n\to n+1}}=1}$.
 
 This move ensures that we sample configurations with different real-space electron patterns.
   \item\texttt{change\_r\_global}\\
   We randomly choose one real-space state $|\Psi_{i,l_0}^n\rangle$ [in all the time-dependent computations we perform, $l_0\neq 1$ due to $|\Psi_{i,1}^n\rangle\equiv|\psi(0)\rangle$ at each Monte Carlo step $n$] and replace it by a new state $|\Psi^{n+1}_{i,l_0}\rangle\neq|\Psi^n_{i,l_0}\rangle$. While the effects of this ``global'' move can be mimicked by multiple applications of its ``local'' version \texttt{change\_r\_local}, we found this move very useful in evenly sampling the configuration space, especially in time-dependent FPQMC simulations.
  \item\texttt{spin\_flip}---\textit{used only in equilibrium FPQMC simulations because it does not separately conserve the number of spin-up and spin-down electrons, yet it conserves the total electron number}\\
  We randomly choose spin $\sigma$ and attempt to increase/decrease the number of electrons of spin $\sigma$/$\overline{\sigma}$ by one.
  
  In each imaginary-time slice $l=1,\dots,N_\tau$, we choose an electron of spin $\sigma$ at position $\mathbf{r}$ from the real-space state $|\Psi_{i,l}^n\rangle$ and construct the real-space state $|\Psi_{i,l}^{n+1}\rangle$ by changing the electron's position $\mathbf{r}\to\mathbf{s}$ and spin $\sigma\to\overline{\sigma}$. The ratio of the proposal probabilities in the real space can be directly computed as $\displaystyle{\left(\frac{W^\mathrm{prop}_{n+1\to n}}{W^\mathrm{prop}_{n\to n+1}}\right)_{i,l}=\frac{N_\sigma(N_c-N_{\overline{\sigma}})}{(1+N_{\overline{\sigma}})(N_c-N_\sigma+1)}}$, where $N_\sigma$ and $N_{\overline{\sigma}}$ are numbers of electrons of spin $\sigma$ and $\overline{\sigma}$ in $|\Psi^n_{i,l}\rangle$.
  
  This move is very useful once the FPQMC starts sampling configurations whose total electron number (the sum of the numbers of spin-up and spin-down electrons) fluctuates around the value predicted by the fixed temperature and chemical potential. One value of the total electron number may be realized via many different combinations of numbers of spin-up and spin-down electrons, and it is precisely this move that enables efficient sampling through all these combinations.
\end{enumerate}  
\subsection{Updates that do not conserve the number of particles}
The updates \texttt{add\_particle} and \texttt{remove\_particle} are used only in equilibrium FPQMC simulations, when the number of particles fluctuates according to the fixed chemical potential and temperature.

The move \texttt{add\_particle}/\texttt{remove\_particle} adds/removes one electron from the configuration. These two moves are inverses of one another and their acceptance rates are mutually equal. The spin $\sigma$ of the electron added to/removed from the imaginary-time slice $l=1$ fixes that an electron added to/removed from the remaining imaginary-time slices $l=2,\dots,N_\tau$ must have the same spin $\sigma$.

Let $N_\sigma$ denote the number of electrons of spin $\sigma$ in state $|\Psi_{i,l}^n\rangle$ (before the update). In the first imaginary-time slice $l=1$, an electron can be added to any of $N_c-N_\uparrow+N_c-N_\downarrow$ empty single-particle states, while an electron can be removed from any of $N_\uparrow+N_\downarrow$ occupied single-particle states. Concerning the inverse move, an electron can be removed from any of $N_\uparrow+N_\downarrow+1$ occupied single-particle states, while an electron can be added to any of $2N_c-N_\uparrow-N_\downarrow+1$ empty single-particle states. We thus have
\begin{equation}
    \left(\frac{W^\mathrm{prop}_{n+1\to n}}{W^\mathrm{prop}_{n\to n+1}}\right)_{i,l=1}^\mathrm{add}=\frac{2N_c-N_\uparrow-N_\downarrow}{N_\uparrow+N_\downarrow+1},\quad
    \left(\frac{W^\mathrm{prop}_{n+1\to n}}{W^\mathrm{prop}_{n\to n+1}}\right)_{i,l=1}^\mathrm{rmv}=\frac{N_\uparrow+N_\downarrow}{2N_c-N_\uparrow-N_\downarrow+1}.
\end{equation}

In all other imaginary-time slices $l=2,\dots,N_\tau$, we have
\begin{equation}
    \left(\frac{W^\mathrm{prop}_{n+1\to n}}{W^\mathrm{prop}_{n\to n+1}}\right)_{i,l\geq 2}^\mathrm{add}=\frac{N_c-N_\sigma}{N_\sigma+1},\quad \left(\frac{W^\mathrm{prop}_{n+1\to n}}{W^\mathrm{prop}_{n\to n+1}}\right)_{i,l\geq 2}^\mathrm{rmv}=\frac{N_\sigma}{N_c-N_\sigma+1}.
\end{equation}

\subsection{Fast determinant updates}
The determinant $\mathcal{D}_\beta(\mathcal{C},\Delta\tau)$ is a product of $2N_\tau$ determinants of imaginary-time single-particle propagators on a lattice. Since all the updates can be seen as a single row/column change or addition/removal of a single row/column, changes in individual determinants may be efficiently computed using the formulae for fast determinant updates. These formulae, which provide the determinant ratio before and after the update, deal with the inverses of the corresponding matrices. We store these inverses in memory and recompute them from scratch each time a Monte Carlo update is accepted.

\subsection{Extraction of Monte Carlo results}
The average sign of FPQMC simulations is relatively large, so that, after the initial equilibration phase, the physical quantities we compute depend quite weakly on the number of Monte Carlo steps completed. All relevant quantities are measured at every Monte Carlo step, and the individual-step data are grouped into bins of length $L_b$, where $L_b\sim 10^4-10^5$, depending on the total number of steps completed. To provide the best possible estimate of a quantity, we discard the first 80\% of the simulation. The average of the binned data in the last 20\% of the simulation is taken as the Monte Carlo estimate of the quantity of interest. The statistical error is estimated as the root-mean-square deviation of the binned data in the final 20\% of the simulation from the above-computed average value. Such an estimate of the error is appropriate for statistically independent data. While we have not performed a systematic binning analysis, we may expect that the bin length we chose is sufficiently large that the data from different bins may be considered as statistically independent.

\pagebreak
\section{ABQMC method in real time: Monte Carlo updates}
\label{Sec:ABQMC_real_time_updates}
Here, we present the Monte Carlo updates we use to move through the configuration space of our real-time ABQMC method to compute the survival probability of an initial (pure) state. We only need updates that conserve the number of particles of each spin orientation.

Within the alternating-basis method, which employs both coordinate-space and momentum-space many-body states, we perform individual Monte Carlo updates on one of the two sets of states. The updates changing coordinate-space states are the updates \texttt{change\_r\_local} and \texttt{change\_r\_global} that we present in Sec.~\ref{Sec:MC_updates}. The updates that change momentum-space states are designed so as to respect the momentum-conservation law. While the proposal probabilities $W_{n\to n+1}^\mathrm{prop}$ for coordinate-space updates can be determined relatively straightforwardly (even analytically), see Sec.~\ref{Sec:MC_updates}, their determination for the momentum-space updates may be quite challenging due to the explicit momentum conservation. For all such moves, we can give no analytical expression for $W_{n\to n+1}^\mathrm{prop}$, and we have to devise computer algorithms capable of precisely enumerating all possible propositions that comply with the momentum (and also particle-number) conservation.
\begin{enumerate}
\item \texttt{add\_q}\\
The momentum $\mathbf{K}$ of each of the states $|\Psi_{k,l}^n\rangle$, $l=1,\dots,N_{\tau/t}$, is increased by a randomly chosen momentum $\mathbf{q}\neq 0$, thereby obtaining new states $|\Psi_{k,l}^{n+1}\rangle$, $l=1,\dots,N_{\tau/t}$, with momentum $\mathbf{K}+\mathbf{q}$. The real-space states are not changed, i.e., $|\Psi_{i,l}^n\rangle=|\Psi_{i,l}^{n+1}\rangle$. In each momentum-space state, momentum $\mathbf{q}$ is added to an electron of spin $\sigma$ that carries momentum $\mathbf{k}_{j,l}^\sigma$ under the condition that the state $(\sigma,\mathbf{k}_{j,l}^\sigma+\mathbf{q})$ is unoccupied in $|\Psi_{k,l}^n\rangle$. All possible momentum-accepting states $(\sigma,\mathbf{k}_{j,l}^\sigma)$ have to be explicitly enumerated and their number is denoted as $p_l^{+\mathbf{q}}$.  
 
 The inverse move starts from states $|\Psi^{n+1}_{k,l}\rangle$, $l=1,\dots,N_{\tau/t}$, with momentum $\mathbf{K}+\mathbf{q}$, and adds momentum $-\mathbf{q}$ to each of them. All possible momentum-accepting states in $|\Psi_{k,l}^{n+1}\rangle$ have to be explicitly enumerated and their number is denoted as $p_l^{-\mathbf{q}}$.
 
 The ratio of the proposal probabilities is then $\displaystyle{\frac{W^\mathrm{prop}_{n+1\to n}}{W^\mathrm{prop}_{n\to n+1}}=\prod_l\frac{p_l^{+\mathbf{q}}}{p_l^{-\mathbf{q}}}}.$
 
 This move ensures that we sample configurations from sectors featuring different electronic momenta.
 
 \item\texttt{exchange\_q}\\We randomly choose one momentum-space state $|\Psi_{k,l_0}^n\rangle$ in which we select two electrons of spins $\sigma_1$ and $\sigma_2$ and momenta $\mathbf{k}_{j_1}^{\sigma_1}$ and $\mathbf{k}_{j_2}^{\sigma_2}$, which are ordered so that $k_{j_1,y}+N_yk_{j_1,x}>k_{j_2,y}+N_yk_{j_2,x}$. The momenta are subsequently changed using $\mathbf{k}_{j_1}^{\sigma_1}\to\mathbf{k}_{j_1}^{\sigma_1}+\mathbf{q}$, $\mathbf{k}_{j_2}^{\sigma_2}\to\mathbf{k}_{j_2}^{\sigma_2}-\mathbf{q}$ ($\mathbf{q}\neq 0$). As a consequence, the net momentum of $|\Psi_{k,l_0}^n\rangle$ remains unchanged. Obviously, the move may be realized only when the states $(\sigma_1,\mathbf{k}_{j_1}^{\sigma_1}+\mathbf{q})$ and $(\sigma_2,\mathbf{k}_{j_2}^{\sigma_2}-\mathbf{q})$ are both unoccupied in $|\Psi_{k,l_0}^n\rangle$.
 
 We note that enumerating all possible momenta $\mathbf{q}\neq 0$ that may be transferred between the electrons is relatively simple for $\sigma_1\neq\sigma_2$ ($\sigma_2=\overline{\sigma_1}$), when the two electrons can be distinguished by their spins. It is then enough to go through all empty states $(\sigma_1,\mathbf{k}')$ to which the electron $(\sigma_1,\mathbf{k}_{j_1}^{\sigma_1})$ can be moved and to determine the corresponding momentum transfer $\mathbf{q}=\mathbf{k}'-\mathbf{k}_{j_1}^{\sigma_1}$. We then ask if the state $(\overline{\sigma_1},\mathbf{k}_{j_2}^{\overline{\sigma}_1}-\mathbf{q})$ is empty; in the affirmative case, we memorize the current $\mathbf{q}$ as one possible momentum transfer. The inverse move proceeds in a completely analogous manner by explicitly enumerating possible back-transfers.
 
 On the other hand, two electrons of the same spin are indistinguishable, and special care should be exercised to avoid double counting. For given $\mathbf{k}_{j_1}$ and $\mathbf{k}_{j_2}$ (we now omit $\sigma_1=\sigma_2$), 
 possible values of $\mathbf{q}$ follow from the construction that is schematically summarized in Fig.~\ref{Fig:exchange_q_solution}.
 \begin{figure}[htbp!]
  \centering
  \includegraphics[scale=2.0]{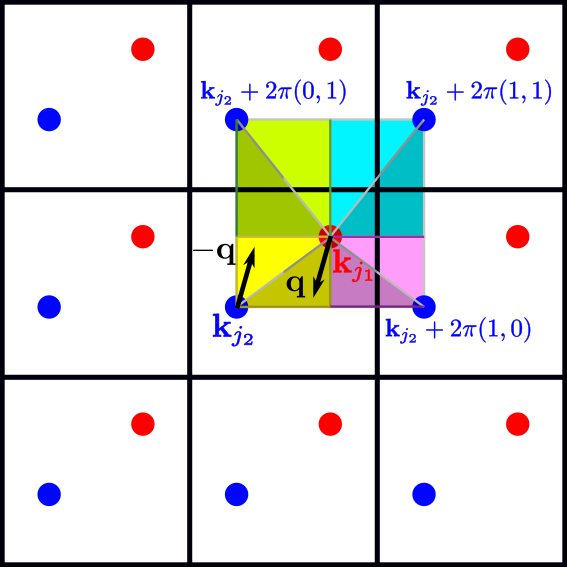}
  \caption{Construction used to correctly enumerate all possible momenta $\mathbf{q}\neq 0$ that can be exchanged between two electrons of equal spins that carry momenta $\mathbf{k}_{j_1}$ and $\mathbf{k}_{j_2}$. The momenta $\mathbf{q}$ that may be added to $\mathbf{k}_{j_1}$ (and subtracted from $\mathbf{k}_{j_2}$) are such that the final state $\mathbf{k}_{j_1}+\mathbf{q}$ is found in one of the four shaded triangles, while $\mathbf{k}_{j_2}-\mathbf{q}$ is found in one of the four unshaded triangles.}
  \label{Fig:exchange_q_solution}
 \end{figure}
 We make use of the periodic boundary conditions to construct a new unit cell (in the momentum space) such that the electron of momentum $\mathbf{k}_{j_1}$ is in its ``center'', while its vertices are at $\mathbf{k}_{j_2}$ and its periodic copies $\mathbf{k}_{j_2}+2\pi(1,0)$, $\mathbf{k}_{j_2}+2\pi(0,1)$, and $\mathbf{k}_{j_2}+2\pi(1,1)$. The ``central point'' and the four vertices partition the unit cell into four rectangular regions that are colored yellow (bottom left), green (top left), cyan (top right), and magenta (bottom right). The vectors $\mathbf{q}$ that may be added to $\mathbf{k}_{j_1}$ and subtracted from $\mathbf{k}_{j_2}$ are to be selected so that the final states $\mathbf{k}_{j_1}+\mathbf{q}$ and $\mathbf{k}_{j_2}-\mathbf{q}$ belong to just one half of each of the regions, the two halves being separated by the line connecting the ``central'' point and the vertices. The halves from which possible final states $\mathbf{k}_{j_1}+\mathbf{q}$, and thus possible momentum transfers $\mathbf{q}$, are selected is shaded. Choosing the momentum transfer such that the final state $\mathbf{k}_{j_1}+\mathbf{q}$ belongs to the other (unshaded) half is equivalent to assigning momentum $\mathbf{k}_{j_1}$ to the blue electron and $\mathbf{k}_{j_2}$ to the red electron, i.e., to exchanging momentum labels $\mathbf{k}_{j_1}$ and $\mathbf{k}_{j_2}$, which produces the setup equivalent to that presented in Fig.~\ref{Fig:exchange_q_solution}. The momentum transfer $\mathbf{q}$ along the edges of the four shaded triangles should be counted only once because of the periodic boundary conditions. For example, if we enumerate possible $\mathbf{q}$s along the vertical edge of the green shaded triangle, then we should not enumerate possible $\mathbf{q}$s along the vertical edge of the cyan shaded triangle. Moreover, if any of the lines connecting the ``center'' and the four edges contains any other lattice point, possible $\mathbf{q}$s along that line are subjected to the condition $|\mathbf{q}|\leq|\mathbf{k}_{j_1}-\mathbf{k}_{j_2}-2\pi(a_x,a_y)|/2$, where $(a_x,a_y)\in\{(0,0),(1,0),(0,1),(1,1)\}$. While this construction is appropriate for $k_{j_1,x}\neq k_{j_2,x}$ and $k_{j_1,y}\neq k_{j_2,y}$, further discussion is needed when either $k_{j_1,x}=k_{j_2,x}$ or $k_{j_1,y}=k_{j_2,y}$. In the inverse move, we start from the two electrons carrying momenta $\mathbf{k}'_{j_1}=\mathbf{k}_{j_1}+\mathbf{q}$ and $\mathbf{k}'_{j_2}=\mathbf{k}_{j_2}-\mathbf{q}$, we order them so that $k'_{j_1,y}+N_yk'_{j_1,x}>k'_{j_2,y}+N_yk'_{j_2,x}$, and repeat the above-described procedure.
 
 This move ensures that we sample configurations belonging to the sector of the chosen total electron momentum.
\end{enumerate}

\newpage

\section{Detailed performance of the FPQMC method applied to evaluate the equation of state}

\subsection{$U/J=4$, $T/J=1.0408$}

\begin{figure}[htbp!]
    \centering
    \includegraphics{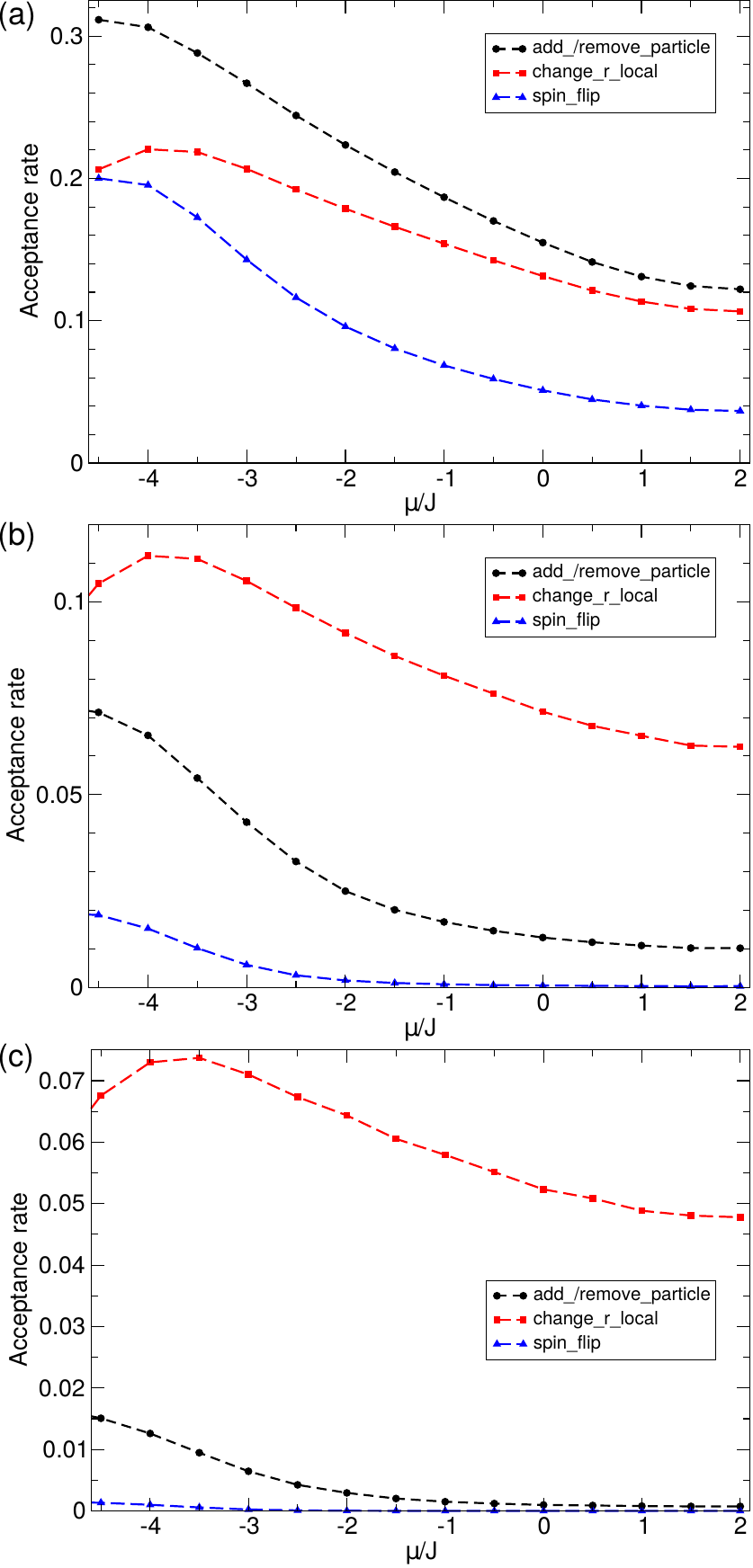}
    \caption{Acceptance rates of individual moves as a function of the chemical potential. FPQMC simulations are performed on a $4\times 4$ square-lattice cluster using (a) $N_\tau=2$, (b) $N_\tau=4$, and (c) $N_\tau=6$ imaginary-time slices. The remaining parameters are: $U/J=4$, $T/J=1.0408$. Acceptance rates generally decrease with $N_\tau$ and with the filling (chemical potential).}
    \label{Fig:merge_ar_U_4}
\end{figure}

\pagebreak

\begin{figure}[htbp!]
    \centering
    \includegraphics[scale=0.5]{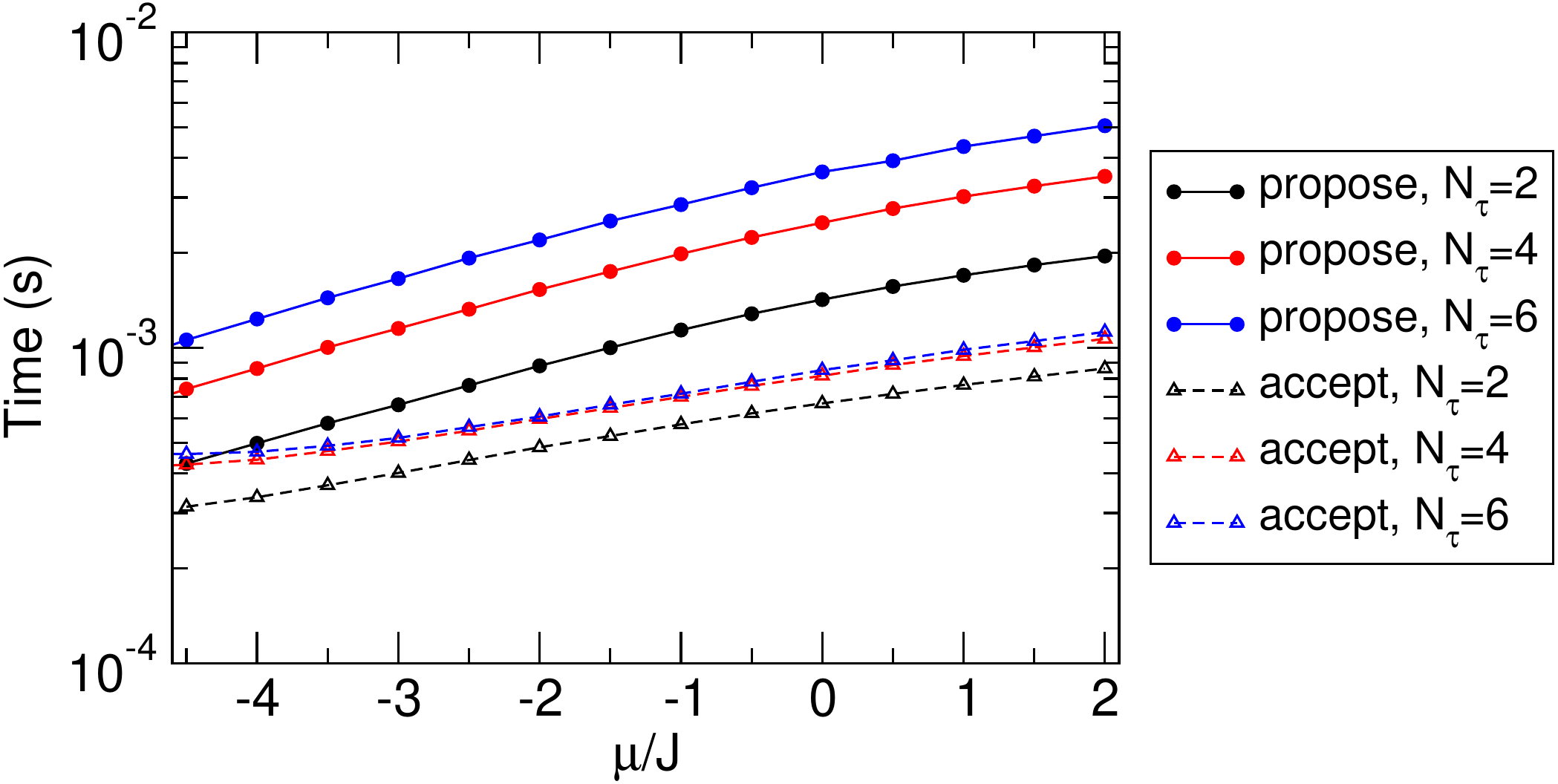}
    \caption{Average time needed to propose (full symbols connected by solid lines) and accept (empty symbols connected by dashed lines) a Monte Carlo update as a function of the chemical potential. FPQMC simulations are performed on a $4\times 4$ square-lattice cluster with different numbers of imaginary-time slices $N_\tau$. The remaining parameters are: $U/J=4$, $T/J=1.0408$.}
    \label{Fig:analyze_times_U_4}
\end{figure}

\pagebreak

\subsection{$U/J=24$, $T/J=1.0408$}

\begin{figure}[htbp!]
    \centering
    \includegraphics{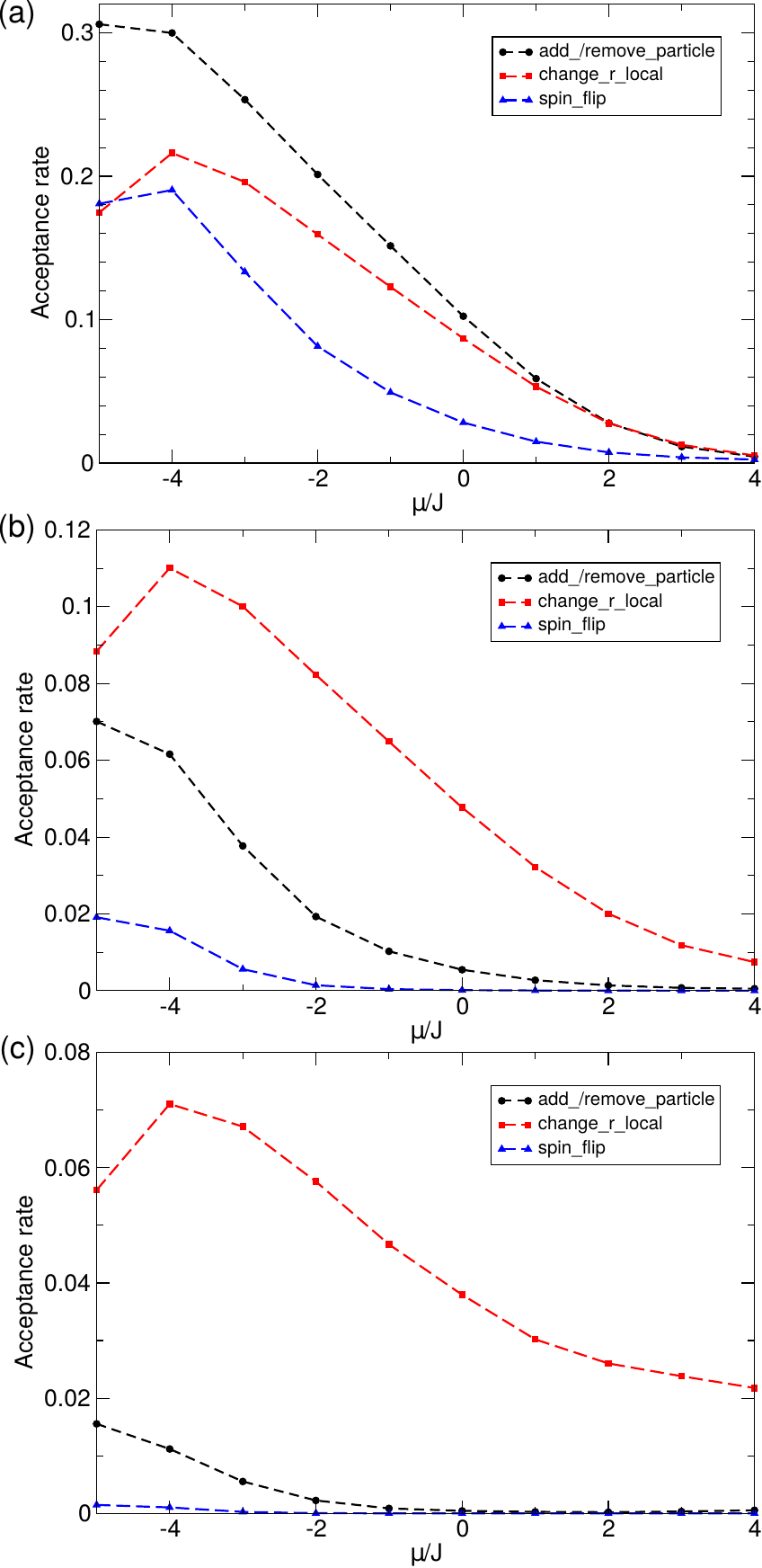}
    \caption{Acceptance rates of individual moves as a function of the chemical potential. FPQMC simulations are performed on a $4\times 4$ square-lattice cluster using (a) $N_\tau=2$, (b) $N_\tau=4$, and (c) $N_\tau=6$ imaginary-time slices. The remaining parameters are: $U/J=24$, $T/J=1.0408$. Acceptance rates generally decrease with $N_\tau$ and with the filling (chemical potential).}
    \label{Fig:merge_ar_U_24}
\end{figure}

\pagebreak

\begin{figure}[htbp!]
    \centering
    \includegraphics[scale=0.5]{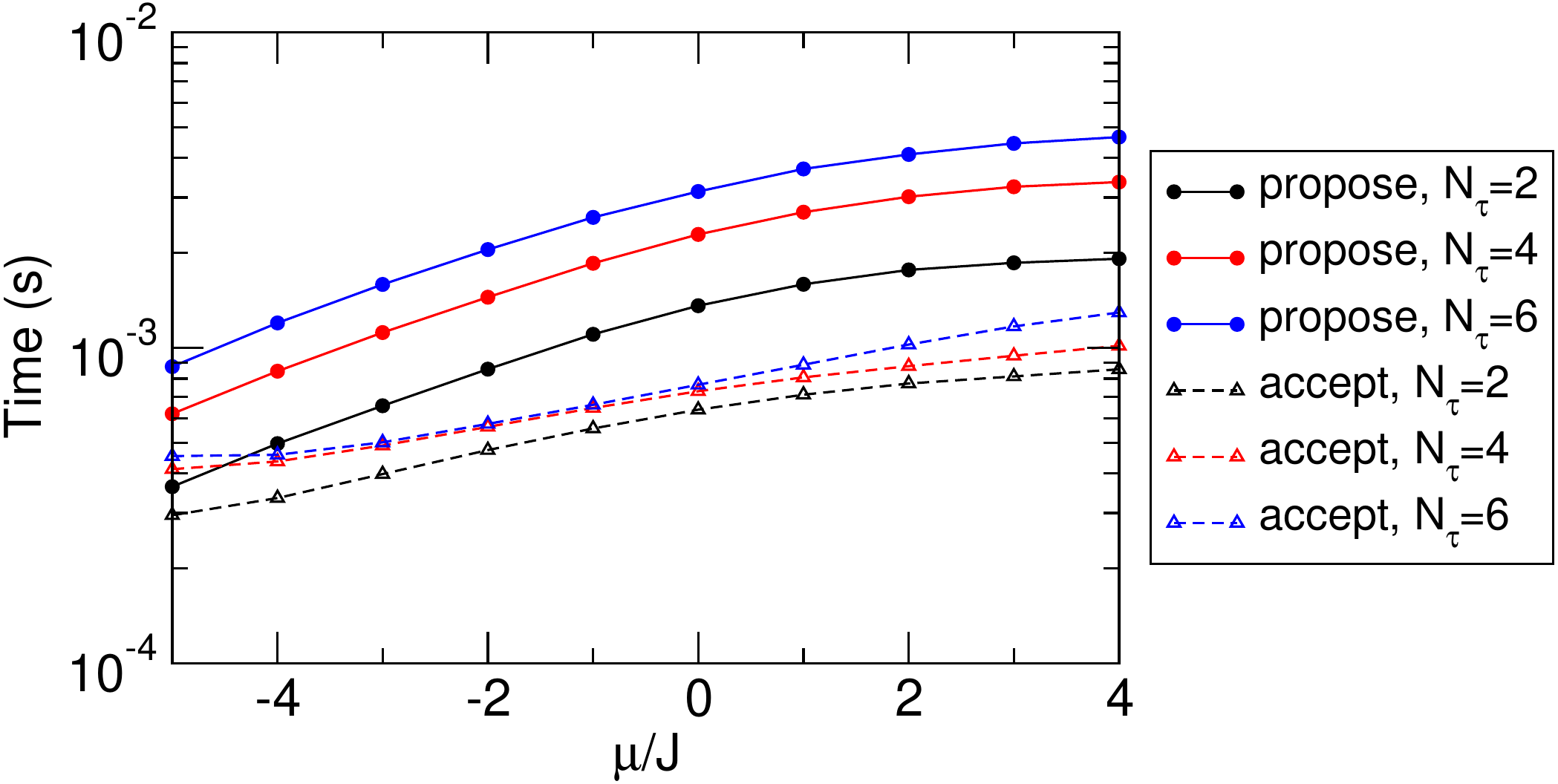}
    \caption{Average time needed to propose (full symbols connected by solid lines) and accept (empty symbols connected by dashed lines) an MC update as a function of the chemical potential. FPQMC simulations are performed on a $4\times 4$ square-lattice cluster with different numbers of imaginary-time slices $N_\tau$. The remaining parameters are: $U/J=24$, $T/J=1.0408$.}
    \label{Fig:analyze_times_U_24}
\end{figure}

\pagebreak

\section{ABQMC method applied to equilibrium situations}
Here, we derive the ABQMC method in equilibrium situations (Sec.~\ref{SSec:eq_abqmc_basic}) and present our implementation of the method (Sec.~\ref{SSec:eq_abqmc_updates}). Since the ABQMC method is equally adept at calculating quantities in real and momentum space, we complement the results presented in Fig.~7 of the main text by presenting the momentum distribution at different fillings (Sec.~\ref{SSec:eq_abqmc_results}). We conclude in Sec.~\ref{SSec:eq_abqmc_performance} by discussing acceptance rates and proposal/acceptance times of MC updates introduced in Sec.~\ref{SSec:eq_abqmc_updates}.



\subsection{ABQMC method in equilibrium: Basic equations}
\label{SSec:eq_abqmc_basic}
Dividing the imaginary-time interval $[0,\beta]$ into $N_\tau$ slices of length $\Delta\tau\equiv\beta/N_\tau$, employing the lowest-order STD, and inserting the spectral decompositions of $H_0$ and $H_\mathrm{int}$, we find the following ABQMC approximant for the partition function at temperature $T=1/\beta$:
\begin{equation}
\label{Eq:Z_most_general}
    Z\approx\sum_{\mathcal{C}} \mathcal{D}(\mathcal{C})e^{-\Delta\tau\varepsilon(\mathcal{C})}.
\end{equation}
The configuration
\begin{equation}
\mathcal{C}=\left\{|\Psi_{k,1}\rangle,\dots,|\Psi_{k,N_\tau}\rangle,|\Psi_{i,1}\rangle,\dots,|\Psi_{i,N_\tau}\rangle\right\}    
\end{equation}
resides on the contour depicted in Fig.~1(a) of the main text and consists of $N_\tau$ Fock states $|\Psi_{k,l}\rangle$ ($l=1,\dots,N_\tau$) in the momentum representation and $N_\tau$ Fock states $|\Psi_{i,l}\rangle$ in the coordinate representation. $\mathcal{D}(\mathcal{C})$ is the product of $2N_\tau$ Slater determinants
\begin{equation}
\label{Eq:D_C_eq}
    \mathcal{D}(\mathcal{C})\equiv\prod_{l=1}^{N_\tau}\langle\Psi_{i\oplus 1,l}|\Psi_{k,l}\rangle\langle\Psi_{k,l}|\Psi_{i,l\ominus 1}\rangle
\end{equation}
that stem from the sequence of basis alternations between the momentum and coordinate eigenbasis. The symbol $\varepsilon(\mathcal{C})$ stands for
\begin{equation}
\label{Eq:total-energy-C}
    \varepsilon(\mathcal{C})\equiv\sum_{l=1}^{N_\tau}\left[\varepsilon_0(\Psi_{k,l})+\varepsilon_\mathrm{int}(\Psi_{i,l})\right].
\end{equation}
The equilibrium expectation value an observable $A_a$ diagonal in either coordinate ($a=i$) or momentum ($a=k$) representation reads as
\begin{equation}
\label{Eq:A_a_expt_value}
    \langle A_a\rangle\approx\frac{1}{Z}\sum_\mathcal{C}\mathcal{D}(\mathcal{C})e^{-\Delta\tau\varepsilon(\mathcal{C})}\frac{1}{N_\tau}\sum_{l=1}^{N_\tau}\mathcal{A}_a(\Psi_{a,l}),
\end{equation}
where
\begin{equation}
\label{Eq:expect_value_A_slice}
    \mathcal{A}_a(\Psi_{a,l})\equiv\langle\Psi_{a,l}|A_a|\Psi_{a,l}\rangle.
\end{equation}
The evaluation of Eq.~\eqref{Eq:A_a_expt_value} using the importance-sampling MC procedure is complicated by the fact that $\mathcal{D}(\mathcal{C})$ is a complex number defined by its modulus and phase. While the modulus can be included in the weight of configuration $\mathcal{C}$, the phase gives rise to the so-called phase problem, which is generally much harder to curb than the ordinary sign problem. However, since the STD preserves the equality $Z=Z^*$,~\footnote{This follows from the cyclic invariance under the trace.} Eq.~\eqref{Eq:Z_most_general} can be replaced by
\begin{equation}
\label{Eq:Z-real-weights}
    Z\approx\sum_{\mathcal{C}} \mathrm{Re}\{\mathcal{D}(\mathcal{C})\}\:e^{-\Delta\tau\varepsilon(\mathcal{C})},
\end{equation} and we remain with the ordinary sign problem. Equation~\eqref{Eq:A_a_expt_value} should then be replaced by
\begin{equation}
\label{Eq:A_a_final}
    \langle A_a\rangle\approx\frac{\sum_\mathcal{C}\mathrm{Re}\{\mathcal{D}(\mathcal{C})\}\:e^{-\Delta\tau\varepsilon(\mathcal{C})}\frac{1}{N_\tau}\sum_{l=1}^{N_\tau}\mathcal{A}_a(\Psi_{a,l})}{\sum_\mathcal{C}\mathrm{Re}\{\mathcal{D}(\mathcal{C})\}\:e^{-\Delta\tau\varepsilon(\mathcal{C})}}.
\end{equation}
Defining the weight $w(\mathcal{C})$ of configuration $\mathcal{C}$ as
\begin{equation}
    w(\mathcal{C})\equiv|\mathrm{Re}\{\mathcal{D}(\mathcal{C})\}|e^{-\Delta\tau\varepsilon(\mathcal{C})},
\end{equation}
Eq.~\eqref{Eq:A_a_final} is rewritten as
\begin{equation}
\label{Eq:A-a-importance-sampling}
    \langle A_a\rangle\approx\frac{\left\langle\mathrm{sgn}(\mathcal{C})\frac{1}{N_\tau}\sum_{l=1}^{N_\tau}\mathcal{A}_a(\Psi_{a,l})\right\rangle_w}{\left\langle\mathrm{sgn}(\mathcal{C})\right\rangle_w}
\end{equation}
where $\langle\dots\rangle_w$ denotes the weighted average over all $\mathcal{C}$ with respect to the weight $w(\mathcal{C})$; $\mathrm{sgn}(\mathcal{C})\equiv\mathrm{Re}\{\mathcal{D}(\mathcal{C})\}/|\mathrm{Re}\{\mathcal{D}(\mathcal{C})\}|$ is the sign of configuration $\mathcal{C}$, while $|\langle\mathrm{sgn}\rangle|\equiv|\langle\mathrm{sgn}(\mathcal{C})\rangle_w|$ is the average sign of the ABQMC simulation.

\subsection{ABQMC method in equilibrium: Monte Carlo updates}
\label{SSec:eq_abqmc_updates}
Apart from previously introduced updates \texttt{change\_r\_local} and \texttt{change\_r\_global} in real space (Sec.~\ref{Sec:MC_updates}) and \texttt{add\_q} and \texttt{exchange\_q} in momentum space (Sec.~\ref{Sec:ABQMC_real_time_updates}), we need updates that insert/remove a particle, which are different from their counterparts in Sec.~\ref{Sec:MC_updates}.

The ABQMC move \texttt{add\_particle}/\texttt{remove\_particle} adds/removes one electron from the configuration. These two moves are inverses of one another. While analogous moves are relatively simply implemented in the FPQMC algorithm, here, special care is to be exercised because of the particle-number and momentum conservation. In more detail, the spin $\sigma$ of the electron added to/removed from the imaginary-time slice $l=1$ fixes that an electron added to/removed from the remaining imaginary-time slices $l=2,\dots,N_\tau$ must have the same spin $\sigma$. Furthermore, the momentum $\mathbf{q}$ of the electron added to/removed from the imaginary-time slice $l=1$ fixes that momentum change upon addition/removal of an electron in the remaining imaginary-time slices $l=2,\dots,N_\tau$ must be precisely $\mathbf{q}$. This requirement may be realized in many different ways. One trivial possibility is to add/remove the electron to/from the state $(\sigma,\mathbf{q})$ if this state is empty/occupied. On the other hand, we may add/remove the electron to/from state $(\sigma,\mathbf{k})$, in which case we should find another electron $(\sigma_\mathrm{comp},\mathbf{k}_\mathrm{comp})$ (of arbitrary spin $\sigma_\mathrm{comp}$) to compensate for the difference in the momentum change from $\pm\mathbf{q}$, where $+$/$-$ sign is for electron addition/removal. The electron in the state $(\sigma_\mathrm{comp},\mathbf{k}_\mathrm{comp})$ that may receive the momentum difference $\pm(\mathbf{q}-\mathbf{k})$ will be termed the compensating electron. The situation is relatively simple when the spin of the compensating electron is $\overline{\sigma}$ because the added electron and the compensating electron may be distinguished by their spins. When the spins of the added and compensating electrons are both equal to $\sigma$, these two electrons cannot be distinguished in the sense that their roles (added/compensating) may be reverted. Special care should thus be taken to avoid double counting. An elaborate analysis reveals that the double counting is avoided by ordering the momenta of:
\begin{itemize}
 \item[(\texttt{add})] the added electron $\mathbf{k}$ and the compensating electron $\mathbf{k}_\mathrm{comp}+\mathbf{q}-\mathbf{k}$ after the compensation;
 \item[(\texttt{rmv})] the removed electron $\mathbf{k}$ and the compensating electron $\mathbf{k}_\mathrm{comp}$ before the compensation.
\end{itemize}
The ordering is the same as in the update \texttt{exchange\_q}.

While the above discussion regards the momentum-space states $|\Psi_{k,l}^n\rangle\to|\Psi_{k,l}^{n+1}\rangle$, the situation with the real-space states $|\Psi_{i,l}^n\rangle\to|\Psi_{i,l}^{n+1}\rangle$ is far less complicated because only the particle-number conservation should be satisfied. The spin of the electron to be added to/removed from from each $|\Psi_{i,l}^n\rangle$ is determined by the spin of the electron added to/removed from $|\Psi_{k,0}^n\rangle$. The ratio of the backward and forward proposal probabilities for the real-space parts of the configuration may be directly computed as:
\begin{itemize}
 \item[(\texttt{add})] $\displaystyle{\left(\frac{W^\mathrm{prop}_{n+1\to n}}{W^\mathrm{prop}_{n\to n+1}}\right)_{i,l}=\frac{N_c-N_\sigma}{1+N_\sigma}},$
 \item[(\texttt{rmv})] $\displaystyle{\left(\frac{W^\mathrm{prop}_{n+1\to n}}{W^\mathrm{prop}_{n\to n+1}}\right)_{i,l}=\frac{N_\sigma}{N_c-N_\sigma+1}}$,
\end{itemize}
where $N_\sigma$ is the number of electrons of spin $\sigma$ in all the states $|\Psi_{i/k,l}^n\rangle$ (before the update).

There are also some differences in the move \texttt{spin\_flip}, in which we randomly choose spin $\sigma$ and attempt to increase/decrease the number of electrons of spin $\sigma$/$\overline{\sigma}$ by one.
 
In each imaginary-time slice $l=1,\dots,N_\tau$, we explicitly enumerate the momentum states $(\sigma,\mathbf{k})\in|\Psi_{k,l}^n\rangle$ such that $(\overline{\sigma},\mathbf{k})\notin|\Psi_{k,l}^n\rangle$. This is the simplest possible update that changes the spin of an electron and yet keeps the total momentum of the configuration fixed. The state $|\Psi_{k,l}^{n+1}\rangle$ is then obtained from the state $|\Psi_{k,l}^n\rangle$ by removing the electron in the state $(\sigma,\mathbf{k})$ and adding the electron in the state $(\overline{\sigma},\mathbf{k})$. The inverse move proceeds in an analogous manner: we explicitly enumerate momentum the states $(\overline{\sigma},\mathbf{k}')\in|\Psi_{k,l}^{n+1}\rangle$ such that $(\sigma,\mathbf{k}')\notin|\Psi_{k,l}^{n+1}\rangle$. 
 
In each imaginary-time slice $l=1,\dots,N_\tau$, we choose an electron of spin $\sigma$ at position $\mathbf{r}$ from the real-space state $|\Psi_{i,l}^n\rangle$ and construct the real-space state $|\Psi_{i,l}^{n+1}\rangle$ by changing the electron's position $\mathbf{r}\to\mathbf{s}$ and spin $\sigma\to\overline{\sigma}$. The ratio of the proposal probabilities in the real space can be directly computed as $\displaystyle{\left(\frac{W^\mathrm{prop}_{n+1\to n}}{W^\mathrm{prop}_{n\to n+1}}\right)_{i,l}=\frac{N_\sigma(N_c-N_{\overline{\sigma}})}{(1+N_{\overline{\sigma}})(N_c-N_\sigma+1)}}$, where $N_\sigma$ and $N_{\overline{\sigma}}$ are numbers of electrons of spin $\sigma$ and $\overline{\sigma}$ in $|\Psi^n_{i/k,l}\rangle$.
 
\subsection{ABQMC method in equilibrium: Numerical results}
\label{SSec:eq_abqmc_results}

\begin{figure}[htbp!]
    \centering
    \includegraphics{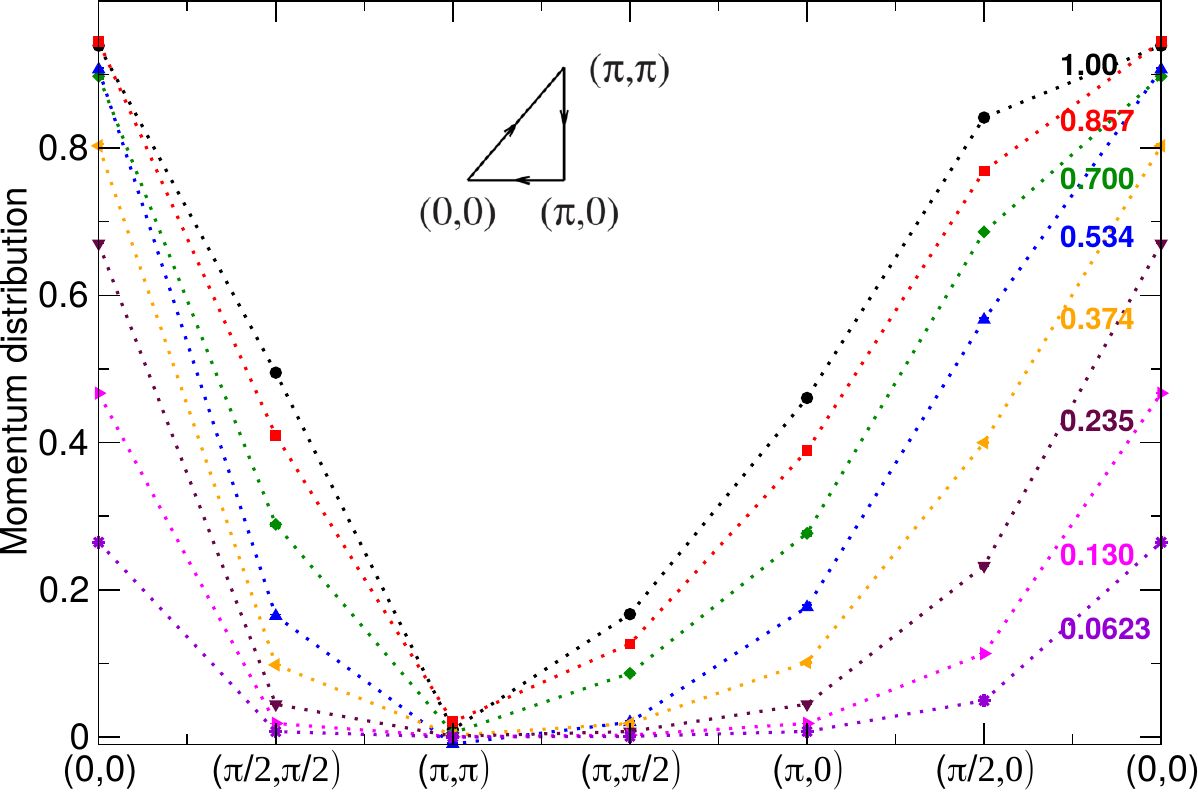}
    \caption{Momentum distribution, $\frac{1}{2}\sum_\sigma\langle n_{\mathbf{k}\sigma}\rangle$, for different fillings $\rho_e$ (half filling is $\rho_e=1$) computed using the equilibrium ABQMC approach on a $4\times 4$ square-lattice cluster. Model parameters are $U/J=4$, $T/J=1.0408$, while $\mu/J$ is varied from 2 to $-5$. The pathway through the irreducible Brillouin zone is summarized in the inset. Cited values of $\rho_e$ are from Ref.~\onlinecite{PhysRevA.84.053611}. Dotted lines serve as guides to the eye. Statistical error bars are generally smaller than symbol size.}
    \label{Fig:moguca_slika_n_k_021122}
\end{figure}

Within the ABQMC method, coordinate and momentum bases are treated symmetrically, meaning that the method should be equally adept at calculating quantities diagonal in these two bases. As an example applications in the momentum space, in Fig.~\ref{Fig:moguca_slika_n_k_021122} we show the momentum distribution, $\frac{1}{2}\sum_\sigma\langle n_{\mathbf{k}\sigma}\rangle$, in the same setup as in Fig.~4 of the main text ($U/J=4$, $T/J=1.0408$).

\pagebreak

\subsection{ABQMC method in equilibrium: Acceptance rates and proposal/acceptance times}
\label{SSec:eq_abqmc_performance}

\subsubsection{$U/J=4$, $T/J=1.0408$}
\begin{figure}[htbp!]
 \centering
 \includegraphics[scale=0.45]{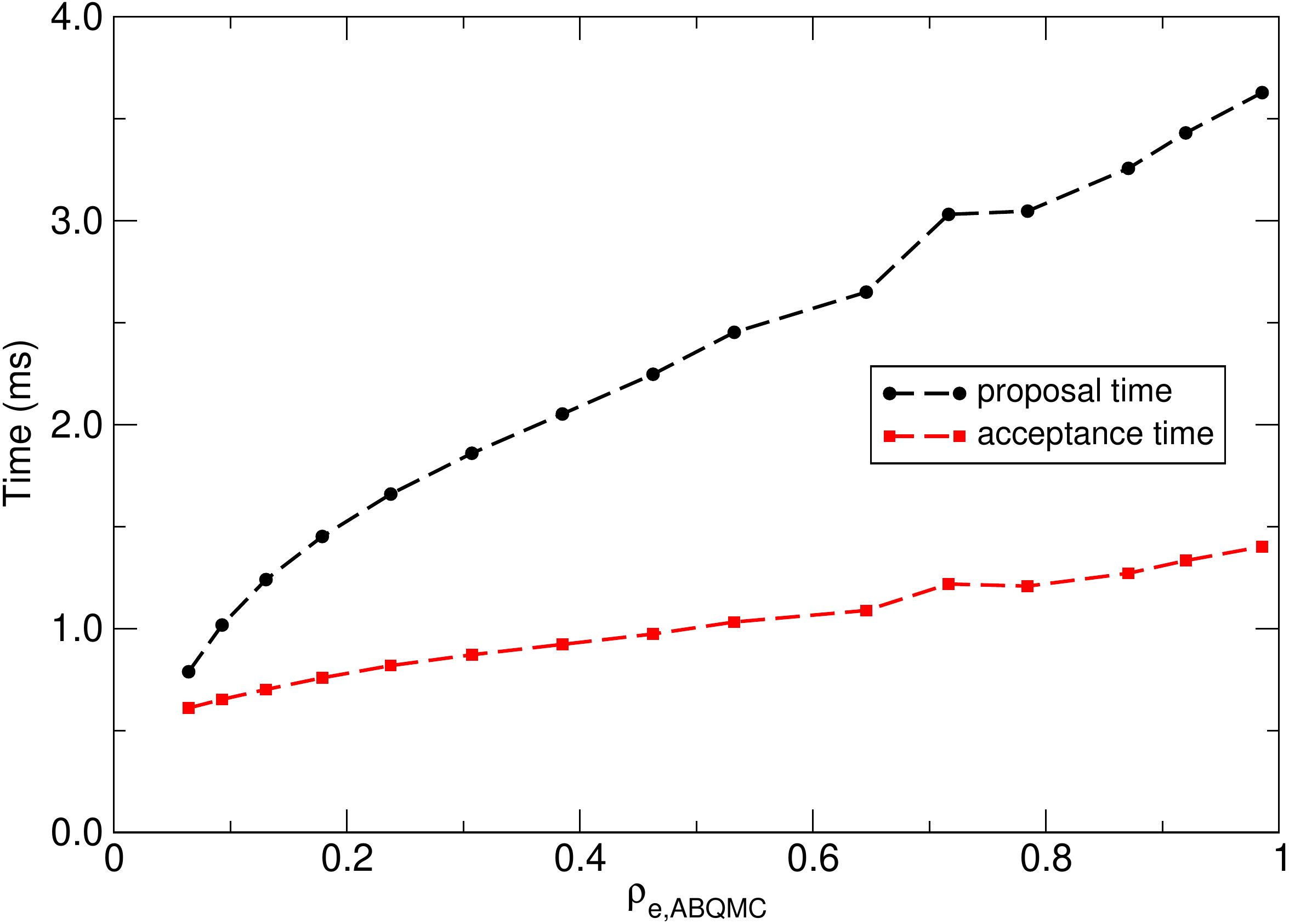}
 \caption{Average time needed to propose (black circles) and accept (red squares) one Monte Carlo update as a function of the ABQMC electron density. The averaging is performed over all updates used in ABQMC simulations to evaluate the equation of state: \texttt{add\_particle}, \texttt{remove\_particle}, \texttt{spin\_flip}, \texttt{add\_q}, \texttt{exchange\_q}, \texttt{change\_r\_local}, and \texttt{change\_r\_global}.}
 \label{Fig:proposal_acceptance_281221}
\end{figure}

\begin{figure}[htbp!]
 \centering
 \includegraphics[scale=0.45]{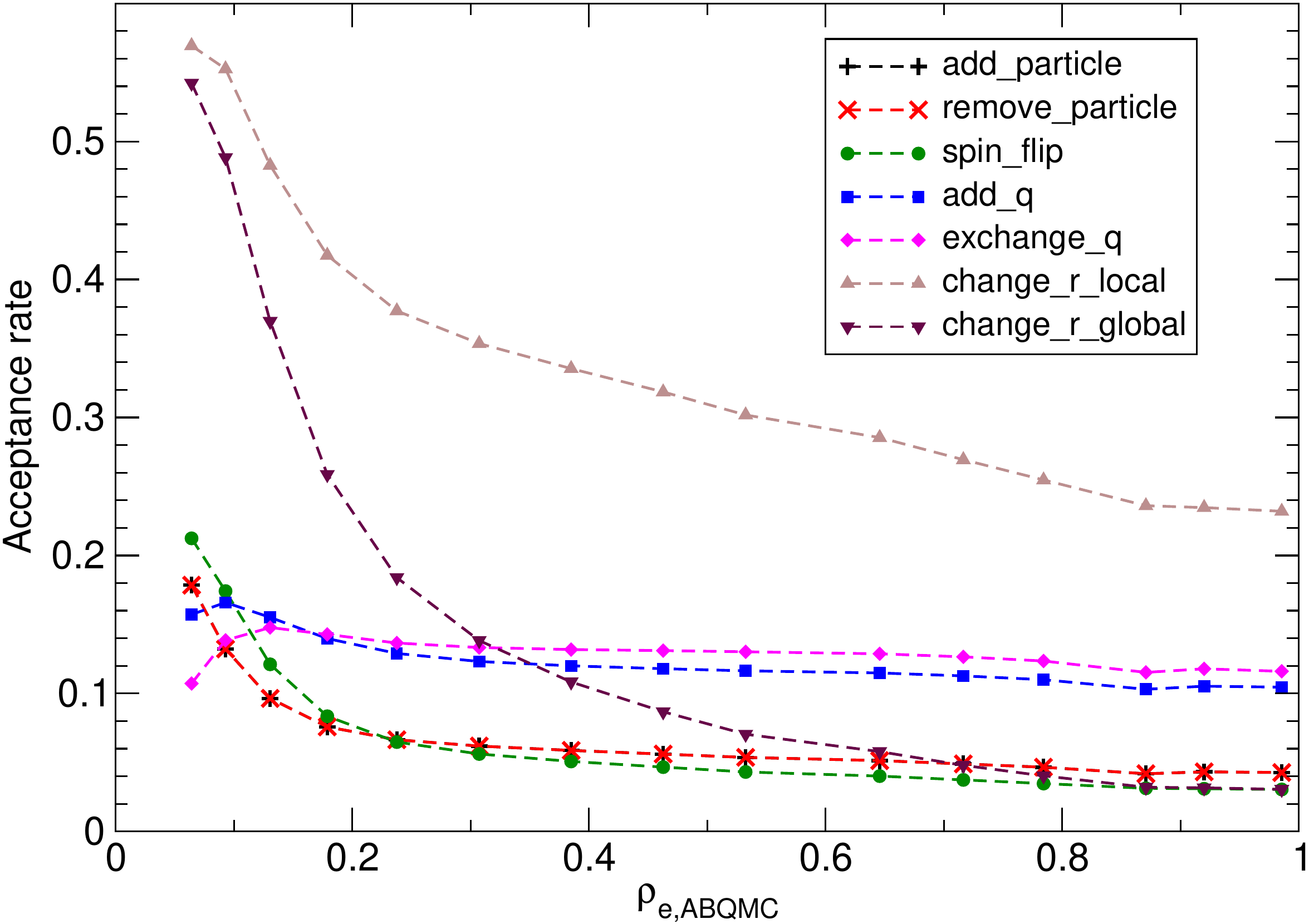}
 \caption{Acceptance rates of individual Monte Carlo updates as a function of the ABQMC electron density. Being inverses of one another, the acceptance rates of moves \texttt{add\_particle} and \texttt{remove\_particle} are mutually equal. The acceptance rates of moves \texttt{add\_particle}, \texttt{remove\_particle}, \texttt{spin\_flip}, \texttt{add\_q}, and \texttt{exchange\_q} exhibit weak dependence on the filling, while moves involving changes in the real-space part of configurations tend to be accepted less frequently as the filling is increased.}
 \label{Fig:acceptance_rates_281221}
\end{figure}

\pagebreak

\subsubsection{$U/J=24$, $T/J=1.0408$}

\begin{figure}[htbp!]
 \centering
 \includegraphics[scale=0.45]{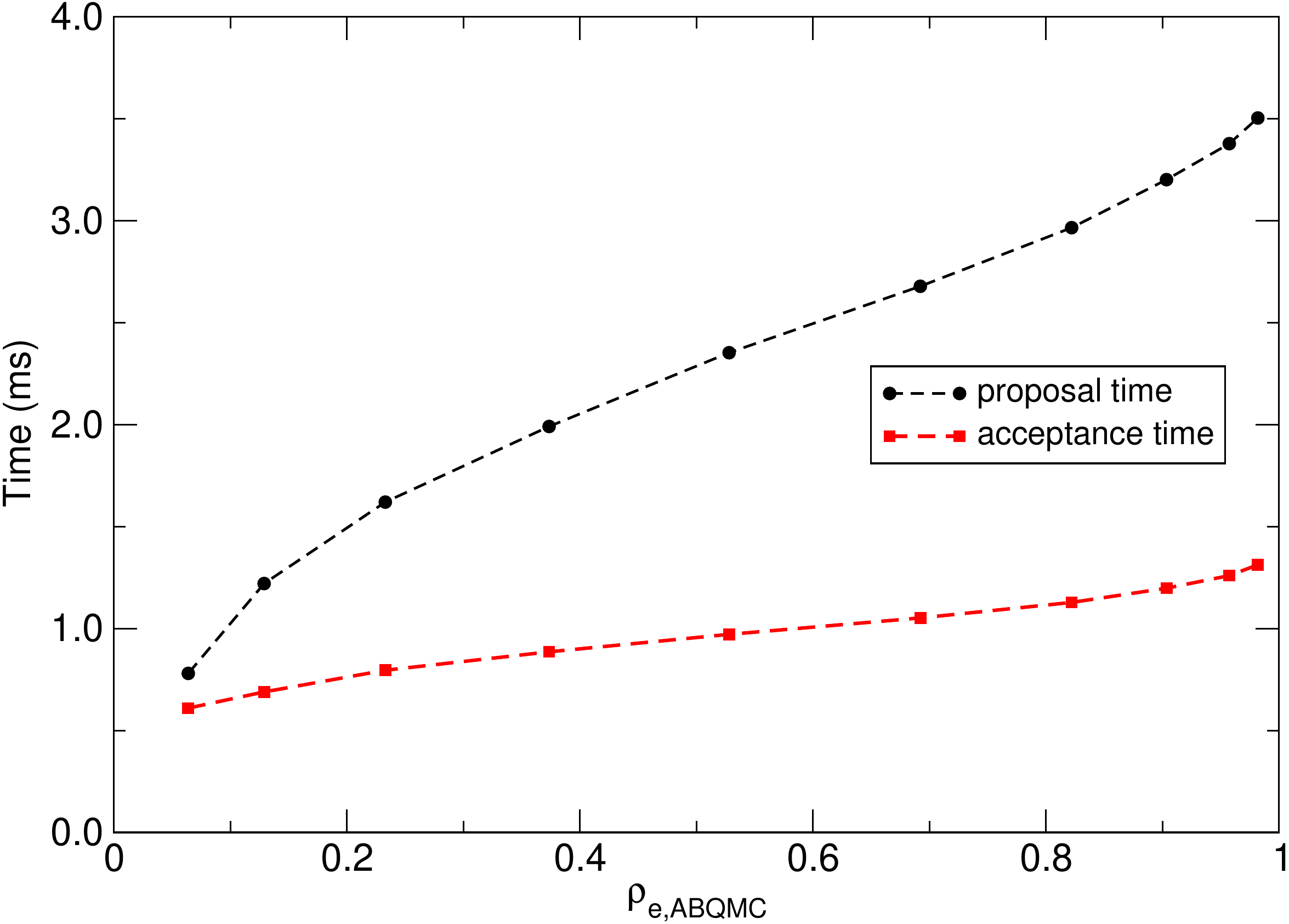}
 \caption{Average time needed to propose (black circles) and accept (red squares) one Monte Carlo update as a function of the ABQMC electron density. The averaging is performed over all updates used in ABQMC simulations to evaluate the equation of state: \texttt{add\_particle}, \texttt{remove\_particle}, \texttt{spin\_flip}, \texttt{add\_q}, \texttt{exchange\_q}, \texttt{change\_r\_local}, and \texttt{change\_r\_global}.}
 \label{Fig:proposal_acceptance_strong_U_301221}
\end{figure}

\begin{figure}[htbp!]
 \centering
 \includegraphics[scale=0.45]{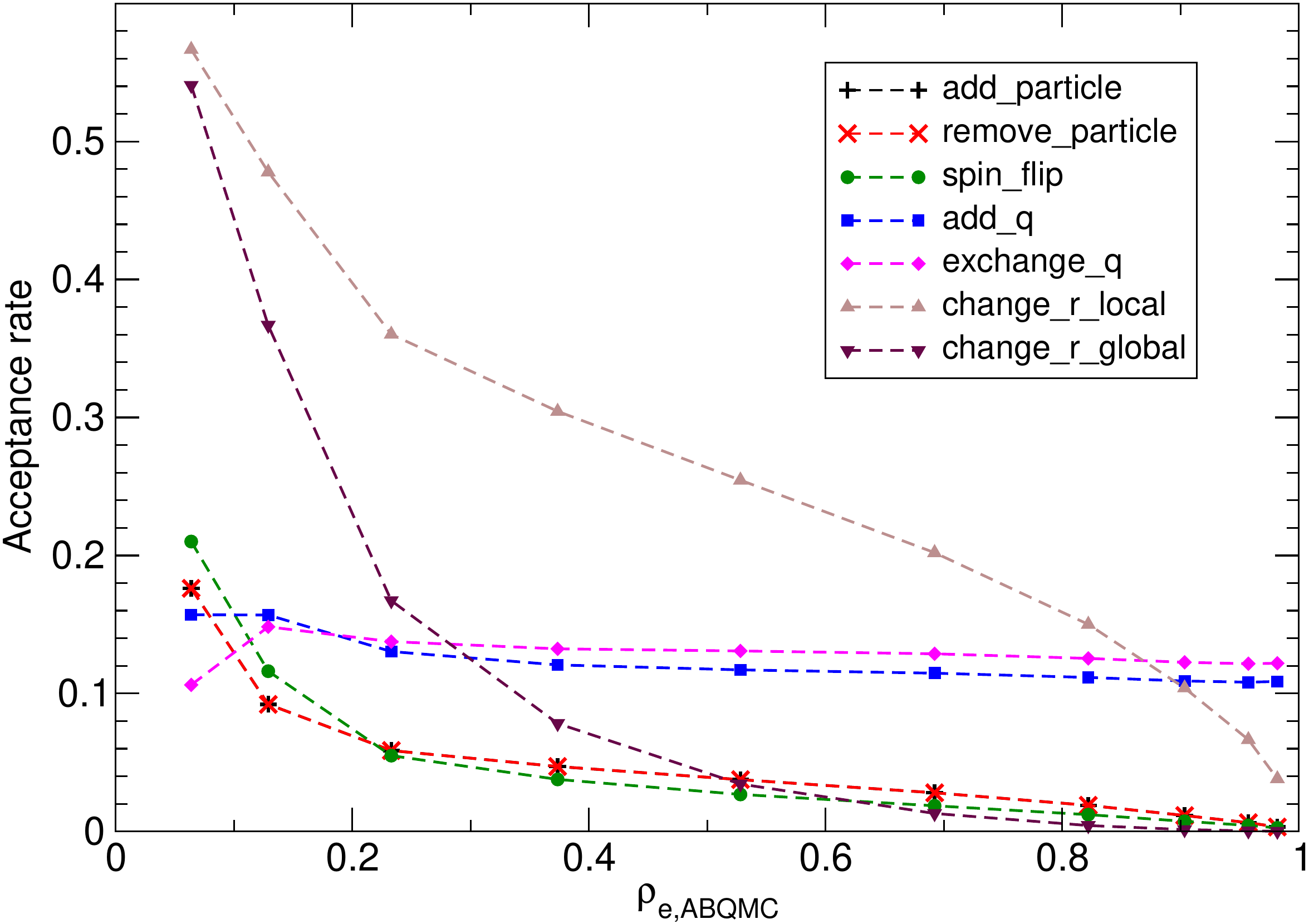}
 \caption{Acceptance rates of individual Monte Carlo updates as a function of the ABQMC electron density. Being inverses of one another, the acceptance rates of moves \texttt{add\_particle} and \texttt{remove\_particle} are mutually equal. The acceptance rates of moves \texttt{add\_q} and \texttt{exchange\_q} exhibit weak dependence on the filling, while moves \texttt{add\_particle}, \texttt{remove\_particle}, \texttt{spin\_flip}, \texttt{change\_r\_local}, and \texttt{change\_r\_global} tend to be accepted less frequently as the filling is increased.}
 \label{Fig:acceptance_rates_strong_U_301221}
\end{figure}

\pagebreak

\section{FPQMC method for time-dependent densities: Additional results}
\begin{figure}[htbp!]
 \centering
 \includegraphics{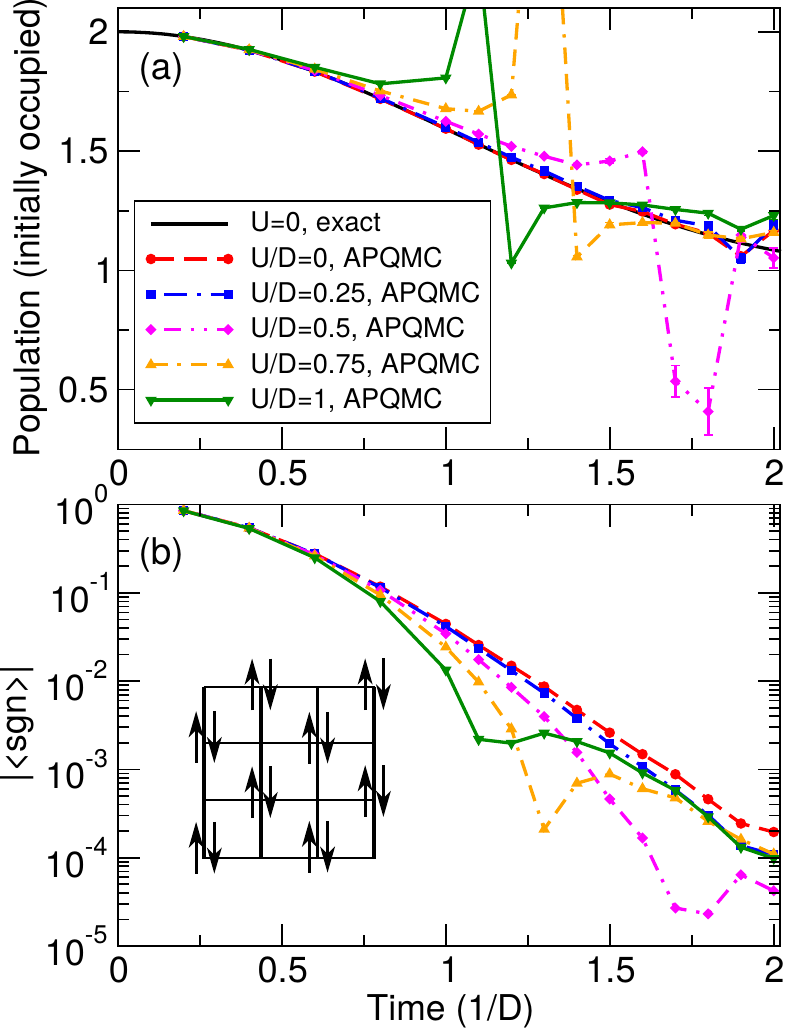}
 \caption{(a) Time-dependent population of sites occupied in the initial CDW state of a $4\times 4$ cluster, which is schematically depicted in the inset of panel (b). FPQMC results using $N_t=2$ real-time slices (4 slices in total) are shown for five different interaction strengths (symbols) and compared with the noninteracting result (solid line). (b) Magnitude of the average sign as a function of time for different interaction strengths. Color code is the same as in panel (a).}
 \label{Fig:Fig_ab_160922}
\end{figure}

Figure~\ref{Fig:Fig_ab_160922}(a) summarizes the evolution of charge densities on initially unoccupied sites of a half-filled $4\times 4$ cluster, on which the electrons are initially arranged as summarized in the inset of Fig.~\ref{Fig:Fig_ab_160922}(b) [the so-called $(\pi,\pi)$ density wave]. Figure~\ref{Fig:Fig_ab_160922}(b) summarizes the extent of the dynamical sign problem, which appears to be somewhat more severe than in the case of the $(\pi,0)$ density wave discussed in Figs.~8(a) and~8(b) of the main text.

\newpage

\begin{figure*}[htbp!]
 \centering
 \includegraphics{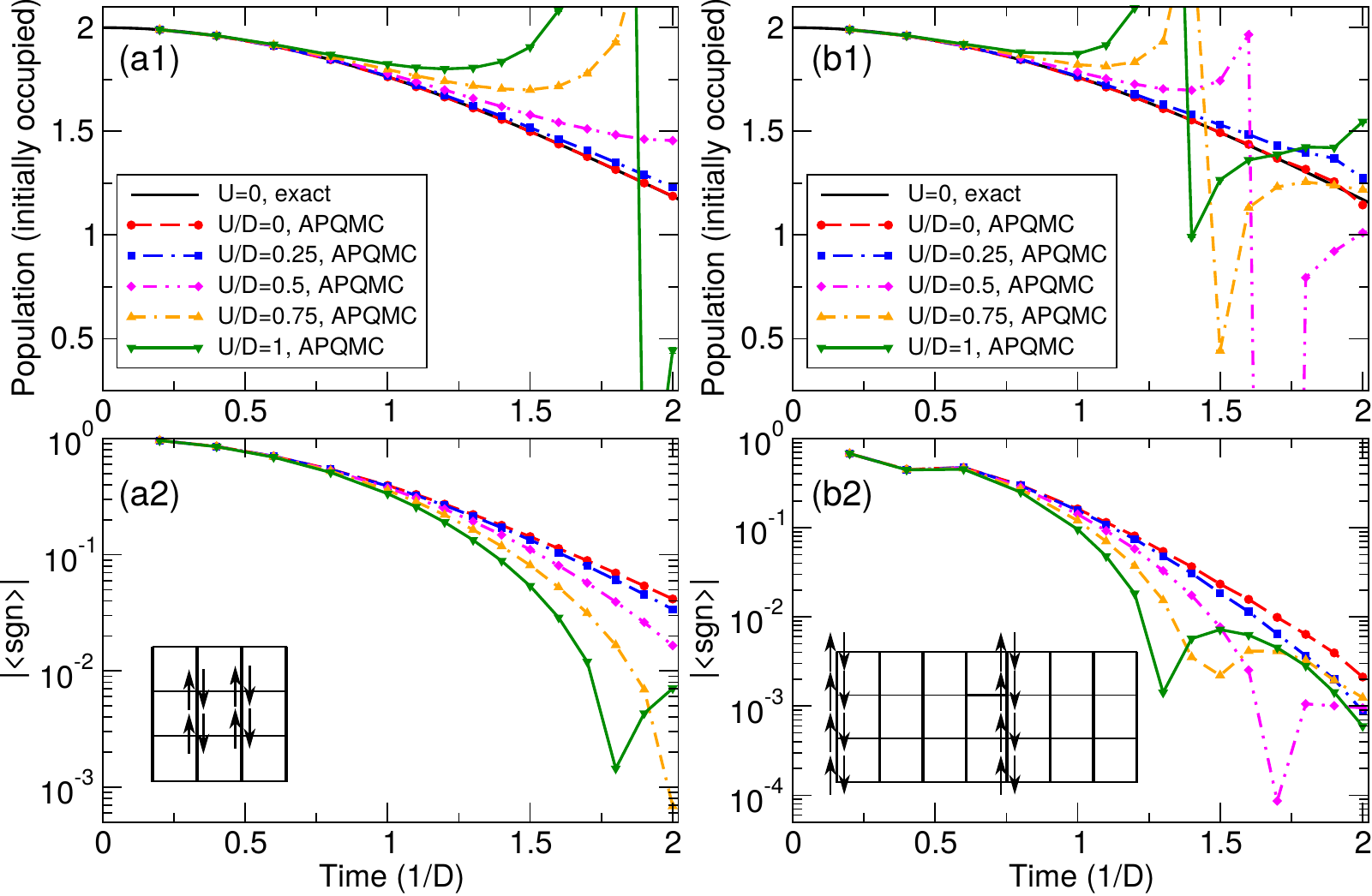}
 \caption{(a1) and (b1) Time-dependent population of sites occupied in the initial state of a $4\times 4$ (a1) and $8\times 4$ (b1) cluster, which is schematically depicted in the inset of panels (a2) and (b2), respectively. FPQMC results using $N_t=2$ real-time slices (4 slices in total) are shown for five different interaction strengths (symbols) and compared with the noninteracting result (solid line). (a2) and (b2) Time-dependent average sign of the FPQMC simulation for different interaction strengths. Color code is the same as in (a1) and (b1), respectively. Statistical errors are generally smaller than the symbol size.}
 \label{Fig:smaller_filling}
\end{figure*}

In equilibrium and in the weak-interaction regime, we concluded that the sign problem becomes more pronounced with the filling, see Fig.~4(b) of the main text. Motivated by this finding, we finally study the dynamics of local densities at a smaller filling, see Figs.~\ref{Fig:smaller_filling}(a1)--\ref{Fig:smaller_filling}(b2), which permits us to perform FPQMC simulations on an $8\times 4$ cluster. While the average sign displayed in Fig.~\ref{Fig:smaller_filling}(a2) is somewhat enhanced with respect to the values reported in Fig.~\ref{Fig:Fig_ab_160922}(b) and Fig.~8(b) of the main text, the simulated dynamics retains the above-described problems. The average sign decreases with the cluster size, compare the average signs for the noninteracting electrons in Figs.~\ref{Fig:smaller_filling}(a2) and~\ref{Fig:smaller_filling}(b2).

\newpage

\section{ABQMC method for time-dependent survival probability: Additional results}
\subsection{Applicability of the ABQMC method to a $4\times 4$ cluster with $N_t=4$ real-time slices}
\begin{figure}[htbp!]
 \centering
 \includegraphics[scale=0.65]{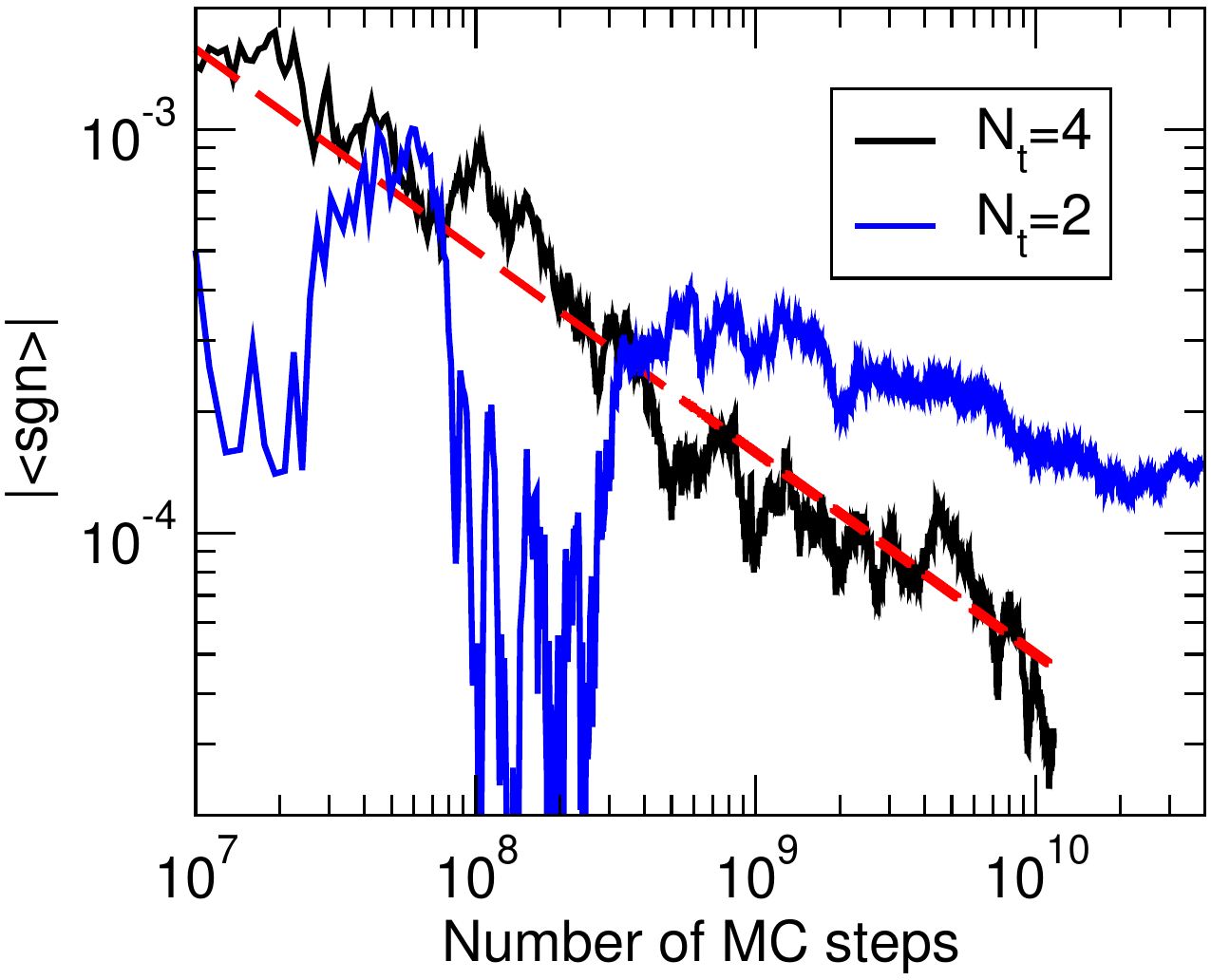}
 \caption{Magnitude of the average sign as a function of the number of MC steps during the ABQMC simulation of the survival probability of the 16-electron CDW/SDW state on a $4\times 4$ square-lattice cluster using $N_t=2$ and $N_t=4$ real-time slices. The overall behavior of $|\langle\mathrm{sgn}\rangle|$ for $N_t=4$ may be fitted very well by the function $5/\sqrt{N_\mathrm{MC}}$ that is shown by the dashed line.}
 \label{Fig:4x4_16el_Ntau_4_161221}
\end{figure}

This section addresses the applicability of the ABQMC method to compute the survival probability of the 16-electron CDW-like/SDW-like state depicted in Fig.~10(a) of the main text when the number of real-time slices is increased from $N_t=2$ to $N_t=4$. In Fig.~\ref{Fig:4x4_16el_Ntau_4_161221} we compare the behavior of the average sign for $N_t=2$, when we make $3.87\times 10^{10}$ steps, and $N_t=4$, when we make $1.16\times 10^{10}$ steps. Increasing the number of real-time slices from 2 to 4 decreases $|\langle\mathrm{sgn}\rangle|$ after $10^{10}$ MC steps by an order of magnitude. With $N_t=4$, the stabilization of the average sign takes much more than $10^{10}$ MC steps, and its overall decrease as the simulation proceeds may be very well described by a power law with the exponent of $-1/2$, see the dashed line in Fig.~\ref{Fig:4x4_16el_Ntau_4_161221}.

\newpage

\subsection{Applicability of the ABQMC method to an $8\times 4$ cluster with $N_t=2$ real-time slices}
\begin{figure}[htbp!]
 \centering
 \includegraphics[scale=1.5]{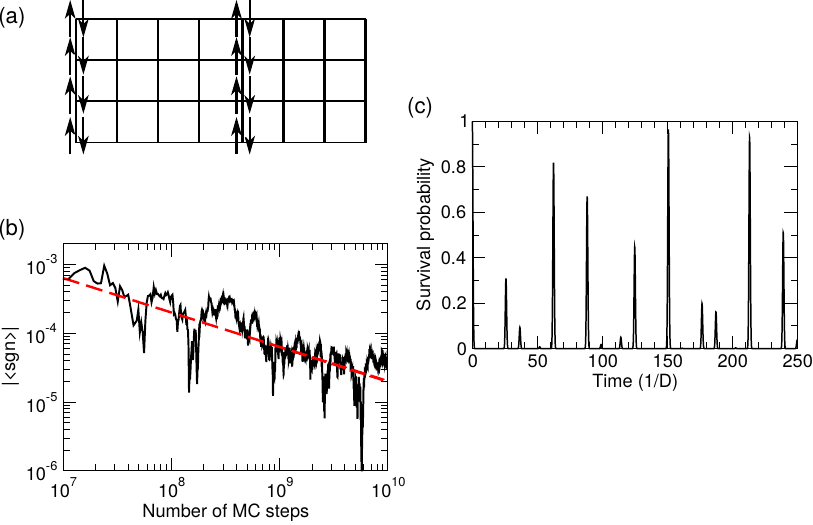}
 \caption{(a) Schematic view of the 16-electron initial state on an $8\times 4$ square-lattice cluster whose survival probability is computed. (b) Magnitude of the average sign as a function of the number of MC steps during the ABQMC simulation of the survival probability of the initial state depicted in (a). $N_t=2$ real-time slices are used. The overall behavior of $|\langle\mathrm{sgn}\rangle|$ may be fitted very well by the function $2/\sqrt{N_\mathrm{MC}}$ that is shown by the dashed line. (c) Time dependence of the survival probability of the initial state depicted in (a) in the noninteracting case $U=0$. The dimensionality of the system's Hilbert space is so large that no perfect revival in $P(t)$ (which is bound to occur since $U=0$) is observed up to $Dt=250$.}
 \label{Fig:8x4_Nel_16_161221_after_inkscape}
\end{figure}

This section addresses the applicability of the ABQMC method to compute the survival probability of a 16-electron initial state on an $8\times 4$ cluster with $N_t=2$ real-time slices. The initial CDW-like state is schematically depicted in Fig.~\ref{Fig:8x4_Nel_16_161221_after_inkscape}(a), while Fig.~\ref{Fig:8x4_Nel_16_161221_after_inkscape}(b) shows the evolution of the average sign during the MC simulation. We perform $10^{10}$ MC steps, during which the magnitude of the average sign shows no signals of stabilization, but steadily decreases in a power-law fashion with the exponent $-1/2$, see the dashed line in Fig.~\ref{Fig:8x4_Nel_16_161221_after_inkscape}(b). Even though the final stages of our MC simulation may suggest that $|\langle\mathrm{sgn}\rangle|$ stabilizes on the level of $\sim 4\times 10^{-5}$, a very noisy behavior of $|\langle\mathrm{sgn}\rangle|$ throughout the simulation casts doubts on such a conclusion. We observe in Fig.~\ref{Fig:8x4_Nel_16_161221_after_inkscape}(b) that $|\langle\mathrm{sgn}\rangle|$ displays pronounced dips whose duration may be as long as a couple of billions of steps, which is in stark contrast with
the rather smooth decrease of $|\langle\mathrm{sgn}\rangle|$ observed, e.g., in
Fig.~\ref{Fig:4x4_16el_Ntau_4_161221}. These dips suggest that there may be problems with the configuration-space sampling. The dimension of the configuration space of our ABQMC method is much larger than the dimension of the Hilbert space of the model, which is astronomically large for the Hubbard model on an $8\times 4$ cluster. To further illustrate the last point, in Fig.~\ref{Fig:8x4_Nel_16_161221_after_inkscape}(c) we present the survival probability of the initial state in the noninteracting case. In contrast to the $4\times 4$ cluster, for which perfect collapses and
revivals in $P(t)$ are observed already on relatively short time scales, there is no such a regular behavior on the $8\times 4$ cluster, see Fig.~\ref{Fig:8x4_Nel_16_161221_after_inkscape}(c). On general grounds, the noninteracting system is bound to display perfect collapses and revivals in $P(t)$, while the time scale on which the pattern in $P(t)$ repeats itself is inversely proportional to the dimension of the system’s Hilbert space.

\newpage

\subsection{ABQMC results for the survival probability near the atomic limit}
\begin{figure}[htbp!]
 \centering
 \includegraphics[width=0.5\columnwidth]{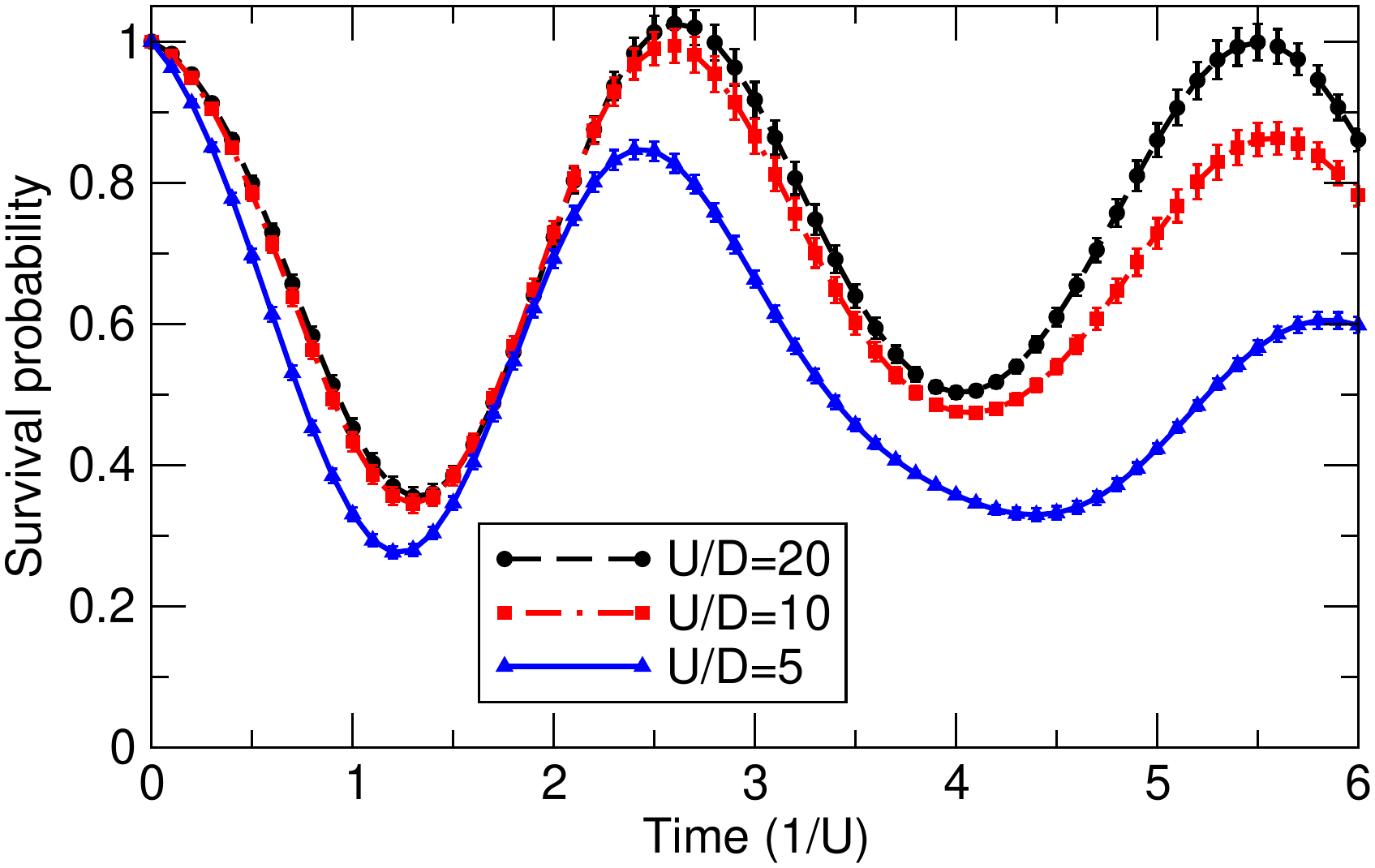}
 \caption{(Color online) Survival-probability dynamics of the 16-electron initial state schematically depicted in Fig.~(a) for three values of the interaction strength that are close to the atomic limit.}
 \label{Fig:fig_from_atomic_251021}
\end{figure}

Figure~\ref{Fig:fig_from_atomic_251021} shows $P(t)$ for the 16-electron initial state schematically depicted in Fig.~10(a) of the main text in regimes that are close to the atomic limit. In these regimes, the natural energy unit is $U$, so that the time is measured in units $1/U$. The time range is chosen on the basis of the results in Figs.~(d1)--(e2), which suggest that the ABQMC method with $N_t=2$ real-time slices produces a qualitatively correct behavior of $P(t)$ up to times $Ut\approx 6$. $P(t)$ exhibits oscillations whose amplitude decreases in time. There is almost no difference between $P(t)$ for $D/U=0.05$ and $0.1$ during the first oscillation, while $P(t)$ for $D/U=0.2$ is at all times below $P(t)$ for stronger $U$. While for the strongest $U$ the maxima reached by $P(t)$ are always close to 1, the maxima for weaker $U$ are smaller than 1 and decrease with time. All these observations can be rationalized by an increased importance of the kinetic over the interaction term as $U/D$ is decreased.

\newpage

\section{ABQMC method to evaluate time-dependent expectation values along the Keldysh--Kadanoff--Baym contour}

We start from
\begin{equation}
\label{Eq:Kadanoff-Baym-general}
 \langle A_a(t)\rangle=\frac{\mathrm{Tr}\left(e^{-\beta H(0)}\:e^{iHt}\:A_a\:e^{-iHt}\right)}{\mathrm{Tr}\left(e^{-\beta H(0)}\right)}
\end{equation}
where $H$ is the Hubbard Hamiltonian, while $H(0)$ additionally contains terms that modulate charge or spin density. For example, to simulate the response recorded in Ref.~\onlinecite{Science.363.379}, the following $H(0)$ may be appropriate:
\begin{equation}
 H(0)=\underbrace{\sum_{\mathbf{k}\sigma}\widetilde{\varepsilon}_\mathbf{k}\:n_{\mathbf{k}\sigma}+U\sum_\mathbf{r}n_{\mathbf{r}\uparrow}n_{\mathbf{r}\downarrow}}_{H}-\sum_{\mathbf{r}\sigma}v_\mathbf{r}n_{\mathbf{r}\sigma}
\end{equation}
where $v_\mathbf{r}$ is the external harmonic potential that modulates electron density, e.g.,
\begin{equation}
 v_\mathbf{r}=V_0\cos(\mathbf{q}\cdot\mathbf{r}).
\end{equation}
Inspired by Ref.~\onlinecite{Science.363.379}, we assume that $\mathbf{q}=q\mathbf{e}_x$, i.e., we assume that the electron density is modulated along one direction with the wavelength $\lambda=2\pi/q$. We also assume that $\lambda\leq N_x$ and that $N_x\:\mathrm{mod}\:\lambda=0$, i.e., the cluster's linear dimension along $x$ axis is spanned by an integer number of wavelengths.

There are at least two manners in which the ABQMC method in the presence of external density-modulating potential (at $t<0$) can be formulated. They differ by the choice of the contributions that are diagonal in the coordinate and momentum representations.
\begin{enumerate}
 \item The part diagonal in the momentum representation is $$[H(0)]_\mathrm{mom}=\sum_\mathbf{k}\varepsilon_\mathbf{k}n_{\mathbf{k}\sigma},$$ while the part diagonal in the coordinate representation $$[H(0)]_\mathrm{coord}=U\sum_\mathbf{r}n_{\mathbf{r}\uparrow}n_{\mathbf{r}\downarrow}-\sum_{\mathbf{r}\sigma}\mu_\mathbf{r}n_{\mathbf{r}\sigma}$$ contains position-dependent chemical potential $\mu_\mathbf{r}=\mu+v_\mathbf{r}$. Such a decomposition provides exact results in the atomic limit ($J=0$) and is expected to provide reasonable results when the electron--electron interaction dominates over the electronic hopping.
 \item The part diagonal in some momentum representation is $$[H(0)]_\mathrm{mom}=\sum_{\mathbf{k}\sigma}\widetilde{\varepsilon}_\mathbf{k}\:n_{\mathbf{k}\sigma}-\frac{V_0}{2}\sum_{\mathbf{k}\sigma}\left(c_{(k_x+q,k_y)\sigma}^\dagger c_{(k_x,k_y)\sigma}+c_{(k_x-q,k_y)\sigma}^\dagger c_{(k_x,k_y)\sigma}\right)$$
 while the part diagonal in the coordinate representation is
 $$[H(0)]_\mathrm{coord}=U\sum_\mathbf{r}n_{\mathbf{r}\uparrow}n_{\mathbf{r}\downarrow}.$$ Such a decomposition provides exact results in the noninteracting limit ($U=0$) and is expected to provide reasonable results when the electronic hopping dominates over the electron--electron interaction. Our further developments will be focused on this decomposition.
\end{enumerate}
The external harmonic potential introduces the coupling between different $\mathbf{k}$ states which results in a reduction of the Brillouin zone along $x$ axis by a factor of $\lambda$. The wave vector $\mathbf{k}$ in the full Brillouin zone $[0,2\pi)\times[0,2\pi)$ is not a good quantum number anymore. Its role is taken by the wave vector $\widetilde{\mathbf{k}}$ in the reduced Brillouin zone $[0,2\pi/\lambda)\times[0,2\pi)$, whose $x$ projection $\widetilde{k}_x$ may assume any of the $n_q=N_x\:\mathrm{div}\:\lambda$ allowed values in the interval $[0,2\pi/\lambda)$ ($p\times 2\pi/N_x$, where $p=0,\dots,n_q-1$) and whose $y$ projection $\widetilde{k}_y$ may assume any of the $N_y$ allowed values in the interval $[0,2\pi)$ ($p\times 2\pi/N_y$, where $p=0,\dots,N_y-1$). In addition to $\widetilde{\mathbf{k}}$, there is another degree of freedom that will be denoted by $\nu_x$ and that may assume values $0,\dots,\lambda-1$. The Hamiltonian $H_\mathrm{mom}$ has a block-diagonal structure and the blocks defined by the wave vector $\widetilde{\mathbf{k}}$ can be diagonalized separately. There are $n_qN_y$ such blocks and the dimension of each of them is $\lambda\times \lambda$.

The Hubbard Hamiltonian $H$ appears in combination $e^{iHt}\dots e^{-iHt}$, meaning that the chemical-potential term $-\mu\sum_{\mathbf{r}\sigma}n_\mathbf{r\sigma}=-\mu\sum_{\mathbf{k}\sigma}n_\mathbf{k\sigma}$ is not effective (we assume that $A_a$ conserves the total particle number, which is a reasonable assumption) and there is no ambiguity regarding the part $H_\mathrm{mom}$ or $H_\mathrm{coord}$ to which it should be associated. Therefore, the partition of the Hubbard Hamiltonian in parts that are diagonal in the momentum and coordinate representation is the same as in the main body of the paper
$$H_\mathrm{mom}=\sum_{\mathbf{k}\sigma}\varepsilon_\mathbf{k}n_{\mathbf{k}\sigma},\quad H_\mathrm{coord}=U\sum_\mathbf{r}n_{\mathbf{r}\uparrow}n_{\mathbf{r}\downarrow}.$$

Let us now exploit symmetries to develop as efficient as possible ABQMC algorithm to evaluate Eq.~\eqref{Eq:Kadanoff-Baym-general}. Since $H(0)$ is not invariant under the bipartite lattice symmetry, we use only the time-reversal symmetry, according to which $\langle A_a(-t)\rangle=\langle A_a(t)\rangle$. Furthermore, we note that both the numerator and the denominator of Eq.~\eqref{Eq:Kadanoff-Baym-general} are purely real. Using the strategy described in the main text, one may derive the ABQMC counterpart of Eq.~\eqref{Eq:Kadanoff-Baym-general}
\begin{equation}
\label{Eq:Kadanoff-Baym-ABQMC}
 \langle A_a(t)\rangle=\frac{\sum_\mathcal{C}\mathrm{Re}\{\mathcal{D}(\mathcal{C})\}\:e^{-\Delta\tau\varepsilon_\mathrm{M}(\mathcal{C})}\:\cos\left\{[\Delta\varepsilon_0(\mathcal{C})+\Delta\varepsilon_\mathrm{int}(\mathcal{C})]\Delta t\right\}\:\mathcal{A}_a(\Psi_{a,l_a})}{\sum_\mathcal{C}\mathrm{Re}\{\mathcal{D}(\mathcal{C})\}\:e^{-\Delta\tau\varepsilon_\mathrm{M}(\mathcal{C})}}
\end{equation}
The configuration $\mathcal{C}$ consists of $2N_t+N_\tau$ many-body states $|\Psi_{k,l}\rangle$ ($l=1,\dots,2N_t+N_\tau$) composed of single-particle momentum eigenstates and $2N_t+N_\tau$ many-body states $|\Psi_{i,l}\rangle$ composed of single-particle coordinate eigenstates. While momenta of states $|\Psi_{k,1}\rangle,\dots,|\Psi_{k,2N_t}\rangle$ are defined in the full Brillouin zone, states $|\Psi_{k,2N_t+1}\rangle,\dots,|\Psi_{k,2N_t+N_\tau}\rangle$ have their momenta defined in the reduced Brillouin zone. The symbols $\mathcal{D}(\mathcal{C}),\:\Delta\varepsilon_0(\mathcal{C}),$ and $\Delta\varepsilon_\mathrm{int}(\mathcal{C})$ are defined as in the main text, while $\varepsilon_\mathrm{M}(\mathcal{C})$ is the sum of energies of $2N_\tau$ states along the Matsubara part of the contour, i.e.,
\begin{equation}
 \varepsilon_\mathrm{M}(\mathcal{C})=\sum_{l=2N_t+1}^{2N_t+N_\tau}\left[\varepsilon_0(\Psi_{k,l})+\varepsilon_\mathrm{int}(\Psi_{i,l})\right].
\end{equation}
The slice on which the expectation value $\mathcal{A}_a(\Psi_{a,l_a})=\langle\Psi_{a,l_a}|A_a|\Psi_{a,l_a}\rangle$ is evaluated depends on the representation $a=i,k$ in which the observable $A_a$ is diagonal, so that $l_i=N_t+1$ and $l_k=N_t$.  

The structure of Eq.~\eqref{Eq:Kadanoff-Baym-ABQMC} is intuitively clear as it is a combination of the ABQMC formula for thermodynamic quantities [Eq.~(17) of the main text] and the ABQMC formula for the time-dependent expectation value along the Keldysh contour [Eq.~(26) of the main text]. However, since we cannot exploit the bipartite lattice symmetry in this setup, the time-dependent part of the numerator in Eq.~\eqref{Eq:Kadanoff-Baym-ABQMC} is $\cos\left\{[\Delta\varepsilon_0(\mathcal{C})+\Delta\varepsilon_\mathrm{int}(\mathcal{C})]\Delta t\right\}$ instead of $\cos[\Delta\varepsilon_0(\mathcal{C})\Delta t]\:\cos[\Delta\varepsilon_\mathrm{int}(\mathcal{C})\Delta t]$. Configuration weight may be chosen as $w(\mathcal{C})=|\mathrm{Re}\{\mathcal{D}(\mathcal{C})\}|\:e^{-\Delta\tau\varepsilon_\mathrm{M}(\mathcal{C})}$ and Eq.~\eqref{Eq:Kadanoff-Baym-ABQMC} is recast as
\begin{equation}
\label{Eq:Kadanoff-Baym-MCMC}
 \langle A_a(t)\rangle=\frac{\left\langle\mathrm{sgn}(\mathcal{C})\:\cos\left\{[\Delta\varepsilon_0(\mathcal{C})+\Delta\varepsilon_\mathrm{int}(\mathcal{C})]\Delta t\right\}\:\mathcal{A}_a(\Psi_{a,l_a})\right\rangle_w}{\left\langle\mathrm{sgn}(\mathcal{C})\right\rangle_w}.
\end{equation}
Markov-chain MC evaluation of Eq.~\eqref{Eq:Kadanoff-Baym-MCMC} suffers from the sign problem that does not depend on time $t$ (it is not dynamical). Still, it becomes more pronounced when the cluster size $N_c$ or the number of slices ($N_t$ and $N_\tau$) are increased. Similarly to equilibrium ABQMC calculations, the weight $w(\mathcal{C})$ depends on all model parameters ($U,\:T,\:\mu,\:V_0,\:t$). Therefore, the calculations for different values of these parameters have to be performed using different Markov chains, which is different from the computation of $P(t)$ or $\langle A_a(t)\rangle$ starting from a pure state $|\psi(0)\rangle$. 

The application of conservation laws on the Kadanoff--Baym contour is somewhat more involved than on simpler contours studied in the main body of the paper. The particle-number conservation demands that all the many-body states constituting configuration $\mathcal{C}$ have the same number of spin-up and spin-down electrons. We discuss the momentum conservation under the assumption that the observable $A_a$ is diagonal in the coordinate representation (e.g., $A_i=\sum_\sigma n_{\mathbf{r}\sigma}$). From the main text, we know that the momentum conservation along the horizontal parts of the contour (Keldysh branch) is broken into two conservation laws that are satisfied separately on the forward and backward branch. In other words, the momenta (in the full Brillouin zone!) of $N_t$ momentum-space states $|\Psi_{k,1}\rangle,\dots,|\Psi_{k,N_t}\rangle$ on the forward branch are all equal to $\mathbf{K}_\mathrm{fwd}$, while the momenta of $N_t$ momentum-space states $|\Psi_{k,N_t+1}\rangle,\dots,|\Psi_{k,2N_t}\rangle$ on the backward branch are all equal to $\mathbf{K}_\mathrm{bwd}$. In contrast to the situation encountered in the main text, $\mathbf{K}_\mathrm{fwd}$ and $\mathbf{K}_\mathrm{bwd}$ are not completely independent because states $|\Psi_{k,1}\rangle$ and $|\Psi_{k,2N_t}\rangle$ are ``in contact'' with states $|\Psi_{k,2N_t+N_\tau}\rangle$ and $|\Psi_{k,2N_t+1}\rangle$ on the Matsubara branch. Therefore, the relation between $\mathbf{K}_\mathrm{fwd}$ and $\mathbf{K}_\mathrm{bwd}$ is determined by the momentum-conservation law along the Matsubara branch, which is formulated in the reduced Brillouin zone. Namely, the momenta (in the reduced Brillouin zone!) of $N_\tau$ momentum-space states $|\Psi_{k,2N_t+1}\rangle,\dots,|\Psi_{k,2N_t+N_\tau}\rangle$ are all equal to $\widetilde{\mathbf{K}}_\mathrm{M}$. The momenta $\mathbf{K}_\mathrm{fwd}$, $\mathbf{K}_\mathrm{bwd}$, and $\widetilde{\mathbf{K}}_\mathrm{M}$ are related as follows:
$$\mathbf{K}_\mathrm{fwd}\cdot\mathbf{e}_y=\mathbf{K}_\mathrm{bwd}\cdot\mathbf{e}_y=\widetilde{\mathbf{K}}_\mathrm{M}\cdot\mathbf{e}_y,$$
$$\frac{\mathbf{K}_\mathrm{fwd}\cdot\mathbf{e}_x}{2\pi/N_x}\:\mathrm{mod}\:\frac{N_x}{\lambda}=\frac{\widetilde{\mathbf{K}}_\mathrm{M}\cdot\mathbf{e}_x}{2\pi/N_x},\quad \frac{\mathbf{K}_\mathrm{bwd}\cdot\mathbf{e}_x}{2\pi/N_x}\:\mathrm{mod}\:\frac{N_x}{\lambda}=\frac{\widetilde{\mathbf{K}}_\mathrm{M}\cdot\mathbf{e}_x}{2\pi/N_x}$$

Due to the more complicated momentum-conservation law, MC updates presented in Sec.~\ref{Sec:MC_updates} have to be amply modified. Instead of describing modified MC updates in detail, we demonstrate the correctness of our implementation by benchmarking it on small clusters. Motivated by Ref.~\onlinecite{Science.363.379}, we limit the discussion to the electron occupation dynamics on individual sites.

\begin{figure}[htbp!]
 \centering
 \includegraphics[scale=0.5]{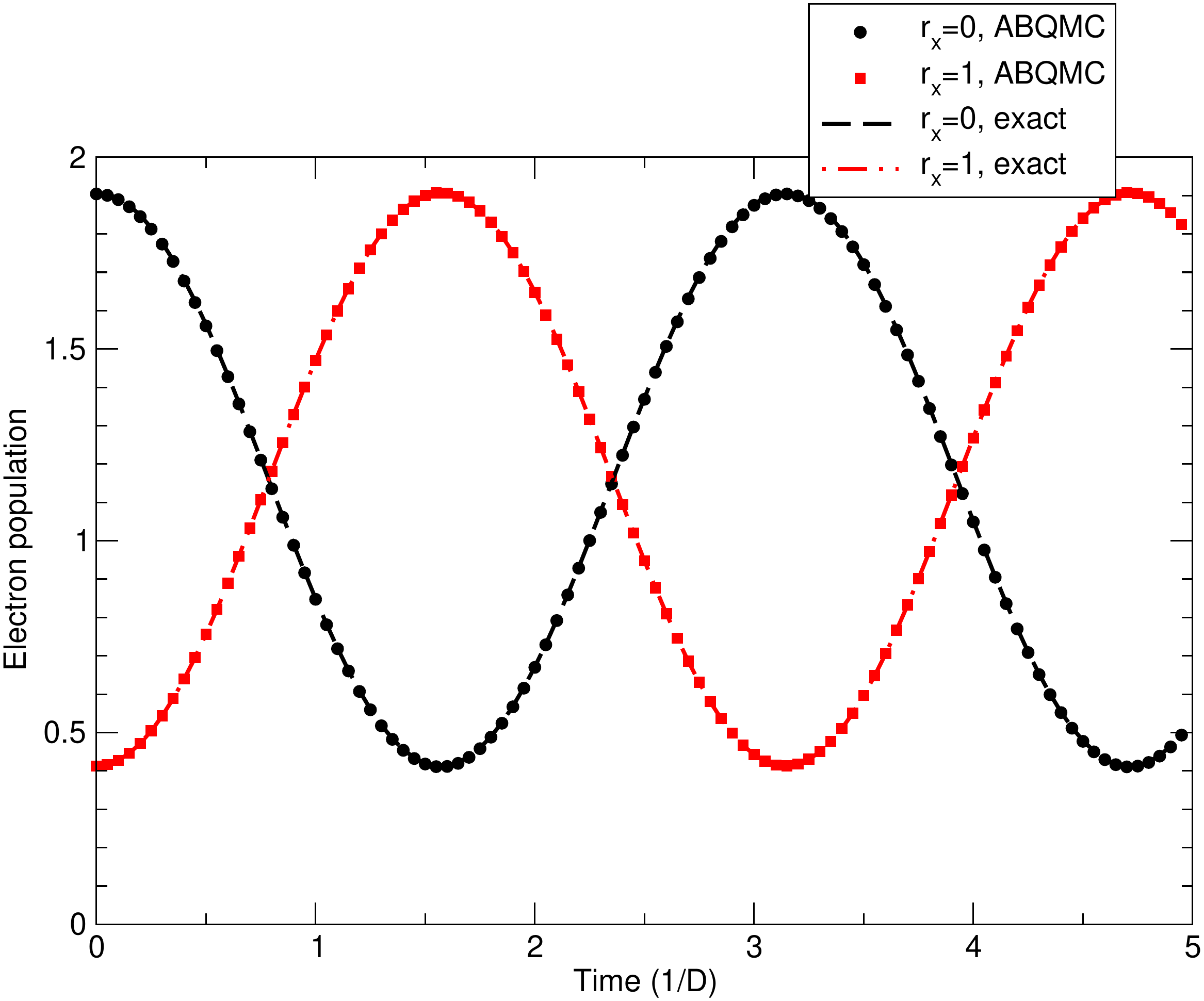}
 \caption{Time-dependent site populations of the Hubbard dimer with the following values of model parameters: $\mu/D=1.3,\:V_0/D=2,\:U=0,T/D=0.57$. The external potential at $t<0$ is $v_{r_x}=V_0\cos(\pi r_x)$ with $r_x=0,1$.}
 \label{Fig:slika_dimer}
\end{figure}

\begin{figure}[htbp!]
 \centering
 \includegraphics[scale=0.5]{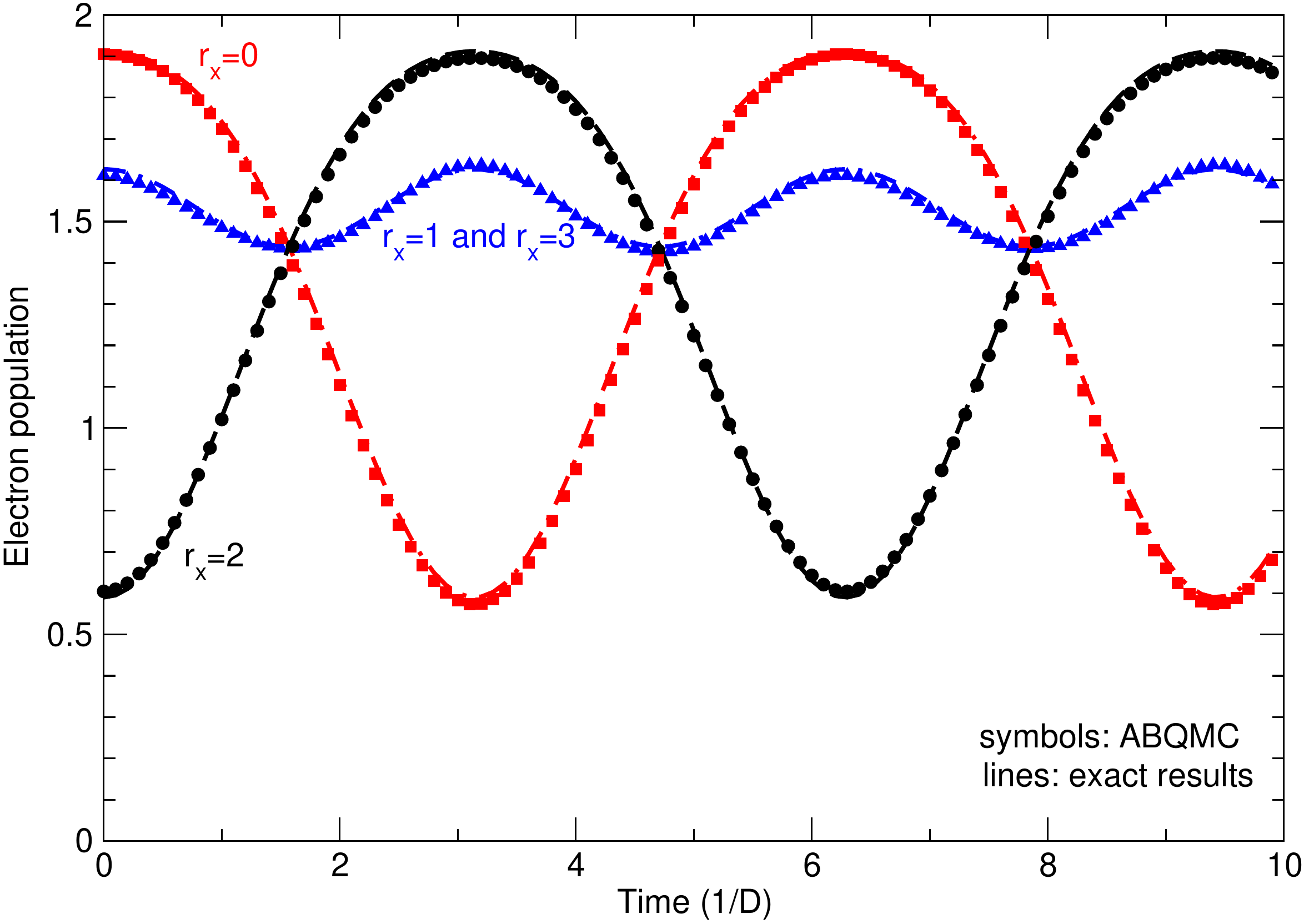}
 \caption{Time-dependent site populations of the Hubbard tetramer with the following values of model parameters: $\mu/D=0.65,\:V_0/D=1,\:U=0,\:T/D=0.285$. The external potential at $t<0$ is $v_{r_x}=V_0\cos(\pi r_x/2)$ with $r_x=0,\:1,\:2,\:3$.}
 \label{Fig:tetramer_lambda_4}
\end{figure}

\begin{figure}[htbp!]
 \centering
 \includegraphics[scale=0.5]{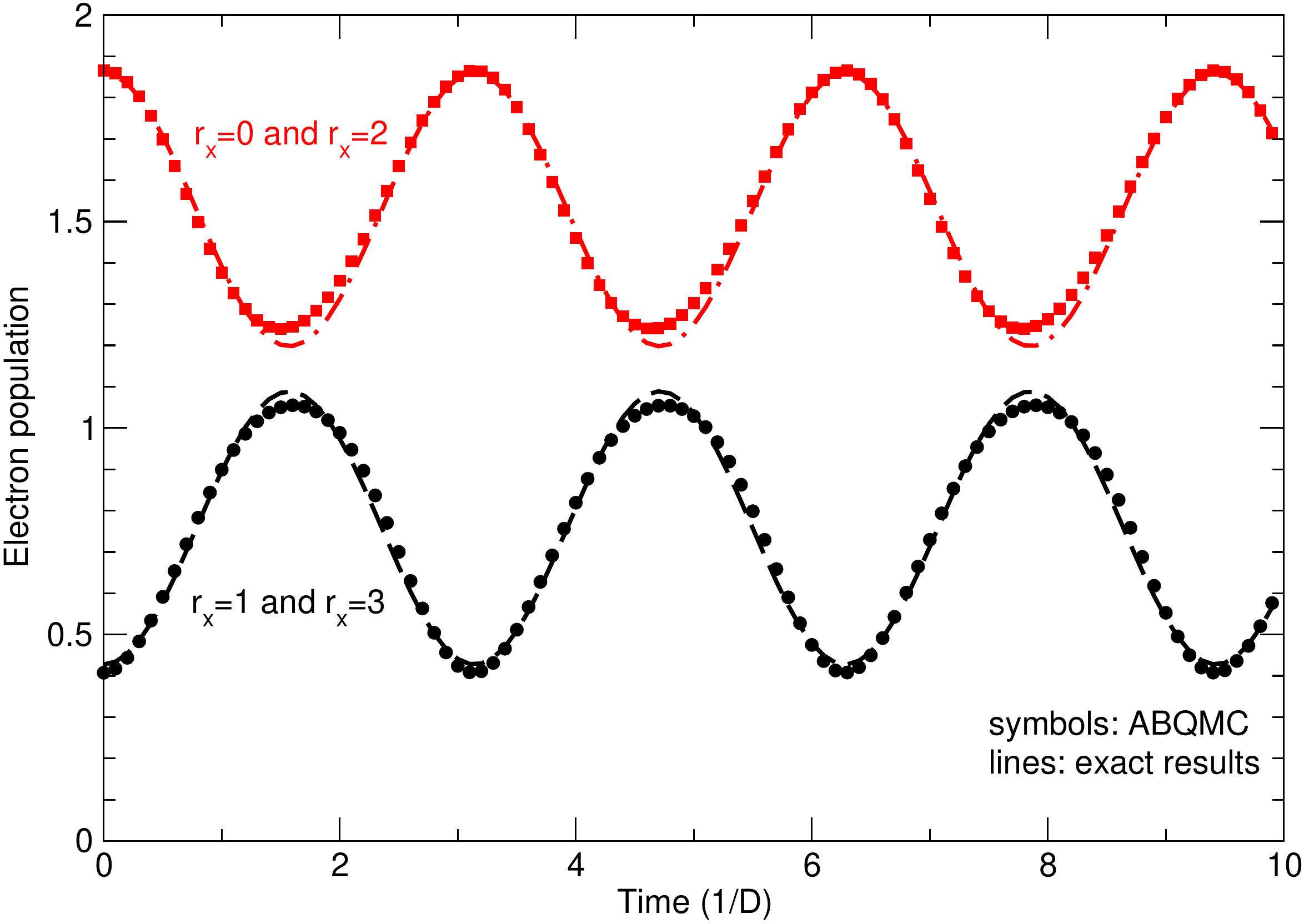}
 \caption{Time-dependent site populations of the Hubbard teteramer with the following values of model parameters: $\mu/D=0.65,\:V_0/D=1,\:U=0,\:T/D=0.285$. The external potential at $t<0$ is $v_{r_x}=V_0\cos(\pi r_x)$ with $r_x=0,\:1,\:2,\:3$.}
 \label{Fig:tetramer_lambda_2}
\end{figure}

We start from the noninteracting electrons, where already $N_\tau=1$ imaginary-time slice and $2N_t=2$ real-time slices ($2N_t+N_\tau=3$ slices in total) are expected to reproduce the exact result. Trivial as they may seem, the benchmarks on the noninteracting case are quite important, because densities of individual sites are expected to display nontrivial oscillatory behavior. The fact that our ABQMC results reproduce these oscillations quite accurately strongly suggests that our implementation is correct. In Fig.~\ref{Fig:slika_dimer} we present results for the Hubbard dimer initially subjected to the external density-modulating field $v_{r_x}=V_0\cos(\pi r_x)$ with $r_x=0,1$. Figure~\ref{Fig:tetramer_lambda_4} displays results for the Hubbard tetramer initially subjected to the external density-modulating field of wavelength $\lambda=4$, $v_{r_x}=V_0\cos(\pi r_x/2)$, with $r_x=0,\:1,\:2,\:3$. Figure~\ref{Fig:tetramer_lambda_2} displays results for the Hubbard tetramer initially subjected to the external density-modulating field of wavelength $\lambda=2$, $v_{r_x}=V_0\cos(\pi r_x)$, with $r_x=0,\:1,\:2,\:3$.  

\newpage

\begin{figure}[htbp!]
 \centering
 \includegraphics[scale=0.5]{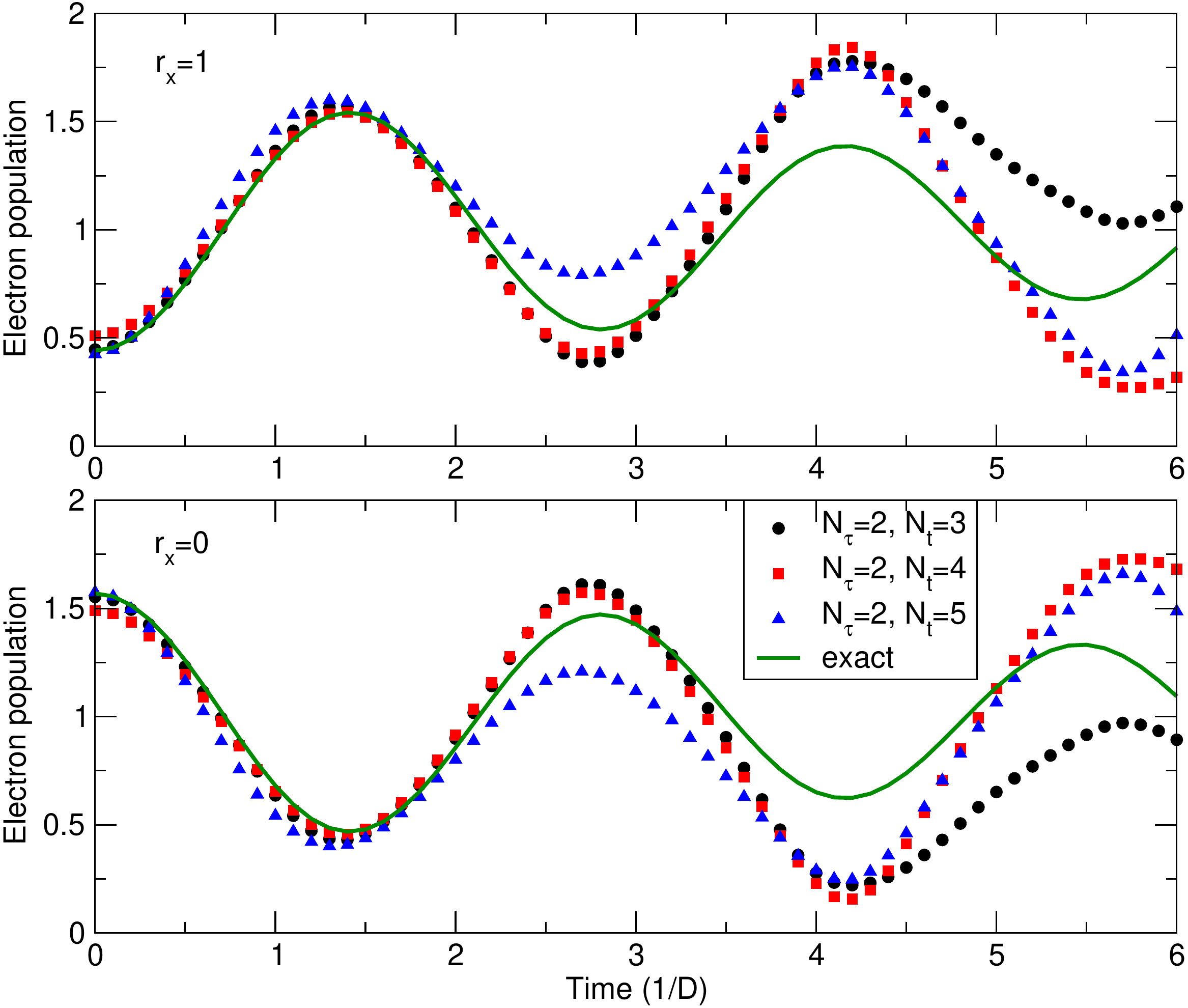}
 \caption{Dynamics of electron populations on individual sites of the Hubbard dimer that at $t<0$ is subjected to the density-modulating potential $v_{r_x}=V_0\cos(\pi r_x)$ with $r_x=0,\:1$. We work in the canonical ensemble with $N_\uparrow=N_\downarrow=1$. The model parameters assume the following values: $V_0/D=1$, $U/D=0.6$, $T/D=0.45$.}
 \label{Fig:dimer_canonical}
\end{figure}

\begin{figure}[htbp!]
 \centering
 \includegraphics[scale=0.5]{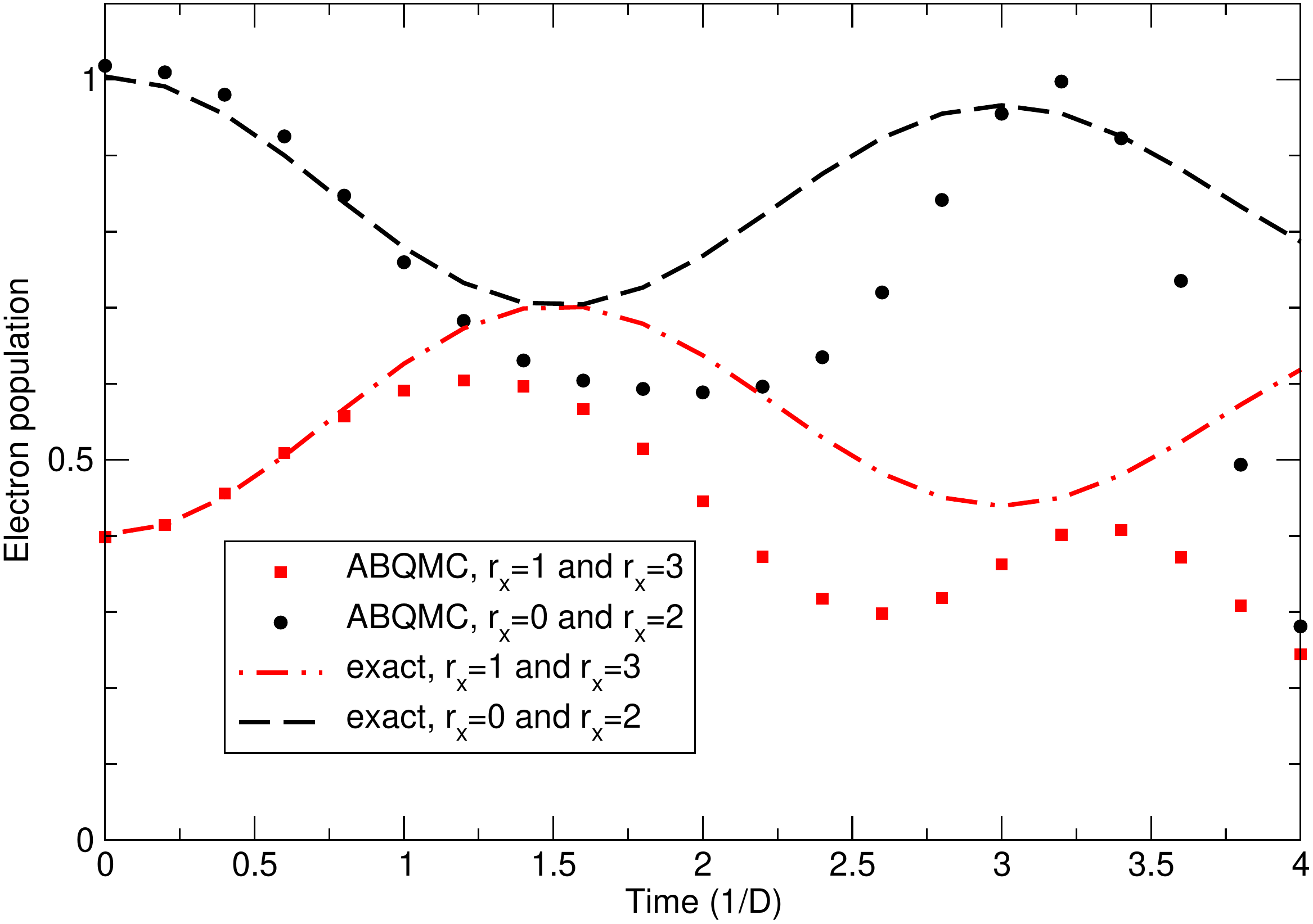}
 \caption{Dynamics of electron populations on individual sites of the Hubbard tetramer that at $t<0$ is subjected to the density-modulating potential $v_{r_x}=V_0\cos(\pi r_x)$ with $r_x=0,\:1,\:2,\:3$. We work in the grand-canonical ensemble with $N_t=1$, $N_\tau=2$ ($2N_t+N_\tau=4$ slices in total), and the following values of model parameters: $\mu/D=-0.325$, $V_0/D=0.5$, $U/D=0.5$, $T/D=0.5$.}
 \label{Fig:tetramer_grand_canonical}
\end{figure}

We conclude this section by applying the ABQMC algorithm to interacting electrons.
We first discuss the Hubbard dimer in the canonical ensemble, where we can obtain converged results with as many as $2N_t+N_\tau=12$ slices in total at half-filling ($N_\uparrow=N_\downarrow=1$). The results are presented in Fig.~\ref{Fig:dimer_canonical}. We observe that ABQMC results qualitatively reproduce the exact result in the whole time window considered. The quantitative agreement is reasonable up to $Dt\approx 2$. Figure~\ref{Fig:tetramer_grand_canonical} presents results for the Hubbard tetramer in the grand-canonical ensemble, where we obtain converged results using only $N_t=1$ and $N_\tau=2$. Further increase in $N_t$ reduces the average sign by orders of magnitude: for $N_t=1$, $N_\tau=2$ we obtain $|\langle\mathrm{sgn}\rangle|=1.2\times 10^{-2}$, while for $N_t=2$, $N_\tau=2$ we find $|\langle\mathrm{sgn}\rangle|\sim 10^{-4}$. While $N_\tau=2$ is enough to reproduce equilibrium populations in the external field at $t=0$, a single real-time slice leads to the quantitative agreement between the ABQMC and exact results only at shortest times.

\newpage

\bibliography{apssamp}